\documentclass[12pt]{article}
\usepackage{amssymb}
\usepackage{amsmath}
\usepackage{graphicx}
\usepackage{indentfirst}
\usepackage{cite}

\allowdisplaybreaks

\begin{document}

\title{Cross sections for inelastic meson-meson scattering \\ 
via quark-antiquark annihilation}
\author{Zhen-Yu Shen$^1$, Xiao-Ming Xu$^1$, and H. J. Weber$^2$}
\date{}
\maketitle \vspace{-1cm}
\centerline{$^1$Department of Physics, Shanghai University, Baoshan, 
Shanghai 200444, China}
\centerline{$^2$Department of Physics, University of Virginia, Charlottesville,
VA 22904, USA}

\begin{abstract}
We study inelastic meson-meson scattering that is governed by quark-antiquark 
annihilation and creation involving a quark and an antiquark annihilating into 
a gluon, and subsequently the gluon creating another quark-antiquark pair. The 
resultant hadronic reactions include for $I=1: \pi \pi \to \rho \rho$, $K \bar 
{K} \to K^* \bar {K}^\ast$, $K \bar{K}^\ast \to K^* \bar{K}^\ast$, $K^\ast 
\bar{K} \to K^* \bar{K}^\ast$, as well as $\pi \pi \to K \bar K$, $\pi \rho \to 
K \bar {K}^\ast$, $\pi \rho \to K^* \bar{K}$, and $K \bar {K} \to \rho \rho$. 
In each reaction, one or two Feynman diagrams are involved in the Born 
approximation. We derive formulas for the unpolarized cross section, the 
transition amplitude, and the transition potential for quark-antiquark 
annihilation and creation. The unpolarized cross sections for the reactions 
are calculated at six temperatures, and prominent temperature dependence is found. 
It is due to differences among mesonic temperature dependence in hadronic matter.
\end{abstract}



\vspace{0.5cm}
\leftline{\bf I. INTRODUCTION}
\vspace{0.5cm}

All possible meson-meson scattering takes place in hadronic matter that is created in
ultrarelativistic heavy-ion collisions. Yet, no hadronic meson beam collision 
experiments are possible. Therefore, we need to study meson-meson scattering 
theoretically. In Refs.~\cite{LX,SX} we have treated in the first Born approximation  
the quark interchange mechanism~\cite{BS} in the endothermic nonresonant reactions  
$\pi\pi \to \rho\rho$ for $I=2$, $KK \to K^* K^*$ for $I=1$, $KK^* \to K^*K^*$ for 
$I=1$, $\pi K \to \rho K^*$ for $I=3/2$, $\pi K^* \to \rho K^*$ for $I=3/2$, 
$\rho K \to \rho K^*$ for $I=3/2$, and $\pi K^* \to \rho K$ for $I=3/2$. In some  
regimes, these reactions are governed by quark interchange. Cross sections for these 
reactions change considerably with temperature of hadronic matter. For example, every
reaction has a rising peak cross section when the temperature becomes critical. Clearly, 
meson-meson scattering in hadronic matter has interesting features that need to be studied.

Meson-meson scattering can be mediated not only by quark interchange but
also by quark-antiquark annihilation and resonances. In the present work, we
concentrate on meson-meson scattering that happens through quark-antiquark 
annihilation. Starting from an effective Lagrangian, cross sections for $\pi \pi \to K 
\bar K$, $\rho \rho \to K \bar K$, $\pi \rho \to K \bar {K}^*$, and
$\pi \rho \to K^* \bar {K}$ in vacuum through one-meson exhange
have been obtained in Ref. \cite{BKWX}. Many other reactions are also possible 
in hadronic matter. Here we consider not only the four reactions 
$\pi \pi \to K \bar K$, $K \bar{K} \to \rho \rho$, $\pi \rho \to K \bar {K}^*$, and
$\pi \rho \to K^\ast \bar {K}$, but also $K \bar {K} \to K^* \bar {K}^\ast$, 
$K \bar {K}^\ast \to K^* \bar {K}^\ast$, $K^* \bar {K} \to K^* \bar {K}^\ast$, 
and $\pi \pi \to \rho \rho$ for $I=1$, which are governed by quark-antiquark 
annihilation. These reactions are important in the evolution of hadronic matter 
created in ultrarelativistic heavy-ion collisions~\cite{UrQMD,CBMTS,LXG}.

In the first Born approximation, we consider the annihilation of a 
quark-antiquark pair into a gluon and the creation of another quark-antiquark
pair from the gluon. Gluon propagation has already been taken into
account in $p\bar p$ annihilation into mesons. Different approximations to
the gluon propagation have been made to derive transition potentials in 
Refs.~\cite{KW,TMM,THT}. These are the $^3S_1$ models. The $p\bar p$ annihilation 
into mesons has also been studied in the $^3P_0$ model where a quark and an 
antiquark annihilate into the vacuum and another quark-antiquark pair is 
created from the vacuum~\cite{DF,MFF}. This nonperturbative mechanism is 
not considered here.  

This paper is organized as follows. In Sect.~II we derive the formulas of
unpolarized cross section for meson-meson reactions that are governed by the
annihilation and creation of a quark-antiquark pair. In Sect.~III we derive a
transition potential for the annihilation and creation of a quark-antiquark
pair. In Sect.~IV  we present transition amplitudes and calculate color matrix
elements, spin matrix elements, and flavor matrix elements. In Sect.~V we
provide mesonic quark-antiquark relative-motion wave functions. In Sect.~VI we
calculate elastic phase shifts for $\pi \pi$ scattering for $I=0$ and
$I=1$, calculate unpolarized cross sections for nine channels of
inelastic meson-meson scattering that is governed by quark-antiquark 
annihilation and creation, and give relevant discussions. In Sect.~VII we 
summarize the present work.

\vspace{0.5cm}
\leftline{\bf II. CROSS-SECTION FORMULAS}
\vspace{0.5cm}

If the quark $q_1$ of meson 
$A(q_1\bar{q}_1)$ has the same flavor as the antiquark $\bar{q}_2$ of meson
$B(q_2\bar{q}_2)$ and/or the antiquark $\bar{q}_1$ and the quark $q_2$ have the
same flavor, the reaction $A+B \to C+D$ may take place through quark-antiquark
annihilation and creation as shown in Fig.~1. Corresponding to the two Feynman 
diagrams, the $S$-matrix element for $A+B \to C+D$ is
\begin{eqnarray}
S_{\rm fi} & = & \delta_{\rm fi} - 2\pi i \delta (E_{\rm f} - E_{\rm i})
\nonumber \\
& &
(<q_3\bar {q}_1,q_2\bar {q}_4 \mid V_{{\rm a}q_1\bar {q}_2} 
\mid q_1\bar {q}_1,q_2\bar {q}_2> +
<q_1\bar {q}_4,q_3\bar {q}_2 \mid V_{{\rm a}\bar {q}_1 q_2}
\mid q_1\bar {q}_1,q_2\bar {q}_2>),
\end{eqnarray}
where $E_{\rm i}$ ($E_{\rm f}$) is the total energy of the two initial (final)
mesons; $V_{{\rm a}q_1\bar{q}_2}$ and $V_{{\rm a}\bar{q}_1q_2}$ are the 
transition potentials for $q_1 + \bar{q}_2 \to q_3 + \bar{q}_4$ and
$\bar{q}_1 + q_2 \to q_3 + \bar{q}_4$, respectively; mesons $C$ and $D$ are
individually
$q_3\bar{q}_1$ and $q_2\bar{q}_4$ in the left diagram and are $q_1\bar{q}_4$
and $q_3\bar{q}_2$ in the right diagram.
The wave function of the initial mesons is 
\begin{equation}
\psi_{q_1\bar {q}_1, q_2\bar {q}_2}=
\frac {e^{i\vec {P}_{q_1\bar {q}_1}\cdot \vec {R}_{q_1\bar {q}_1}}}{\sqrt V} 
\psi_{q_1\bar {q}_1} (\vec {r}_{q_1\bar {q}_1})
\frac {e^{i\vec {P}_{q_2\bar {q}_2}\cdot \vec {R}_{q_2\bar {q}_2}}}{\sqrt V} 
\psi_{q_2\bar {q}_2} (\vec {r}_{q_2\bar {q}_2}),
\end{equation}
where $\vec {P}_{q_1\bar {q}_1}$ ($\vec {P}_{q_2\bar {q}_2}$),
$\vec {R}_{q_1\bar {q}_1}$ ($\vec {R}_{q_2\bar {q}_2}$), and
$\vec {r}_{q_1\bar {q}_1}$ ($\vec {r}_{q_2\bar {q}_2}$) are the total 
momentum, the center-of-mass coordinate, and the relative coordinate of $q_1$
($q_2$) and $\bar {q}_1$ ($\bar {q}_2$), respectively.
$\psi_{ab} (\vec {r}_{ab})$ is the product of the color wave function, the spin
wave function, the flavor wave function, and the relative-motion wave function
of constituents $a$ and $b$. The wave function of the final mesons is
\begin{equation}
\psi_{q_3\bar {q}_1, q_2\bar {q}_4}=
\frac {e^{i\vec {P}_{q_3\bar {q}_1}\cdot \vec {R}_{q_3\bar {q}_1}}}{\sqrt V} 
\psi_{q_3\bar {q}_1} (\vec {r}_{q_3\bar {q}_1})
\frac {e^{i\vec {P}_{q_2\bar {q}_4}\cdot \vec {R}_{q_2\bar {q}_4}}}{\sqrt V} 
\psi_{q_2\bar {q}_4} (\vec {r}_{q_2\bar {q}_4}),
\end{equation}
corresponding to the left diagram, or
\begin{equation}
\psi_{q_1\bar {q}_4, q_3\bar {q}_2}=
\frac {e^{i\vec {P}_{q_1\bar {q}_4}\cdot \vec {R}_{q_1\bar {q}_4}}}{\sqrt V} 
\psi_{q_1\bar {q}_4} (\vec {r}_{q_1\bar {q}_4})
\frac {e^{i\vec {P}_{q_3\bar {q}_2}\cdot \vec {R}_{q_3\bar {q}_2}}}{\sqrt V} 
\psi_{q_3\bar {q}_2} (\vec {r}_{q_3\bar {q}_2}),
\end{equation}
corresponding to the right diagram.
$\vec {P}_{q_3\bar {q}_1}$ ($\vec {P}_{q_2\bar {q}_4}$,
$\vec {P}_{q_1\bar {q}_4}$, $\vec {P}_{q_3\bar {q}_2}$),
$\vec {R}_{q_3\bar {q}_1}$ ($\vec {R}_{q_2\bar {q}_4}$,
$\vec {R}_{q_1\bar {q}_4}$, $\vec {R}_{q_3\bar {q}_2}$), and
$\vec {r}_{q_3\bar {q}_1}$ ($\vec {r}_{q_2\bar {q}_4}$,
$\vec {r}_{q_1\bar {q}_4}$, $\vec {r}_{q_3\bar {q}_2}$) are the total 
momentum, the center-of-mass coordinate, and the relative coordinate of $q_3$
and $\bar {q}_1$ ($q_2$ and  $\bar {q}_4$, $q_1$ and  $\bar {q}_4$,
$q_3$ and  $\bar {q}_2$), respectively. Every meson wave function is normalized
in the volume $V$.

We first deal with 
\begin{eqnarray}
& & <q_3\bar {q}_1,q_2\bar {q}_4 \mid V_{{\rm a}q_1\bar {q}_2}
\mid q_1\bar {q}_1,q_2\bar {q}_2>
             \nonumber    \\
& = & \int d\vec{r}_{q_1} d\vec{r}_{\bar{q}_1} d\vec{r}_{q_2} d\vec{r}_{q_3}
\frac {e^{-i\vec {P}_{q_3\bar {q}_1}\cdot \vec {R}_{q_3\bar {q}_1}}}{\sqrt V} 
\psi_{q_3\bar {q}_1}^+ (\vec {r}_{q_3\bar {q}_1})
\frac {e^{-i\vec {P}_{q_2\bar {q}_4}\cdot \vec {R}_{q_2\bar {q}_4}}}{\sqrt V} 
\psi_{q_2\bar {q}_4}^+ (\vec {r}_{q_2\bar {q}_4})
             \nonumber    \\
& &
V_{{\rm a}q_1\bar{q}_2}
\frac {e^{i\vec {P}_{q_1\bar {q}_1}\cdot \vec {R}_{q_1\bar {q}_1}}}{\sqrt V} 
\psi_{q_1\bar {q}_1} (\vec {r}_{q_1\bar {q}_1})
\frac {e^{i\vec {P}_{q_2\bar {q}_2}\cdot \vec {R}_{q_2\bar {q}_2}}}{\sqrt V} 
\psi_{q_2\bar {q}_2} (\vec {r}_{q_2\bar {q}_2}).
\end{eqnarray}
Let $\vec {R}_{\rm total}$ be the center-of-mass coordinate of the two 
initial or final mesons, $\vec {P}_{\rm i}$ ($\vec {P}_{\rm f}$) the 
total three-dimensional momentum of the two initial (final) mesons, and
$m_{q_1\bar{q}_1}$ ($m_{q_2\bar{q}_2}$, $m_{q_3\bar{q}_1}$, $m_{q_2\bar{q}_4}$,
$m_{q_1\bar{q}_4}$, $m_{q_3\bar{q}_2}$) the mass of 
$q_1\bar{q}_1$ ($q_2\bar{q}_2$, $q_3\bar{q}_1$, $q_2\bar{q}_4$,
$q_1\bar{q}_4$, $q_3\bar{q}_2$).
Denote the relative coordinate and the relative momentum of $q_1\bar {q}_1$ 
and $q_2\bar {q}_2$ by $\vec {r}_{q_1\bar {q}_1,q_2\bar {q}_2}$ and
$\vec {p}_{q_1\bar {q}_1,q_2\bar {q}_2}$, respectively. We have
\begin{equation}
\vec{R}_{q_1\bar{q}_1}=\vec {R}_{\rm total} 
+ \frac {m_{q_2\bar{q}_2}}{m_{q_1\bar{q}_1}+m_{q_2\bar{q}_2}}
\vec{r}_{q_1\bar {q}_1,q_2\bar {q}_2},
\end{equation}
\begin{equation}
\vec{R}_{q_2\bar{q}_2}=\vec {R}_{\rm total} 
- \frac {m_{q_1\bar{q}_1}}{m_{q_1\bar{q}_1}+m_{q_2\bar{q}_2}}
\vec{r}_{q_1\bar {q}_1,q_2\bar {q}_2},
\end{equation}
\begin{equation}
\vec {P}_{q_1\bar{q}_1}=\frac {m_{q_1\bar{q}_1}}
{m_{q_1\bar{q}_1}+m_{q_2\bar{q}_2}} \vec {P}_{\rm i}
+\vec{p}_{q_1\bar {q}_1,q_2\bar {q}_2},
\end{equation}
\begin{equation}
\vec {P}_{q_2\bar{q}_2}=\frac {m_{q_2\bar{q}_2}}
{m_{q_1\bar{q}_1}+m_{q_2\bar{q}_2}} \vec {P}_{\rm i}
-\vec{p}_{q_1\bar {q}_1,q_2\bar {q}_2}.
\end{equation}
Denote the relative coordinate and the 
relative momentum of $q_3\bar {q}_1$ and $q_2\bar {q}_4$ by
$\vec {r}_{q_3\bar {q}_1,q_2\bar {q}_4}$ and 
$\vec {p}_{q_3\bar {q}_1,q_2\bar {q}_4}$, respectively. We have
\begin{equation}
\vec{R}_{q_3\bar{q}_1}=\vec {R}_{\rm total} 
+ \frac {m_{q_2\bar{q}_4}}{m_{q_3\bar{q}_1}+m_{q_2\bar{q}_4}}
\vec{r}_{q_3\bar {q}_1,q_2\bar {q}_4},
\end{equation}
\begin{equation}
\vec{R}_{q_2\bar{q}_4}=\vec {R}_{\rm total} 
- \frac {m_{q_3\bar{q}_1}}{m_{q_3\bar{q}_1}+m_{q_2\bar{q}_4}}
\vec{r}_{q_3\bar {q}_1,q_2\bar {q}_4},
\end{equation}
\begin{equation}
\vec {P}_{q_3\bar{q}_1}=\frac {m_{q_3\bar{q}_1}}
{m_{q_3\bar{q}_1}+m_{q_2\bar{q}_4}} \vec {P}_{\rm f}
+\vec{p}_{q_3\bar {q}_1,q_2\bar {q}_4},
\end{equation}
\begin{equation}
\vec {P}_{q_2\bar{q}_4}=\frac {m_{q_2\bar{q}_4}}
{m_{q_3\bar{q}_1}+m_{q_2\bar{q}_4}} \vec {P}_{\rm f}
-\vec{p}_{q_3\bar {q}_1,q_2\bar {q}_4}.
\end{equation}
From Eqs.~(6), (7), (10), and (11) we obtain
\begin{equation}
\vec{r}_{q_1}=\vec{r}_{q_1\bar{q}_1} 
+\frac {m_{q_2}m_{q_3}}{m_{\bar{q}_1}(m_{q_2}+m_{\bar{q}_4})}
\vec{r}_{q_2\bar{q}_4}
+\frac {m_{q_3}+m_{\bar{q}_1}}{m_{\bar{q}_1}}\vec{R}_{q_3\bar{q}_1}
-\frac {m_{q_3}}{m_{\bar{q}_1}}\vec{R}_{q_2\bar{q}_4},
\end{equation}
\begin{equation}
\vec{r}_{\bar{q}_1}=
\frac {m_{q_2}m_{q_3}}{m_{\bar{q}_1}(m_{q_2}+m_{\bar{q}_4})}
\vec{r}_{q_2\bar{q}_4}
+\frac {m_{q_3}+m_{\bar{q}_1}}{m_{\bar{q}_1}}\vec{R}_{q_3\bar{q}_1}
-\frac {m_{q_3}}{m_{\bar{q}_1}}\vec{R}_{q_2\bar{q}_4},
\end{equation}
\begin{equation}
\vec{r}_{q_2}=
\frac {m_{\bar{q}_4}}{m_{q_2}+m_{\bar{q}_4}}\vec{r}_{q_2\bar{q}_4}
+\vec{R}_{q_2\bar{q}_4},
\end{equation}
\begin{equation}
\vec{r}_{q_3}=
-\frac {m_{q_2}}{m_{q_2}+m_{\bar{q}_4}}\vec{r}_{q_2\bar{q}_4}
+\vec{R}_{q_2\bar{q}_4},
\end{equation}
which lead to
\begin{eqnarray}
d\vec{r}_{q_1} d\vec{r}_{\bar{q}_1} d\vec{r}_{q_2} d\vec{r}_{q_3}
& = & \frac {(m_{q_3}+m_{\bar {q}_1})^3}{m_{\bar {q}_1}^3}
d\vec{r}_{q_1\bar{q}_1} d\vec{r}_{q_2\bar{q}_4} d\vec{R}_{q_3\bar{q}_1} 
d\vec{R}_{q_2\bar{q}_4}
          \nonumber  \\
& = & \frac {(m_{q_3}+m_{\bar {q}_1})^3}{m_{\bar {q}_1}^3}
d\vec{r}_{q_1\bar{q}_1} d\vec{r}_{q_2\bar{q}_4} 
d\vec{r}_{q_3\bar{q}_1,q_2\bar{q}_4} d\vec{R}_{\rm total},
\end{eqnarray}
where constituent $a$ has the mass $m_a$ and the position vector $\vec {r}_a$.
Eq.~(5) becomes
\begin{eqnarray}
<q_3\bar {q}_1,q_2\bar {q}_4 \mid & V_{{\rm a}q_1\bar {q}_2} &
\mid q_1\bar {q}_1,q_2\bar {q}_2>
        \nonumber   \\
& = & \frac {(m_{q_3}+m_{\bar{q}_1})^3}{m_{\bar{q}_1}^3}
\int d\vec{r}_{q_1\bar{q}_1} d\vec{r}_{q_2\bar{q}_4} 
d\vec{r}_{q_3\bar{q}_1,q_2\bar{q}_4} d\vec{R}_{\rm total}
        \nonumber   \\
& &
\frac {\psi_{q_3\bar {q}_1}^+ (\vec {r}_{q_3\bar {q}_1})}{\sqrt V}
\frac {\psi_{q_2\bar {q}_4}^+ (\vec {r}_{q_2\bar {q}_4})}{\sqrt V}
e^{-i\vec {P}_{\rm f} \cdot \vec {R}_{\rm total}
-i\vec {p}_{q_3\bar {q}_1,q_2\bar {q}_4}
\cdot \vec {r}_{q_3\bar {q}_1,q_2\bar {q}_4}}
       \nonumber   \\
& &
V_{{\rm a}q_1\bar{q}_2}
\frac {\psi_{q_1\bar {q}_1} (\vec {r}_{q_1\bar {q}_1})}{\sqrt V}
\frac {\psi_{q_2\bar {q}_2} (\vec {r}_{q_2\bar {q}_2})}{\sqrt V}
e^{i\vec {P}_{\rm i}\cdot \vec {R}_{\rm total}
+i\vec {p}_{q_1\bar {q}_1,q_2\bar {q}_2}
\cdot \vec {r}_{q_1\bar {q}_1,q_2\bar {q}_2}}
       \nonumber   \\
& = & \frac {(m_{q_3}+m_{\bar{q}_1})^3}{m_{\bar{q}_1}^3}
(2\pi)^3 \delta (\vec{P}_{\rm f} - \vec{P}_{\rm i})
\int d\vec{r}_{q_1\bar{q}_1} d\vec{r}_{q_2\bar{q}_4} 
d\vec{r}_{q_3\bar{q}_1,q_2\bar{q}_4} 
        \nonumber   \\
& &
\frac {\psi_{q_3\bar {q}_1}^+ (\vec {r}_{q_3\bar {q}_1})}{\sqrt V}
\frac {\psi_{q_2\bar {q}_4}^+ (\vec {r}_{q_2\bar {q}_4})}{\sqrt V}
V_{{\rm a}q_1\bar{q}_2}
\frac {\psi_{q_1\bar {q}_1} (\vec {r}_{q_1\bar {q}_1})}{\sqrt V}
\frac {\psi_{q_2\bar {q}_2} (\vec {r}_{q_2\bar {q}_2})}{\sqrt V}
       \nonumber   \\
& &
e^{i\vec {p}_{q_1\bar {q}_1,q_2\bar {q}_2}
\cdot \vec {r}_{q_1\bar {q}_1,q_2\bar {q}_2}
-i\vec {p}_{q_3\bar {q}_1,q_2\bar {q}_4}
\cdot \vec {r}_{q_3\bar {q}_1,q_2\bar {q}_4}}
       \nonumber   \\
& = & (2\pi)^3 \delta (\vec{P}_{\rm f} - \vec{P}_{\rm i})
\frac {{\cal M}_{{\rm a}q_1\bar{q}_2}}{V^2\sqrt {2E_A2E_B2E_C2E_D}},
\end{eqnarray}
where $E_A$ ($E_B$, $E_C$, $E_D$) is the energy of meson $A$ ($B$, $C$, $D$), 
and
\begin{eqnarray}
{\cal M}_{{\rm a}q_1\bar {q}_2} & = & 
\frac {(m_{q_3}+m_{\bar{q}_1})^3}{m_{\bar{q}_1}^3}
\sqrt {2E_A2E_B2E_C2E_D}
\int d\vec{r}_{q_1\bar{q}_1} d\vec{r}_{q_2\bar{q}_4} 
d\vec{r}_{q_3\bar{q}_1,q_2\bar{q}_4} 
        \nonumber   \\
& &
\psi_{q_3\bar {q}_1}^+ (\vec {r}_{q_3\bar {q}_1})
\psi_{q_2\bar {q}_4}^+ (\vec {r}_{q_2\bar {q}_4})
V_{{\rm a}q_1\bar{q}_2}
\psi_{q_1\bar {q}_1} (\vec {r}_{q_1\bar {q}_1})
\psi_{q_2\bar {q}_2} (\vec {r}_{q_2\bar {q}_2})
       \nonumber   \\
& &
e^{i\vec {p}_{q_1\bar {q}_1,q_2\bar {q}_2}
\cdot \vec {r}_{q_1\bar {q}_1,q_2\bar {q}_2}
-i\vec {p}_{q_3\bar {q}_1,q_2\bar {q}_4}
\cdot \vec {r}_{q_3\bar {q}_1,q_2\bar {q}_4}}.
\end{eqnarray}

We now address 
\begin{eqnarray}
& & <q_1\bar {q}_4,q_3\bar {q}_2 \mid V_{{\rm a}\bar {q}_1q_2}
\mid q_1\bar {q}_1,q_2\bar {q}_2>
\nonumber   \\
& = & \int d\vec{r}_{q_1} d\vec{r}_{q_2} d\vec{r}_{\bar {q}_2} d\vec{r}_{q_3}
\frac {e^{-i\vec {P}_{q_1\bar {q}_4}\cdot \vec {R}_{q_1\bar {q}_4}}}{\sqrt V} 
\psi_{q_1\bar {q}_4}^+ (\vec {r}_{q_1\bar {q}_4})
\frac {e^{-i\vec {P}_{q_3\bar {q}_2}\cdot \vec {R}_{q_3\bar {q}_2}}}{\sqrt V} 
\psi_{q_3\bar {q}_2}^+ (\vec {r}_{q_3\bar {q}_2})
\nonumber   \\
& &
V_{{\rm a}\bar {q}_1q_2}
\frac {e^{i\vec {P}_{q_1\bar {q}_1}\cdot \vec {R}_{q_1\bar {q}_1}}}{\sqrt V} 
\psi_{q_1\bar {q}_1} (\vec {r}_{q_1\bar {q}_1})
\frac {e^{i\vec {P}_{q_2\bar {q}_2}\cdot \vec {R}_{q_2\bar {q}_2}}}{\sqrt V} 
\psi_{q_2\bar {q}_2} (\vec {r}_{q_2\bar {q}_2}).
\end{eqnarray}
Let us denote the relative coordinate and the relative momentum of $q_1\bar {q}_4$
and $q_3\bar {q}_2$ by $\vec {r}_{q_1\bar {q}_4,q_3\bar {q}_2}$ and
$\vec {p}_{q_1\bar {q}_4,q_3\bar {q}_2}$, respectively. We have
\begin{equation}
\vec{R}_{q_1\bar{q}_4}=\vec {R}_{\rm total} 
+ \frac {m_{q_3\bar{q}_2}}{m_{q_1\bar{q}_4}+m_{q_3\bar{q}_2}}
\vec{r}_{q_1\bar {q}_4,q_3\bar {q}_2},
\end{equation}
\begin{equation}
\vec{R}_{q_3\bar{q}_2}=\vec {R}_{\rm total} 
- \frac {m_{q_1\bar{q}_4}}{m_{q_1\bar{q}_4}+m_{q_3\bar{q}_2}}
\vec{r}_{q_1\bar {q}_4,q_3\bar {q}_2},
\end{equation}
\begin{equation}
\vec {P}_{q_1\bar{q}_4}=\frac {m_{q_1\bar{q}_4}}
{m_{q_1\bar{q}_4}+m_{q_3\bar{q}_2}} \vec {P}_{\rm f}
+\vec{p}_{q_1\bar {q}_4,q_3\bar {q}_2},
\end{equation}
\begin{equation}
\vec {P}_{q_3\bar{q}_2}=\frac {m_{q_3\bar{q}_2}}
{m_{q_1\bar{q}_4}+m_{q_3\bar{q}_2}} \vec {P}_{\rm f}
-\vec{p}_{q_1\bar {q}_4,q_3\bar {q}_2}.
\end{equation}
From Eqs. (6), (7), (22), and (23) we obtain
\begin{equation}
\vec{r}_{q_1}=
-\frac {m_{\bar {q}_2}m_{\bar {q}_4}}{m_{q_1}(m_{q_3}+m_{\bar{q}_2})}
\vec{r}_{q_3\bar{q}_2}
+\frac {m_{q_1}+m_{\bar{q}_4}}{m_{q_1}}\vec{R}_{q_1\bar{q}_4}
-\frac {m_{\bar {q}_4}}{m_{\bar{q}_1}}\vec{R}_{q_3\bar{q}_2},
\end{equation}
\begin{equation}
\vec{r}_{q_2}=-\vec {r}_{q_1\bar {q}_1}
-\frac {m_{\bar {q}_2}m_{\bar {q}_4}}{m_{q_1}(m_{q_3}+m_{\bar{q}_2})}
\vec{r}_{q_3\bar{q}_2}
+\frac {m_{q_1}+m_{\bar{q}_4}}{m_{q_1}}\vec{R}_{q_1\bar{q}_4}
-\frac {m_{\bar {q}_4}}{m_{q_1}}\vec{R}_{q_3\bar{q}_2},
\end{equation}
\begin{equation}
\vec{r}_{\bar {q}_2}=
-\frac {m_{q_3}}{m_{q_3}+m_{\bar {q}_2}}\vec{r}_{q_3\bar{q}_2}
+\vec{R}_{q_3\bar{q}_2},
\end{equation}
\begin{equation}
\vec{r}_{q_3}=
\frac {m_{\bar {q}_2}}{m_{q_3}+m_{\bar{q}_2}}\vec{r}_{q_3\bar{q}_2}
+\vec{R}_{q_3\bar{q}_2},
\end{equation}
which lead to
\begin{eqnarray}
d\vec{r}_{q_1} d\vec{r}_{q_2} d\vec{r}_{\bar{q}_2} d\vec{r}_{q_3}
& = & \frac {(m_{q_1}+m_{\bar{q}_4})^3}{m_{q_1}^3}
d\vec{r}_{q_1\bar{q}_1} d\vec{r}_{q_3\bar{q}_2} d\vec{R}_{q_1\bar{q}_4} 
d\vec{R}_{q_3\bar{q}_2}
       \nonumber       \\
& = & \frac {(m_{q_1}+m_{\bar{q}_4})^3}{m_{q_1}^3}
d\vec{r}_{q_1\bar{q}_1} d\vec{r}_{q_3\bar{q}_2} 
d\vec{r}_{q_1\bar{q}_4,q_3\bar{q}_2} d\vec{R}_{\rm total}.
\end{eqnarray}
Then Eq.~(21) becomes
\begin{eqnarray}
<q_1\bar {q}_4,q_3\bar {q}_2 \mid & V_{{\rm a}\bar{q}_1q_2} &
\mid q_1\bar {q}_1,q_2\bar {q}_2>
        \nonumber   \\
& = & \frac {(m_{q_1}+m_{\bar{q}_4})^3}{m_{q_1}^3}
\int d\vec{r}_{q_1\bar{q}_1} d\vec{r}_{q_3\bar{q}_2} 
d\vec{r}_{q_1\bar{q}_4,q_3\bar{q}_2} d\vec{R}_{\rm total}
        \nonumber   \\
& &
\frac {\psi_{q_1\bar {q}_4}^+ (\vec {r}_{q_1\bar {q}_4})}{\sqrt V}
\frac {\psi_{q_3\bar {q}_2}^+ (\vec {r}_{q_3\bar {q}_2})}{\sqrt V}
e^{-i\vec {P}_{\rm f} \cdot \vec {R}_{\rm total}
-i\vec {p}_{q_1\bar {q}_4,q_3\bar {q}_2}
\cdot \vec {r}_{q_1\bar {q}_4,q_3\bar {q}_2}}
       \nonumber   \\
& &
V_{{\rm a}\bar{q}_1q_2}
\frac {\psi_{q_1\bar {q}_1} (\vec {r}_{q_1\bar {q}_1})}{\sqrt V}
\frac {\psi_{q_2\bar {q}_2} (\vec {r}_{q_2\bar {q}_2})}{\sqrt V}
e^{i\vec {P}_{\rm i}\cdot \vec {R}_{\rm total}
+i\vec {p}_{q_1\bar {q}_1,q_2\bar {q}_2}
\cdot \vec {r}_{q_1\bar {q}_1,q_2\bar {q}_2}}
       \nonumber   \\
& = & \frac {(m_{q_1}+m_{\bar{q}_4})^3}{m_{q_1}^3}
(2\pi)^3 \delta (\vec{P}_{\rm f} - \vec{P}_{\rm i})
\int d\vec{r}_{q_1\bar{q}_1} d\vec{r}_{q_3\bar{q}_2} 
d\vec{r}_{q_1\bar{q}_4,q_3\bar{q}_2} 
        \nonumber   \\
& &
\frac {\psi_{q_1\bar {q}_4}^+ (\vec {r}_{q_1\bar {q}_4})}{\sqrt V}
\frac {\psi_{q_3\bar {q}_2}^+ (\vec {r}_{q_3\bar {q}_2})}{\sqrt V}
V_{{\rm a}\bar{q}_1q_2}
\frac {\psi_{q_1\bar {q}_1} (\vec {r}_{q_1\bar {q}_1})}{\sqrt V}
\frac {\psi_{q_2\bar {q}_2} (\vec {r}_{q_2\bar {q}_2})}{\sqrt V}
       \nonumber   \\
& &
e^{i\vec {p}_{q_1\bar {q}_1,q_2\bar {q}_2}
\cdot \vec {r}_{q_1\bar {q}_1,q_2\bar {q}_2}
-i\vec {p}_{q_1\bar {q}_4,q_3\bar {q}_2}
\cdot \vec {r}_{q_1\bar {q}_4,q_3\bar {q}_2}}
       \nonumber   \\
& = & (2\pi)^3 \delta (\vec{P}_{\rm f} - \vec{P}_{\rm i})
\frac {{\cal M}_{{\rm a}\bar{q}_1q_2}}{V^2\sqrt {2E_A2E_B2E_C2E_D}},
\end{eqnarray}
where
\begin{eqnarray}
{\cal M}_{{\rm a}\bar{q}_1q_2} & = & 
\frac {(m_{q_1}+m_{\bar{q}_4})^3}{m_{q_1}^3}
\sqrt {2E_A2E_B2E_C2E_D}
\int d\vec{r}_{q_1\bar{q}_1} d\vec{r}_{q_3\bar{q}_2} 
d\vec{r}_{q_1\bar{q}_4,q_3\bar{q}_2} 
        \nonumber   \\
& &
\psi_{q_1\bar {q}_4}^+ (\vec {r}_{q_1\bar {q}_4})
\psi_{q_3\bar {q}_2}^+ (\vec {r}_{q_3\bar {q}_2})
V_{{\rm a}\bar{q}_1q_2}
\psi_{q_1\bar {q}_1} (\vec {r}_{q_1\bar {q}_1})
\psi_{q_2\bar {q}_2} (\vec {r}_{q_2\bar {q}_2})
       \nonumber   \\
& &
e^{i\vec {p}_{q_1\bar {q}_1,q_2\bar {q}_2}
\cdot \vec {r}_{q_1\bar {q}_1,q_2\bar {q}_2}
-i\vec {p}_{q_1\bar {q}_4,q_3\bar {q}_2}
\cdot \vec {r}_{q_1\bar {q}_4,q_3\bar {q}_2}}.
\end{eqnarray}

Meson $i$ $(i=A, B, C, D)$ has the mass $m_i$, the four-momentum 
$P_i=(E_{i},\vec{P}_i)$, and the angular momentum $J_i$ with
its magnetic projection quantum number $J_{iz}$.
The unpolarized cross section for $A+B \rightarrow C+D$ is
\begin{eqnarray}
\sigma^{\rm unpol}(\sqrt {s},T) & = & \frac {(2\pi)^4}
{4\sqrt {(P_A \cdot P_B)^2 - m_A^2m_B^2}}
     \int \frac {d^3P_C}{(2\pi)^32E_C} \frac {d^3P_D}{(2\pi)^32E_D}
\frac {1}{(2J_A+1)(2J_B+1)}
\nonumber     \\
& &
\sum\limits_{J_{Az}J_{Bz}J_{Cz}J_{Dz}}
\mid {\cal M}_{{\rm a}q_1\bar{q}_2}+{\cal M}_{{\rm a}\bar{q}_1q_2} \mid^2 
\delta (E_{\rm f} - E_{\rm i})
\delta (\vec {P}_{\rm f} - \vec {P}_{\rm i}),     
\end{eqnarray}
where $s$ is the Mandelstam variable given by
$s=(E_A+E_B)^2-(\vec{P}_A+\vec{P}_B)^2$, and $T$ is the temperature.
In the center-of-mass frame of $A$ and $B$ the three-dimensional momenta of
mesons $A$ and $C$ are $\vec P$ and $\vec {P}^\prime$, respectively. Denote
the angle between $\vec{P}$ and $\vec{P}'$ by $\theta$. The unpolarized cross
section is given by
\begin{eqnarray}
\sigma^{\rm unpol}(\sqrt {s},T) & = & \frac {1}{(2J_A+1)(2J_B+1)}
\frac{1}{32\pi s}\frac{|\vec{P}^{\prime }(\sqrt{s})|}{|\vec{P}(\sqrt{s})|}
              \nonumber    \\
& & \int_{0}^{\pi }d\theta \sum\limits_{J_{Az}J_{Bz}J_{Cz}J_{Dz}}
\mid {\cal M}_{{\rm a}q_1\bar{q}_2}+{\cal M}_{{\rm a}\bar{q}_1q_2} \mid^2 
\sin \theta.
\end{eqnarray}

\vspace{0.5cm}
\leftline{\bf III. TRANSITION POTENTIAL}
\vspace{0.5cm}

In the reaction $A+B \to C+D$ the quark of an initial meson annihilates with
the antiquark of another meson to produce a gluon, and 
subsequently this gluon creates a quark
to form a final meson and an antiquark to form another final meson. The 
quark-antiquark annihilation and creation, 
$q (p_1) + \bar{q} (-p_2) \to q^\prime (p_3) + \bar{q}^\prime (-p_4)$,
is shown in Fig.~2. The initial quark wave function is
\begin{equation}
\psi_q (\vec {p}_1,s_{qz})=\left(
\begin{array}{c}
G_1(\vec {p}_1)   \\
\frac {\vec {\sigma} \cdot \vec {p}_1}{2m_q} G_1(\vec {p}_1)
\end{array}
\right) \chi_{s_{qz}},
\end{equation}
and the final quark wave function is
\begin{equation}
\psi_{q'} (\vec {p}_3,s_{q'z})=\left(
\begin{array}{c}
G_3(\vec {p}_3)   \\
\frac {\vec {\sigma} \cdot \vec {p}_3}{2m_{q'}} G_3(\vec {p}_3)
\end{array}
\right) \chi_{s_{q^\prime z}},
\end{equation}
where $\vec \sigma$ are the Pauli matrices;
$\chi_{s_{qz}}$ and $\chi_{s_{q^\prime z}}$ are the spin wave functions
with the magnetic projection quantum numbers, $s_{qz}$ and $s_{q^\prime z}$, of
the quark spin, respectively. The initial antiquark wave function is
\begin{equation}
\psi_{\bar q} (\vec {p}_2,s_{\bar{q}z})=\left(
\begin{array}{c}
\frac {\vec {\sigma} \cdot \vec {p}_2}{2m_{\bar q}} G_2(\vec {p}_2)    \\
G_2(\vec {p}_2)
\end{array}
\right) \chi_{s_{\bar {q}z}},
\end{equation}
and the final antiquark wave function is
\begin{equation}
\psi_{\bar {q}'} (\vec {p}_4,s_{\bar{q}^\prime z})=\left(
\begin{array}{c}
\frac {\vec {\sigma} \cdot \vec {p}_4}{2m_{\bar {q}'}} G_4(\vec {p}_4)    \\
G_4(\vec {p}_4)
\end{array}
\right) \chi_{s_{\bar{q}^\prime z}},
\end{equation}
where $m_{\bar{q}}=m_q$, $m_{\bar{q}^\prime}=m_{q^\prime}$, and
$\chi_{s_{\bar {q}z}}$ and $\chi_{s_{\bar{q}^\prime z}}$ are the spin 
wave functions with the magnetic projection quantum numbers, $s_{\bar{q}z}$ 
and $s_{\bar{q}^\prime z}$, of the antiquark spin, respectively.
In Eqs.~(35)-(38) the color and flavor wave functions are suppressed.
According to the Feynman rules given in Ref.~\cite{Muta}, the amplitude for
the Feynman diagram in Fig.~2 is written as
\begin{equation}
{\cal M}_{\rm a}  = 
\frac {g_{\rm s}^2}{q^2}\bar{\psi}_{q^\prime}(\vec{p}_3,s_{q^\prime z}) 
\gamma_\tau T_e \psi_{\bar {q}^\prime}(\vec{p}_4,s_{\bar{q}^\prime z})
\bar {\psi}_{\bar q}(\vec{p}_2,s_{\bar{q}z}) \gamma_\tau T_e
\psi_q(\vec{p}_1,s_{qz}),
\label{39}
\end{equation}
where $g_{\rm s}$  is the gauge coupling constant,
$T_e$ $(e=1, \cdot \cdot \cdot, 8)$ are the $SU(3)$ color generators, and
$T_eT_e=\frac {\vec {\lambda} (34)}{2} \cdot \frac {\vec {\lambda} (21)}{2}$
with $\vec{\lambda}$ the Gell-Mann matrices. Repeated
color and space-time indices ($\tau$) are summed. Using these quark and 
antiquark wave functions, the amplitude to order of the inverse of the 
(constituent) quark mass squared is
\begin{eqnarray}
{\cal M}_{\rm a} & = &
\frac {g_{\rm s}^2}{k^2} \left[
\chi_{s_{q^\prime z}}^+ \chi_{s_{\bar{q}z}}^+ T_e T_e G_3 (\vec{p}_3) 
G_2 (\vec{p}_2)
\frac {\vec {\sigma}(34) \cdot \vec{k} \vec {\sigma}(21) \cdot \vec{k}}
{4m_{q^\prime} m_q} G_4 (\vec{p}_4) G_1 (\vec{p}_1) 
\chi_{s_{\bar{q}^\prime z}} \chi_{s_{qz}}   \right.
                   \nonumber     \\
& - &
\chi_{s_{q^\prime z}}^+ \chi_{s_{\bar{q}z}}^+ T_e T_e G_3 (\vec{p}_3) 
G_2 (\vec{p}_2)   \left(
\vec{\sigma} (34) \cdot \vec{\sigma} (21) +
\frac {\vec {\sigma}(21) \cdot \vec{p}_2 \vec {\sigma}(34) 
\cdot \vec {\sigma}(21) \vec {\sigma}(21) \cdot \vec{p}_1}{4m_q^2} \right.
                   \nonumber     \\
& + &
\left.  \left.
\frac {\vec {\sigma}(34) \cdot \vec{p}_3 \vec {\sigma}(34) 
\cdot \vec {\sigma}(21) \vec {\sigma}(34) \cdot \vec{p}_4}
{4m_{q^\prime}^2}  \right)   G_4 (\vec{p}_4) G_1 (\vec{p}_1)
\chi_{s_{\bar{q}^\prime z}} \chi_{s_{qz}}  \right].
\end{eqnarray}
From the expression of the amplitude we obtain
the transition potential for $q (p_1) + \bar{q} (-p_2) \to q^\prime (p_3)
+ \bar{q}^\prime (-p_4)$,
\begin{eqnarray}
V_{{\rm a}q\bar{q}}(\vec {k}) & = & 
\frac {g_{\rm s}^2}{k^2}\frac {\vec {\lambda}(34)}{2} \cdot
\frac {\vec {\lambda}(21)}{2} \left( 
\frac {\vec {\sigma}(34) \cdot \vec{k} \vec {\sigma}(21) \cdot \vec{k}}
{4m_{q^\prime} m_q} -\vec {\sigma}(34) \cdot \vec {\sigma}(21)  \right.
                       \nonumber     \\
& - &
\left. \frac {\vec {\sigma}(21) \cdot \vec{p}_2 \vec {\sigma}(34) 
\cdot \vec {\sigma}(21) \vec {\sigma}(21) \cdot \vec{p}_1}{4m_q^2} 
-\frac {\vec {\sigma}(34) \cdot \vec{p}_3 \vec {\sigma}(34) 
\cdot \vec {\sigma}(21) \vec {\sigma}(34) \cdot \vec{p}_4}
{4m_{q^\prime}^2} \right), ~~~~~~
\end{eqnarray}
where $\vec{\lambda}(34)$ ($\vec{\lambda}(21)$) mean that they have matrix
elements between the color wave functions of the final (initial) quark and 
the final (initial) antiquark, and
$\vec{\sigma}(34)$ ($\vec{\sigma}(21)$) mean that they have matrix elements
between the spin wave functions of the final (initial) quark and 
the final (initial) antiquark.

\vspace{0.5cm}
\leftline{\bf IV. TRANSITION AMPLITUDE}
\vspace{0.5cm}

In order to obtain the unpolarized cross section, we calculate the transition
amplitudes, ${\cal M}_{{\rm a}q_1\bar{q}_2}$ and 
${\cal M}_{{\rm a}\bar{q}_1q_2}$. We take the Fourier transform of the meson 
wave functions, $V_{{\rm a}q_1\bar{q}_2}$ in Eq.~(20), and 
$V_{{\rm a}\bar{q}_1q_2}$ in Eq.~(32):
\begin{equation}
\psi_{q_1\bar{q}_1}(\vec{r}_{q_1\bar{q}_1}) =
\int \frac {d^3p_{q_1\bar{q}_1}}{(2\pi)^3} \psi_{q_1\bar{q}_1}
(\vec {p}_{q_1\bar{q}_1})
e^{i\vec {p}_{q_1\bar{q}_1} \cdot \vec {r}_{q_1\bar{q}_1}},
\end{equation}
\begin{equation}
\psi_{q_2\bar{q}_2}(\vec{r}_{q_2\bar{q}_2}) =
\int \frac {d^3p_{q_2\bar{q}_2}}{(2\pi)^3} \psi_{q_2\bar{q}_2}
(\vec {p}_{q_2\bar{q}_2})
e^{i\vec {p}_{q_2\bar{q}_2} \cdot \vec {r}_{q_2\bar{q}_2}},
\end{equation}
\begin{equation}
\psi_{q_3\bar{q}_1}(\vec{r}_{q_3\bar{q}_1}) =
\int \frac {d^3p_{q_3\bar{q}_1}}{(2\pi)^3} \psi_{q_3\bar{q}_1}
(\vec {p}_{q_3\bar{q}_1})
e^{i\vec {p}_{q_3\bar{q}_1} \cdot \vec {r}_{q_3\bar{q}_1}},
\end{equation}
\begin{equation}
\psi_{q_2\bar{q}_4}(\vec{r}_{q_2\bar{q}_4}) =
\int \frac {d^3p_{q_2\bar{q}_4}}{(2\pi)^3} \psi_{q_2\bar{q}_4}
(\vec {p}_{q_2\bar{q}_4})
e^{i\vec {p}_{q_2\bar{q}_4} \cdot \vec {r}_{q_2\bar{q}_4}},
\end{equation}
\begin{equation}
\psi_{q_1\bar{q}_4}(\vec{r}_{q_1\bar{q}_4}) =
\int \frac {d^3p_{q_1\bar{q}_4}}{(2\pi)^3} \psi_{q_1\bar{q}_4}
(\vec {p}_{q_1\bar{q}_4})
e^{i\vec {p}_{q_1\bar{q}_4} \cdot \vec {r}_{q_1\bar{q}_4}},
\end{equation}
\begin{equation}
\psi_{q_3\bar{q}_2}(\vec{r}_{q_3\bar{q}_2}) =
\int \frac {d^3p_{q_3\bar{q}_2}}{(2\pi)^3} \psi_{q_3\bar{q}_2}
(\vec {p}_{q_3\bar{q}_2})
e^{i\vec {p}_{q_3\bar{q}_2} \cdot \vec {r}_{q_3\bar{q}_2}},
\end{equation}
\begin{equation}
V_{{\rm a}q_1\bar{q}_2}(\vec{r}_{q_3}-\vec{r}_{q_1}) =
\int \frac {d^3k}{(2\pi)^3} V_{{\rm a}q_1\bar{q}_2} (\vec {k})
e^{i\vec {k} \cdot (\vec{r}_{q_3}-\vec{r}_{q_1})},
\end{equation}
\begin{equation}
V_{{\rm a}\bar{q}_1q_2}(\vec{r}_{q_3}-\vec{r}_{q_2}) =
\int \frac {d^3k}{(2\pi)^3} V_{{\rm a}\bar{q}_1q_2} (\vec {k})
e^{i\vec {k} \cdot (\vec{r}_{q_3}-\vec{r}_{q_2})},
\end{equation}
where $\vec {p}_{ab}$ is the relative momentum of constituents $a$ and $b$.
The transition amplitudes are then calculated in momentum space by the
following expressions:
\begin{eqnarray}
{\cal M}_{{\rm a}q_1\bar {q}_2} & = &
\sqrt {2E_A2E_B2E_C2E_D}
\int \frac {d^3p_{q_1\bar{q}_1}}{(2\pi)^3}\frac {d^3p_{q_2\bar{q}_2}}{(2\pi)^3}
                       \nonumber         \\
& &
\psi^+_{q_3\bar {q}_1} (\vec {p}_{q_3\bar {q}_1})
\psi^+_{q_2\bar {q}_4} (\vec {p}_{q_2\bar {q}_4})
V_{{\rm a}q_1\bar{q}_2}(\vec{k})
\psi_{q_1\bar {q}_1} (\vec {p}_{q_1\bar {q}_1})
\psi_{q_2\bar {q}_2} (\vec {p}_{q_2\bar {q}_2}),
\end{eqnarray}
\begin{eqnarray}
{\cal M}_{{\rm a}\bar{q}_1q_2} & = &
\sqrt {2E_A2E_B2E_C2E_D}
\int \frac {d^3p_{q_1\bar{q}_1}}{(2\pi)^3}\frac {d^3p_{q_2\bar{q}_2}}{(2\pi)^3}
                       \nonumber         \\
& &
\psi^+_{q_1\bar {q}_4} (\vec {p}_{q_1\bar {q}_4})
\psi^+_{q_3\bar {q}_2} (\vec {p}_{q_3\bar {q}_2})
V_{{\rm a}\bar{q}_1q_2}(\vec{k})
\psi_{q_1\bar {q}_1} (\vec {p}_{q_1\bar {q}_1})
\psi_{q_2\bar {q}_2} (\vec {p}_{q_2\bar {q}_2}).
\end{eqnarray}
For convenience sake we also use the 
notation $\psi_A$, $\psi_B$, $\psi_C$, and $\psi_D$: 
$\psi_A = \psi_{q_1\bar{q}_1}$, $\psi_B = \psi_{q_2\bar{q}_2}$,
$\psi_C = \psi_{q_3\bar{q}_1} = \psi_{q_1\bar{q}_4}$, and
$\psi_D = \psi_{q_2\bar{q}_4} = \psi_{q_3\bar{q}_2}$. The wave functions of
mesons $A$, $B$, $C$, and $D$ are individually given by
\begin{equation}
\psi_A =\phi_{A\rm rel} \phi_{A\rm color} \phi_{A\rm flavor} \chi_{S_A S_{Az}},
\end{equation}
\begin{equation}
\psi_B =\phi_{B\rm rel} \phi_{B\rm color} \phi_{B\rm flavor} \chi_{S_B S_{Bz}},
\end{equation}
\begin{equation}
\psi_C =\phi_{C\rm rel} \phi_{C\rm color} \phi_{C\rm flavor} \chi_{S_C S_{Cz}},
\end{equation}
\begin{equation}
\psi_D =\phi_{D\rm rel} \phi_{D\rm color} \phi_{D\rm flavor} \chi_{S_D S_{Dz}},
\end{equation}
where $\phi_{A\rm rel}$
($\phi_{B\rm rel}$, $\phi_{C\rm rel}$, $\phi_{D\rm rel}$),
$\phi_{A\rm color}$ ($\phi_{B\rm color}$, $\phi_{C\rm color}$, 
$\phi_{D\rm color}$), $\phi_{A\rm flavor}$ ($\phi_{B\rm flavor}$, 
$\phi_{C\rm flavor}$, $\phi_{D\rm flavor}$), and $\chi_{S_A S_{Az}}$ 
($\chi_{S_B S_{Bz}}$, $\chi_{S_C S_{Cz}}$, $\chi_{S_D S_{Dz}}$)
are the quark-antiquark relative-motion wave function, the color wave function,
the flavor wave function, and the spin wave function of meson $A$ ($B$, $C$, 
$D$), respectively. The spin of meson $A$ ($B$, $C$, $D$) is $S_A$ ($S_B$, 
$S_C$, $S_D$) with its magnetic projection quantum number $S_{Az}$ 
($S_{Bz}$, $S_{Cz}$, $S_{Dz}$).
The transition amplitudes contain color, spin, and 
flavor matrix elements. The color matrix element is
\begin{equation}
{\cal M}_{{\rm a}q_1\bar{q}_2\rm c}=\phi_{C\rm color}^+ \phi_{D\rm color}^+
\frac {\vec{\lambda}(34)}{2} \cdot \frac {\vec{\lambda}(21)}{2}
\phi_{A\rm color} \phi_{B\rm color} =\frac {4}{9},
\end{equation}
for the left diagram of Fig.~1, and
\begin{equation}
{\cal M}_{{\rm a}\bar{q}_1q_2\rm c}=\phi_{C\rm color}^+ \phi_{D\rm color}^+
\frac {\vec{\lambda}(34)}{2} \cdot \frac {\vec{\lambda}(12)}{2}
\phi_{A\rm color} \phi_{B\rm color} =\frac {4}{9},
\end{equation}
for the right diagram of Fig.~1. The transition potential in Eq.~(41) includes
the Pauli matrices $\vec{\sigma} (21)$ and $\vec{\sigma} (34)$. 
The spin matrix elements of the Pauli matrices involved in the left diagram 
are listed in Tables 1-14. The matrix elements of $\sigma_1(21)$, 
$\sigma_3(21)$, $\sigma_1(34)$, $\sigma_3(34)$, $\sigma_1(21)\sigma_1(34)$,
$\sigma_1(21)\sigma_3(34)$, $\sigma_2(21)\sigma_2(34)$, 
$\sigma_3(21)\sigma_1(34)$, and $\sigma_3(21)\sigma_3(34)$ are real, and the
matrix elements of $\sigma_2(21)$, $\sigma_2(34)$, $\sigma_1(21)\sigma_2(34)$,
$\sigma_2(21)\sigma_1(34)$, $\sigma_2(21)\sigma_3(34)$, and
$\sigma_3(21)\sigma_2(34)$ are pure imaginary or zero. The spin matrix
elements for the right diagram in Fig. 1 can be obtained from the ones for the 
left diagram. The real (imaginary) spin matrix elements of
${\cal M}_{{\rm a}\bar{q}_1q_2}$ equal the real spin matrix elements 
(equal the negative of the imaginary spin matrix elements) of 
${\cal M}_{{\rm a}q_1\bar{q}_2}$, while none, two, or four of the initial and
final mesons have zero spins. The real (imaginary) spin matrix elements of
${\cal M}_{{\rm a}\bar{q}_1q_2}$ equal the negative of the real spin matrix 
elements (equal the imaginary spin matrix elements) of 
${\cal M}_{{\rm a}q_1\bar{q}_2}$, when one or three of the initial and
final mesons have zero spins.

The flavor states, $\phi_{A\rm flavor}$ and 
$\phi_{B\rm flavor}$, are coupled to the flavor state $\mid AB, I, I_z>$
with the total isospin $I$ and its magnetic projection quantum number $I_z$.
$\phi_{C\rm flavor}$ and $\phi_{D\rm flavor}$ are coupled to 
$\mid CD, I, I_z>$. The flavor matrix element is
\begin{equation}
{\cal M}_{{\rm a}q_1\bar{q}_2\rm f}=<CD, I, I_z \mid P_{q_1 + \bar{q}_2 \to
q_3 +\bar{q}_4} \mid AB, I, I_z>,
\end{equation}
for the left diagram of Fig.~1, and
\begin{equation}
{\cal M}_{{\rm a}\bar{q}_1q_2\rm f}=<CD, I, I_z \mid P_{\bar{q}_1 + q_2 \to
q_3 +\bar{q}_4} \mid AB, I, I_z>,
\end{equation}
for the right diagram of Fig.~1. The symbol $P_{q_1 + \bar{q}_2 \to
q_3 +\bar{q}_4}$ ($P_{\bar{q}_1 + q_2 \to q_3 +\bar{q}_4}$) is the operator
that implements $q_1 + \bar{q}_2 \to q_3 +\bar{q}_4$ 
($\bar{q}_1 + q_2 \to q_3 +\bar{q}_4$) in flavor space.
${\cal M}_{{\rm a}\bar{q}_1q_2\rm f}$ may  
differ from ${\cal M}_{{\rm a}q_1\bar{q}_2\rm f}$.
In Table 15 the flavor matrix elements for the reactions are listed: 
\begin{eqnarray}
\nonumber
\pi \pi \to \rho \rho, \quad K \bar {K} \to K^* \bar {K}^\ast , \quad
K \bar{K}^\ast \to K^* \bar{K}^\ast, \quad K^\ast \bar{K} \to K^* \bar{K}^\ast,
\\  \nonumber
\pi \pi \to K \bar K, \quad \pi \rho \to K \bar {K}^\ast, 
\quad \pi \rho \to K^* \bar{K}, \quad K \bar {K} \to \rho \rho.
\end{eqnarray}
We use the following notation,
$ K= \left( \begin{array}{c} K^+ \\ K^0 \end{array} \right) $,
$\bar{K}= \left( \begin{array}{c} \bar{K}^0 \\ K^- \end{array} \right)$,
$ K^*= \left( \begin{array}{c} K^{*+} \\ K^{*0} \end{array} \right) $, and
$\bar{K}^*= \left( \begin{array}{c}\bar{K}^{*0} \\ K^{*-} \end{array} \right)$.
Let us give an example that shows how to obtain the flavor matrix elements
listed in Table 15. The example is $\pi \pi \to K \bar K$ for $I=1$.
The initial and final flavor states are given by
\begin{equation}
\mid \pi \pi , I=1 ,I_z=-1 > = 
\frac {1}{\sqrt 2} (\mid \pi^0 > \mid \pi^- > - \mid \pi^- > \mid \pi^0 >),
\end{equation}
\begin{equation}
\mid \pi \pi , I=1 ,I_z=0 > = 
\frac {1}{\sqrt 2} (\mid \pi^+> \mid \pi^- > - \mid \pi^-> \mid \pi^+ >),
\end{equation}
\begin{equation}
\mid \pi \pi , I=1 ,I_z=1 > = 
\frac {1}{\sqrt 2} (\mid \pi^+> \mid \pi^0 > - \mid \pi^0> \mid \pi^+ >),
\end{equation}
\begin{equation}
\mid K\bar{K}, I=1, I_z=-1 > = \mid K^0 > \mid K^- >, 
\end{equation}
\begin{equation}
\mid K \bar{K} , I=1 ,I_z=0 > = 
\frac {1}{\sqrt 2} (\mid K^+ > \mid K^- > + \mid K^0 > \mid \bar{K}^0 >),
\end{equation}
\begin{equation}
\mid K\bar{K}, I=1, I_z=1 > = \mid K^+ > \mid  \bar{K}^0 >.
\end{equation}
The flavor wave functions of the pion and kaon are
$\mid \pi^+ > = -\mid u \bar{d} >$,
$\mid \pi^0 >= \frac {1}{\sqrt 2} ( \mid u \bar{u} > - \mid d \bar{d} >)$,
$\mid \pi^- > = \mid d\bar {u} >$, 
$\mid K^+ > = \mid u \bar{s}>$, $\mid K^0> =\mid d\bar{s} >$,
$\mid \bar{K}^0 > = -\mid s\bar{d}>$, $\mid K^- > = \mid s\bar{u}>$.
The flavor matrix elements for $\pi \pi \to K \bar{K}$ for $I=1$ are
${\cal M}_{{\rm a}q_1\bar{q}_2{\rm f}}=0$ and
${\cal M}_{{\rm a}\bar{q}_1q_2{\rm f}}=-1$,
which are independent of $I_z$. This indicates that only the right diagram in
Fig.~1 contributes to $\pi \pi \to K \bar{K}$ for $I=1$.

\vspace{0.5cm}
\leftline{\bf V. QUARK-ANTIQUARK RELATIVE-MOTION WAVE FUNCTIONS}
\vspace{0.5cm}

The mesonic quark-antiquark relative-motion wave functions in coordinate space
are solutions of the 
Schr\"odinger equation with the potential in the medium~\cite{JSX}:
\begin{equation}
V(\vec {r}) = V_{\rm si}(\vec {r}) + V_{\rm ss}(\vec {r}),
\end{equation}
where $\vec r$ is the relative coordinate of the quark and the antiquark inside
the meson. $V_{\rm si}$ is the central spin-independent potential,
\begin{equation}
V_{\rm {si}}(\vec {r}) =
D \left[ 1.3- \left( \frac {T}{T_{\rm c}} \right)^4 \right]
\tanh (Ar) - \frac {8\pi}{25} \frac {v(\lambda r)}{r} \exp (-Er),
\end{equation}
where $D=0. 7$ GeV, $T_{\rm c}=0.175$ GeV, 
$A=1.5[0.75+0.25 (T/{T_{\rm c}})^{10}]^6$ GeV, $E=0. 6$ GeV,
$\lambda=\sqrt{25/16\pi^2 \alpha'}$ with $\alpha'=1.04$ GeV$^{-2}$, and
$v(x)$ is
\begin{eqnarray}
v(x)=\frac
{100}{3\pi} \int^\infty_0 \frac {dQ}{Q} \left[\rho (\vec {Q}^2) -\frac {K}{\vec
{Q}^2}\right] \sin \left(\frac {Q}{\lambda}x\right),
\end{eqnarray}
where $K=3/16\pi^2\alpha'$
and $\rho (\vec {Q} ^2)$ is given by Buchm\"uller and Tye~\cite{BT}.
At short distances the quark interaction is described by 
perturbative QCD in vacuum, and 
one-gluon exchange plus perturbative one- and two-loop corrections gives the
quark-antiquark potential $-\frac {8\pi}{25} \frac {v(\lambda r)}{r}$~\cite{BT}. 
Medium screening sets in at distances $r \geq 0.3$ fm. Including
medium effects lattice QCD calculations have provided the
numerical quark-antiquark
potential at intermediate and long distances~\cite{KLP}, which depends on
the temperature of the medium. The potential $V_{\rm si}(\vec {r})$ fits rather 
well $-\frac {8\pi}{25} \frac {v(\lambda r)}{r}$ at short distances and
the numerical potential at intermediate and long distances.

Starting with Feynman diagrams for elastic particle-particle scattering,
one gets a relativistic particle-particle potential. Application of the 
Foldy-Wouthuysen canonical transformation to the two-particle relativistic
Hamiltonian with the relativistic potential leads to a nonrelativistic
particle-particle potential that includes a central spin-independent potential,
a spin-spin interaction, and other terms~\cite{Chraplyvy}. 
This standard procedure for obtaining the nonrelativistic particle-particle 
potential is the same for heavy or light particles. In other words,
the particle-particle potential is valid no matter how light the particles are.
This procedure has been successfully applied and tested to get the electron-electron
potential arising from propagation of a space-like photon~\cite{Stroscio}, 
the short-distance quark-quark potential arising from propagation of a 
space-like gluon~\cite{RGG}, and so on. Thus, the first term in Eq.~(67), 
obtained in lattice calculations involving a heavy quark and a heavy antiquark, 
is reasonably applied to a light constituent quark and a light constituent antiquark 
as well.

The second term in Eq.~(66) is the spin-spin interaction that arises from
perturbative one-gluon exchange plus one- and two-loop corrections~\cite{Xu}
and includes relativistic effects~\cite{BS,GI}:
\begin{eqnarray}
V_{\rm ss}(\vec {r})=
\frac {64\pi^2}{75}\frac{d^3}{\pi^{3/2}}\exp(-d^2r^2) \frac {\vec {s}_a \cdot 
\vec {s} _b} {m_am_b}
- \frac {16\pi}{75} \frac {1} {r}
\frac {d^2v(\lambda r)}{dr^2} \frac {\vec {s}_a \cdot \vec {s}_b}{m_am_b} ,
\end{eqnarray}
where $\vec {s}_a$ is the spin of constituent $a$, 
and the quantity $d$ is given by
\begin{eqnarray}
d^2=\sigma_{0}^2\left[\frac{1}{2}+\frac{1}{2}\left(\frac{4m_a m_b}{(m_a+m_b)^2}
\right)^4\right]+\sigma_{1}^2\left(\frac{2m_am_b}{m_a+m_b}\right)^2,
\end{eqnarray}
where $\sigma_0=0.15$ GeV and $\sigma_1=0.705$.

With up quark mass 0.32 GeV, down quark mass 0.32 GeV, strange 
quark mass 0.5 GeV, and charm quark mass 1.51 GeV,
the Schr\"odinger equation with the potential given in Eq.~(66) at $T=0$  is 
solved to reproduce the experimental masses of $\pi$, $\rho$, $K$, $K^*$, 
$J/\psi$, $\psi'$, $\chi_{c}$, $D$, $D^*$, $D_s$, and $D^*_s$ mesons~\cite{PDG}. 
The elastic $\pi \pi$ scattering for $I=2$ is governed by the
quark-interchange process. The experimental data of $S$-wave phase shifts 
for the scattering in vacuum~\cite{Colton,Durusoy,Losty,Hoogland} are 
reproduced in the Born approximation
together with the pionic quark-antiquark relative-motion 
wave functions obtained from the Schr\"odinger equation~\cite{JSX}.

\vspace{0.5cm}
\leftline{\bf VI. NUMERICAL CROSS SECTIONS AND DISCUSSIONS }
\vspace{0.5cm}

We are now ready to calculate elastic phase
shifts for $\pi \pi$ scattering in vacuum. From the transition amplitudes we
get the reduced $T$-matrix element,
\begin{equation}
T_{\rm fi}=
\frac {{\cal M}_{{\rm a}q_1\bar{q}_2}+{\cal M}_{{\rm a}\bar{q}_1q_2}}
{(2\pi)^3\sqrt{2E_A2E_B2E_C2E_D}}.
\end{equation}
The phase shift for $A+B \to A+B$ is
\begin{equation}
\exp (i\delta_l) \sin{\delta_l} = -\frac {2\pi^2\mid \vec{P} \mid E_AE_B}
{E_A+E_B}
\int_{-1}^1 dx T_{\rm fi} P_l (x),
\end{equation}
where $x=\cos \theta$, and $P_l$ are the Legendre polynonimals. The elastic
phase shift is calculated by the expression,
\begin{equation}
\delta_l = - \frac{1}{2}\arcsin {\rm Re} \left[ \frac {4\pi^2\mid \vec{P} \mid 
E_AE_B}{E_A+E_B} \int_{-1}^1 dx T_{\rm fi} P_l (x) \right].
\end{equation}
The $S$-wave $I=0$ and $P$-wave $I=1$ elastic phase shifts for $\pi \pi$ 
scattering in vacuum are shown respectively in Figs.~3 and 4 and compared with
the experimental data~\cite{PABFF,HJWBD,EM,SHLPP,REFGM,FP,BBMPS,AKMMP,GKPRY}.

When the $S$-wave $I=0$ elastic phase shift is larger than 30 degree, it 
increases rapidly with increasing $\sqrt s$. This behavior
is already known from results of chiral perturbation theory~\cite{GL,GMei} and 
the result of the master formula approach~\cite{SYZ}.
The experimental data of $P$-wave $I=1$ elastic phase shifts for 
$\pi\pi$ scattering in vacuum cannot be reproduced. This is because resonances
are not included in the present approach. This is similar to the tree 
calculation in chiral perturbation theory~\cite{GL}. Including one- and 
two-loop corrections in chiral perturbation theory and imposing 
unitarity, the resonances appear and the experimental data are reproduced. 
This has been shown in approaches based on the results of chiral 
perturbation theory, for example, the master formula approach~\cite{SYZ},
the Pad\'e method~\cite{DHT}, the
large-$N_f$ expansion~\cite{Einhorn}, the $N/D$ method~\cite{CM}, the
inverse amplitude method~\cite{DP}, the $K$-matrix method~\cite{ZB}, the
current algebra unitarization~\cite{BBO}, the Roy equations~\cite{AB}, 
the coupled-channel Lippmann-Schwinger equations~\cite{OO}, the Bethe-Salpeter 
approach~\cite{NA}, and the approaches based on effective meson 
Lagrangians~\cite{JPHS,HHKR,Chen,GMou,DGL,AK,AOORR,Li}.

We consider the following inelastic meson-meson scattering processes that are 
governed by quark-antiquark annihilation and creation:
\begin{eqnarray}
\nonumber
I=1~ \pi \pi \to \rho \rho, \quad K \bar {K} \to K^* \bar {K}^\ast , \quad
K \bar{K}^\ast \to K^* \bar{K}^\ast, \quad K^\ast \bar{K} \to K^* \bar{K}^\ast,
\\  \nonumber
I=1 ~ \pi \pi \to K \bar K, \quad \pi \rho \to K \bar {K}^\ast, 
\quad \pi \rho \to K^* \bar{K}, \quad K \bar {K} \to \rho \rho.
\end{eqnarray}
The quark-antiquark relative-motion wave functions, $\phi_{A\rm rel}$,
$\phi_{B\rm rel}$, $\phi_{C\rm rel}$, and $\phi_{D\rm rel}$, are obtained
from the Schr\"odinger equation with the potential given in Eq.~(66). With the
transition potential and the wave functions in Eqs.~(52)-(55),
we calculate the transition amplitudes, ${\cal M}_{{\rm a}q_1\bar{q}_2}$
and ${\cal M}_{{\rm a}\bar{q}_1q_2}$. According to Eq.~(34)
we calculate unpolarized cross sections at the six temperatures 
$T/T_{\rm c} =0$, $0.65$, $0.75$, $0.85$, $0.9$, and $0.95$. In Figs. 5-13
we plot the unpolarized cross sections for the nine channels:
\begin{eqnarray}
\nonumber
I=1~ \pi \pi \to \rho \rho, \quad I=1~ K \bar {K} \to K^* \bar {K}^\ast,
\quad I=0~ K \bar {K} \to K^* \bar {K}^\ast,
\\ \nonumber
I=1~ K \bar {K}^\ast \to K^* \bar {K}^\ast, \quad 
I=0~ K \bar {K}^\ast \to K^* \bar {K}^\ast, \quad I=1~ \pi \pi \to K \bar K,
\\ \nonumber
I=1~ \pi \rho \to K \bar {K}^\ast, \quad I=1~ \pi \rho \to K^\ast \bar{K},
\quad I=1~ K \bar {K} \to \rho \rho.
\end{eqnarray}
Depending on temperature, a reaction is either endothermic or exothermic.
The numerical cross sections for endothermic reactions are parametrized as
\begin{eqnarray}
\sigma^{\rm unpol}(\sqrt {s},T)
&=&a_1 \left( \frac {\sqrt {s} -\sqrt {s_0}} {b_1} \right)^{c_1}
\exp \left[ c_1 \left( 1-\frac {\sqrt {s} -\sqrt {s_0}} {b_1} \right) \right] 
\nonumber \\
&&+ a_2 \left( \frac {\sqrt {s} -\sqrt {s_0}} {b_2} \right)^{c_2}
\exp \left[ c_2 \left( 1-\frac {\sqrt {s} -\sqrt {s_0}} {b_2} \right) \right],
\end{eqnarray}
where $\sqrt{s_0}$ is the threshold energy, and $a_1$, $b_1$, $c_1$, $a_2$, 
$b_2$, and $c_2$ are parameters. The numerical cross sections
for exothermic reactions are parametrized as
\begin{eqnarray}
\sigma^{\rm unpol}(\sqrt {s},T)
&=&\frac{\vec{P}^{\prime 2}}{\vec{P}^2}
\left\{a_1 \left( \frac {\sqrt {s} -\sqrt {s_0}} {b_1} \right)^{c_1}
\exp \left[ c_1 \left( 1-\frac {\sqrt {s} -\sqrt {s_0}} {b_1} \right) \right] 
\right.
\nonumber \\
&&+ \left.
a_2 \left( \frac {\sqrt {s} -\sqrt {s_0}} {b_2} \right)^{c_2}
\exp \left[ c_2 \left( 1-\frac {\sqrt {s} -\sqrt {s_0}} {b_2} \right) \right]
\right\}.
\end{eqnarray}
The parameter values are listed in Tables 16-18. 
As in Ref.~\cite{SX}, we also list the quantities $d_0$ and $\sqrt{s_z}$;  
$d_0$ is the separation between the peak's location on 
the $\sqrt s$-axis and the threshold energy, and $\sqrt{s_z}$ is 
the square root of the Mandelstam variable at which the cross section is 
1/100 of the peak cross section. 

In the temperature region that is covered by hadronic matter produced in 
ultrarelativistic heavy-ion collisions, the central spin-independent potential
given in Eq.~(67) at long distances
becomes independent of distance and exhibits a plateau. Confinement is marked
by the plateau. With increasing temperature the height of
the plateau decreases, confinement becomes weaker and weaker,
and quark-antiquark bound states become looser and looser. Increasing radii of 
mesons $A$ and $B$ cause increasing cross sections for meson-meson reactions.
Weakening confinement causes the combination of final quarks and
antiquarks in forming mesons $C$ and $D$ to be more difficult, thus decreasing 
cross sections for meson-meson reactions. The two factors determine the decrease 
or the increase of peak cross sections of endothermic reactions shown in Figs.~5-13. 
For example, from $T/T_{\rm c}=0$ to 0.85 and near the threshold energy the increase of 
the cross
section due to increasing radii of initial mesons cannot overcome the
decrease of the cross section due to weakening confinement. Thus, the peak
cross section of $K\bar {K} \to K^\ast \bar{K}^\ast$ decreases. In contrast,
the peak cross section of $K\bar {K} \to K^\ast \bar{K}^\ast$ increases as
temperature goes from $T/T_{\rm c}=0.85$ to 0.95. In vacuum
the $\rho$ mass is larger than the kaon mass. When the temperature increases,
the $\rho$ mass decreases faster than the kaon mass. 
At $T \simeq 0.785T_{\rm c}$ the
$\rho$ mass equals the kaon mass. Below this temperature the reaction 
$K\bar{K} \to \rho \rho$ for $I=1$ in Fig.~13 is endothermic; 
otherwise, it is exothermic. When temperature increases from
$T/T_{\rm c}=0.6$ to 1, the $\pi$, $\rho$, $K$, and $K^\ast$ masses decrease.
The threshold energies shown in Figs. 5-13 thus decrease.
The ratio of the peak cross section at $T/T_{\rm c}$=0.75 to the peak cross
section at $T/T_{\rm c}$=0 is 0.08, 0.34, 0.25, 0.16, 0.19, 0.11, 0.05, and 
0.13 for $\pi \pi \to \rho \rho$ for $I=1$, $K \bar {K} \to K^* \bar {K}^\ast$
for $I=1$, $K \bar {K} \to K^* \bar {K}^\ast$ for $I=0$, 
$K \bar {K}^\ast \to K^* \bar {K}^\ast$ for $I=1$,
$K \bar {K}^\ast \to K^* \bar {K}^\ast$ for $I=0$,
$\pi \rho \to K \bar {K}^\ast$ for $I=1$, $\pi \rho \to K^\ast \bar{K}$ for 
$I=1$, and $K \bar {K} \to \rho \rho$ for $I=1$, respectively. Clearly, the
cross sections have remarkable dependence on temperature.

It is shown in Table 15 that the two diagrams in Fig.~1 contribute to 
$K \bar{K} \to K^\ast \bar{K}^\ast$ for $I=0$ and 
$K \bar{K}^* \to K^\ast \bar{K}^\ast$ for $I=0$, 
and only the right diagram contributes to $K \bar{K} \to K^\ast \bar{K}^\ast$ 
for $I=1$ and 
$K \bar{K}^* \to K^\ast \bar{K}^\ast$ for $I=1$. The peak cross section of
$K \bar{K} \to K^\ast \bar{K}^\ast$ ($K \bar{K}^* \to K^\ast \bar{K}^\ast$)
for $I=0$ at a given temperature is more than 2 times the one for $I=1$. In
addition, cross sections for $K^\ast \bar{K} \to K^* \bar{K}^\ast$ equal the
cross sections for $K \bar{K}^* \to K^\ast \bar{K}^\ast$.

It is shown in Table 15 that only the right diagram in Fig.~1 contributes to
the reactions, $\pi \pi \to K \bar{K}$, $\pi \rho \to K \bar{K}^*$, 
$\pi \rho \to K^* \bar{K}$, and $K \bar{K} \to \rho \rho$. 
Since the flavor matrix element for the channel $I=0$ is $-\sqrt {6}/2$ 
times the one for $I=1$, the cross section
for the channel $I=0$ is 1.5 times the one for $I=1$. Therefore,
we only plot the unpolarized cross sections for $\pi \pi \to K \bar{K}$ for
$I=1$ in Fig.~10, for $\pi \rho \to K \bar{K}^*$ for $I=1$ in Fig.~11, 
for $\pi \rho \to K^* \bar{K}$ for $I=1$ in Fig.~12, 
and for $K \bar{K} \to \rho \rho$ for $I=1$ in Fig.~13. The cross section
for $\pi \rho \to K\bar{K}^*$ at $T=0$ leads to the isospin-averaged
cross section that has a maximum value of about 0.36 mb which is close to 
the one obtained from an effective meson Lagrangian in Refs.~\cite{BKWX,CBMTS}.
Nevertheless, the cross section for $K\bar{K} \to \rho \rho$ at $T=0$ provides
about 1 mb as the maximum value of the isospin-averaged cross section, which
is not close to 3.5 mb given in Refs.~\cite{BKWX,CBMTS}.

The unpolarized cross section for $\pi \pi \to \rho \rho$ for $I=2$ has been
shown in Fig.~2 of Ref.~\cite{SX}. The channel is governed by the 
quark-interchange process. The reaction $\pi \pi \to \rho \rho$ for $I=1$
studied in the present work is governed by quark-antiquark annihilation
and creation.
The unpolarized cross sections for both channels increase rapidly when  
$\sqrt s$ increases from the threshold energy. When $\sqrt s$ increases
from the magnitude that corresponds to the peak cross section, the cross 
section
for the channel $I=2$ decreases rapidly, but the one for $I=1$ decreases 
slowly. The difference is related to the quark-antiquark relative-motion
wave functions of final mesons. The wave functions are exponentially decreasing
functions of the quark-antiquark relative momentum. When $\sqrt s$ is far away 
from the threshold energy, the relative momentum of an interchanged quark and 
an antiquark, which form a final meson
in the channel $I=2$, is usually large~\cite{LX}, but the relative
momentum of the quark created from the gluon and an antiquark, which form a 
final meson in the channel $I=1$, can still be small. 
The squared transition amplitude for the reaction with the quark-interchange
process at such a value of $\sqrt s$ is very small in comparison to the one
near the threshold energy, and by contrast the squared transition amplitude for
the
reaction with the quark-antiquark annihilation is comparable to the one near 
the threshold energy. Therefore, the cross section has the behavior of rapid
decrease for $I=2$ and of slow decrease for $I=1$, and the
difference between the cross sections for the two channels is large.

\vspace{0.5cm}
\leftline{\bf VII. SUMMARY }
\vspace{0.5cm}

We have derived the unpolarized cross section for inelastic meson-meson 
scattering in quark degrees of freedom. The reactions are governed by 
quark-antiquark annihilation and creation. The reactions include
$\pi \pi \to \rho \rho$ for $I=1$, $K \bar {K} \to K^* \bar {K}^\ast$,
$K \bar{K}^\ast \to K^* \bar{K}^\ast$, $K^\ast \bar{K} \to K^* \bar{K}^\ast$,
$\pi \pi \to K \bar K$ for $I=1$, $\pi \rho \to K \bar {K}^\ast$, 
$\pi \rho \to K^* \bar{K}$, and $K \bar {K} \to \rho \rho$. The transition
potential corresponding to quark-antiquark annihilation and creation has
been derived in perturbative QCD. Some reactions involve only one 
Feynman diagram at tree level, 
and the others two. The transition amplitudes including color,
spin, and flavor matrix elements are given
upon integrating over the relative momenta of the quark and the
antiquark of the two initial mesons. The experimental data of $S$-wave and
$P$-wave elastic phase shifts for 
$\pi \pi$ scattering near the threshold energy can be accounted for by 
quark-antiquark annihilation and creation in the Born approximation. 
Numerical unpolarized cross sections have been obtained at the six temperatures
and have shown remarkable temperature dependence. The dependence arises
from the quark-antiquark relative-motion wave functions of the initial and
final mesons. The numerical cross sections are parametrized for future use in 
the evolution of hadronic matter.

\vspace{0.5cm}
\leftline{\bf ACKNOWLEDGEMENTS}
\vspace{0.5cm}

We thank W. Weise for introducing us to his work and B. S. Zou for helpful
discussions on meson-meson reactions.
This work was supported by the National Natural Science Foundation of China 
under Grant No. 11175111.

\newpage
\begin{figure}[htbp]
  \centering
    \includegraphics[width=60mm,height=85mm,angle=0]{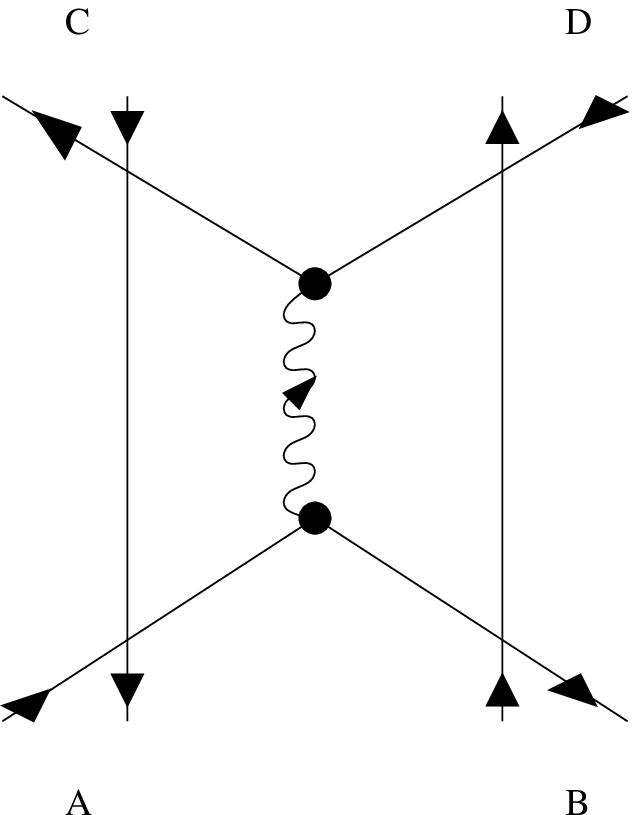}
      \hspace{2cm}
    \includegraphics[width=60mm,height=85mm,angle=0]{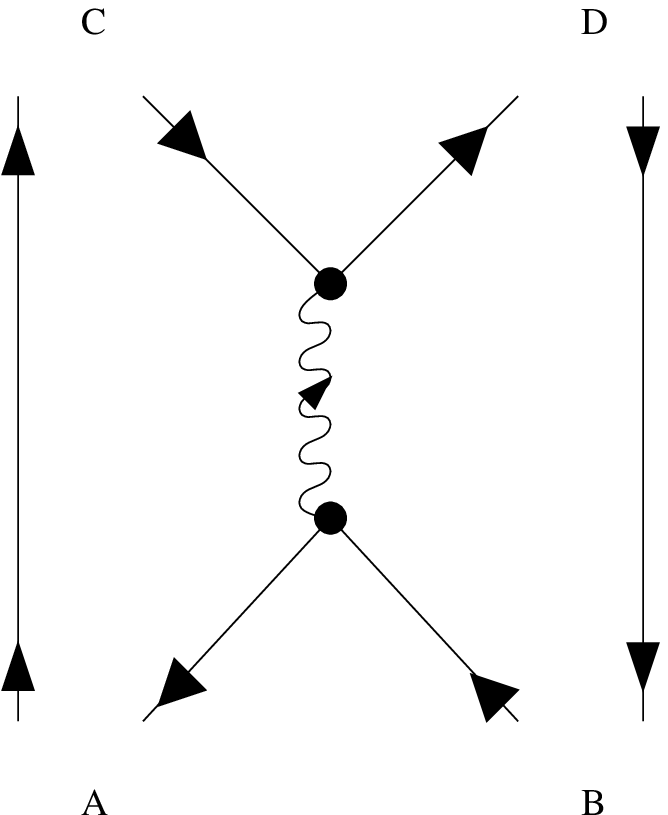}
\caption{Left diagram with $q_1 + \bar {q}_2 \to q_3 + \bar{q}_4$ and right 
diagram with $\bar{q}_1 + q_2 \to q_3 + \bar{q}_4$ for $A+B \to C+D$. The 
vertical lines represent spectator quarks (antiquarks) of the mesons $A,B,C,D.$}
\label{fig1}
\end{figure}

\newpage
\begin{figure}[htbp]
  \centering
    \includegraphics[width=60mm,height=85mm,angle=0]{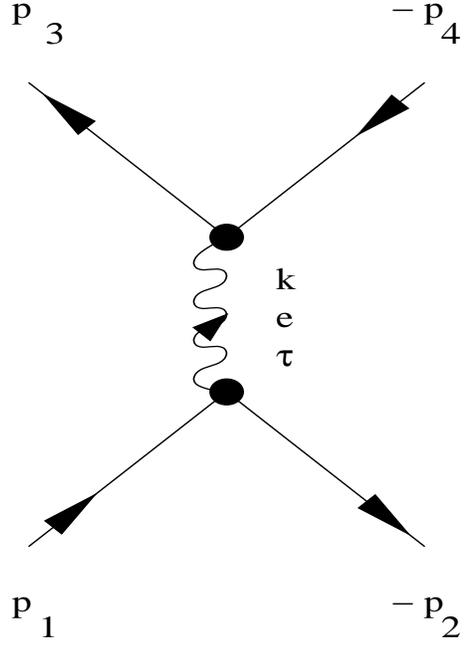}
\caption{Quark-antiquark annihilation and creation: 
$q (p_1) + \bar{q} (-p_2) \to q^\prime (p_3) + \bar{q}^\prime (-p_4)$, 
where $k$ denotes the gluon four-momentum, $e$ its color index and $\tau$ 
its space-time index (cf.~Eq.~(\ref{39})).}
\label{fig2}
\end{figure}

\newpage
\begin{figure}[htbp]
  \centering
    \includegraphics[scale=0.65]{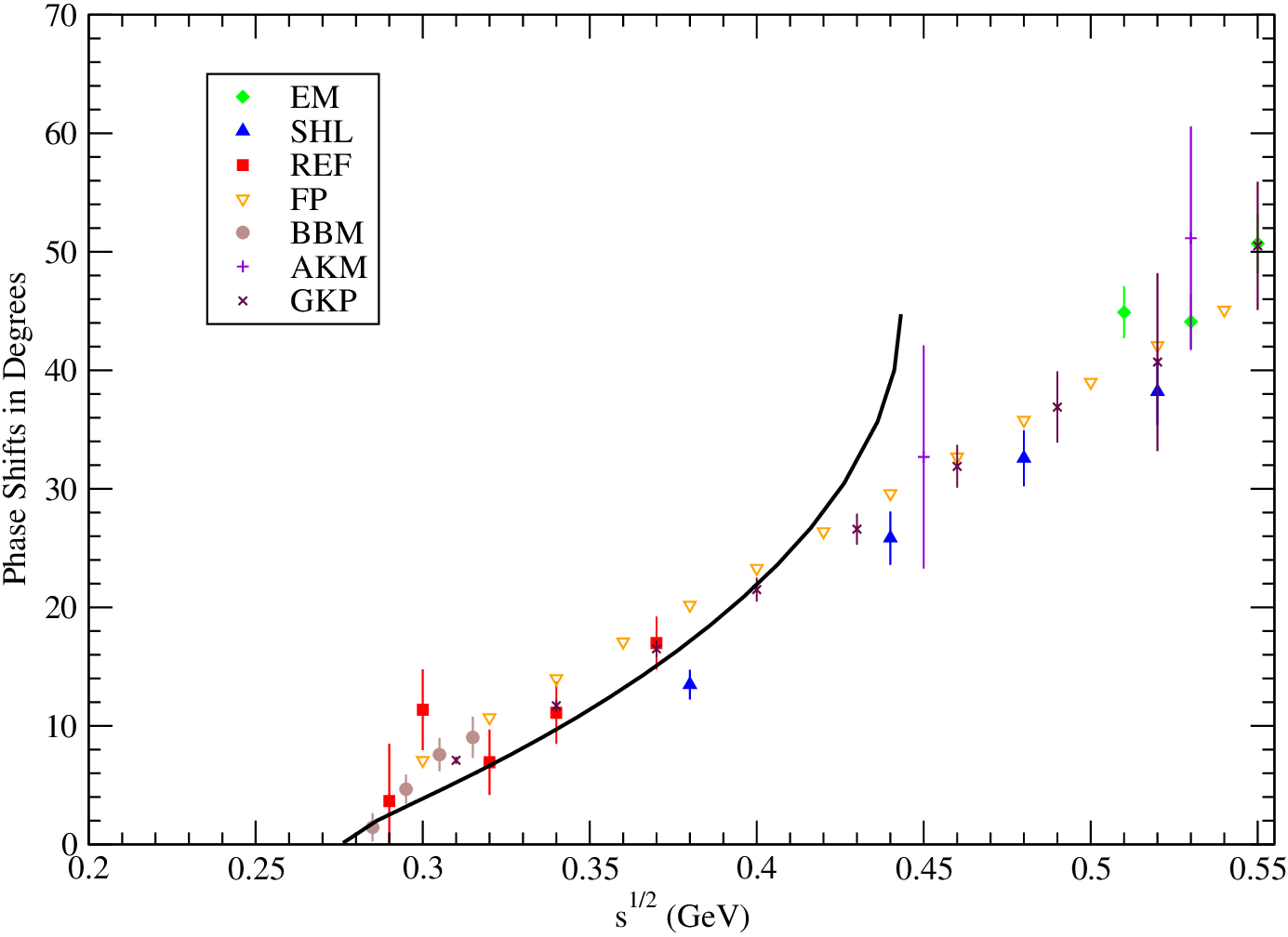}
\caption{$S$-wave $I=0$ elastic phase shifts for $\pi \pi$ scattering. The 
solid curve is the present theoretical result. Experimental data: $\Diamond$,
Ref. \cite{EM}; $\triangle$, Ref. \cite{SHLPP}; $\Box$, Ref. \cite{REFGM}; 
$\bigtriangledown$, Ref. \cite{FP}; $\bigcirc$, Ref. \cite{BBMPS}; 
+, Ref. \cite{AKMMP}; $\times$, Ref. \cite{GKPRY}.}
\label{fig3}
\end{figure}

\newpage
\begin{figure}[htbp]
  \centering
    \includegraphics[scale=0.65]{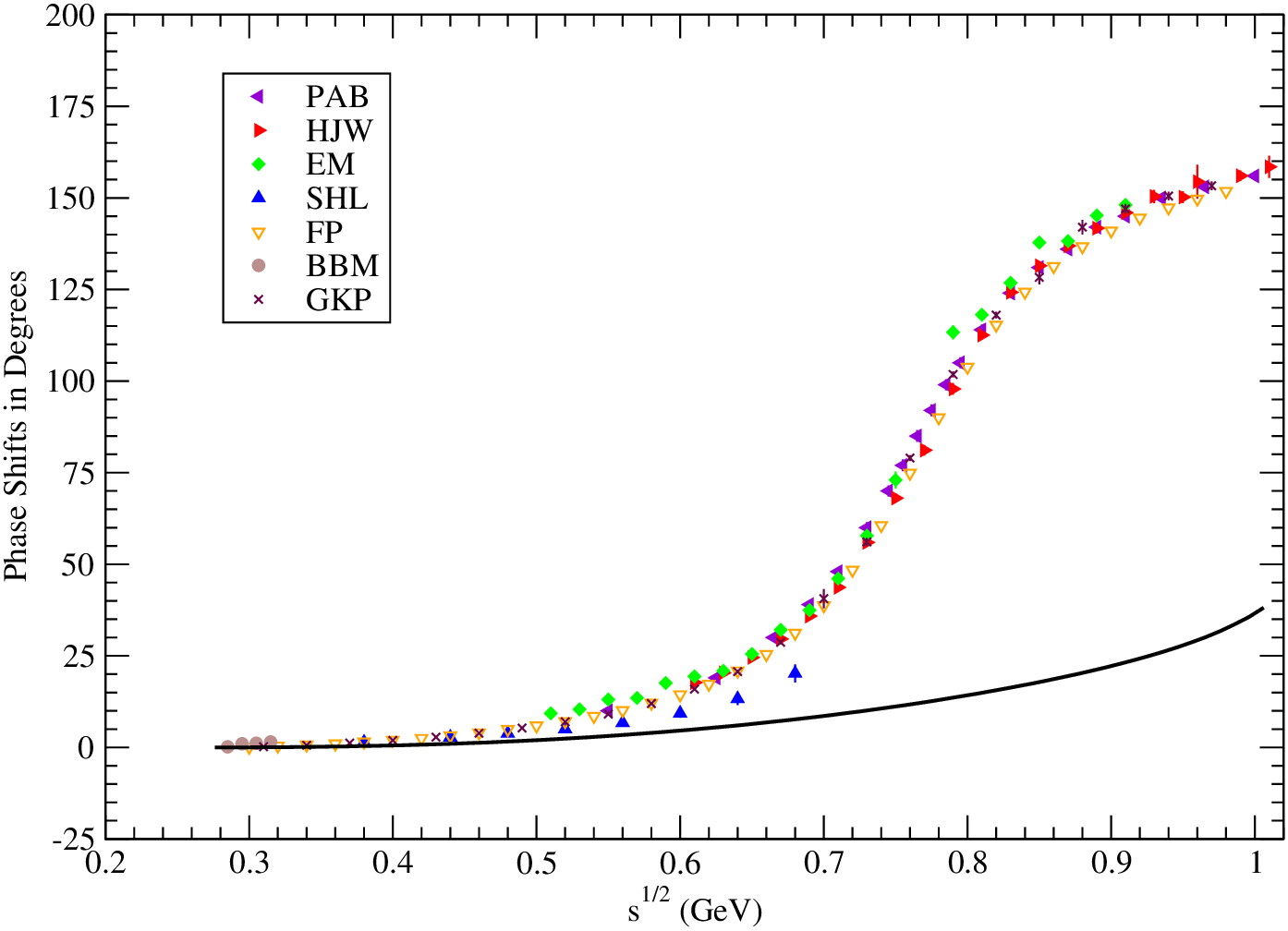}
\caption{$P$-wave $I=1$ elastic phase shifts for $\pi \pi$ scattering. The 
solid curve is the present theoretical result. Experimental data: 
$\lhd$, Ref. \cite{PABFF}; $\rhd$, Ref. \cite{HJWBD}; $\Diamond$, Ref. 
\cite{EM}; $\bigtriangleup$, Ref. \cite{SHLPP}; $\bigtriangledown$, 
Ref. \cite{FP}; $\bigcirc$, Ref. \cite{BBMPS}; $\times$, Ref. \cite{GKPRY}.}
\label{fig4}
\end{figure}

\newpage
\begin{figure}[htbp]
  \centering
    \includegraphics[scale=0.65]{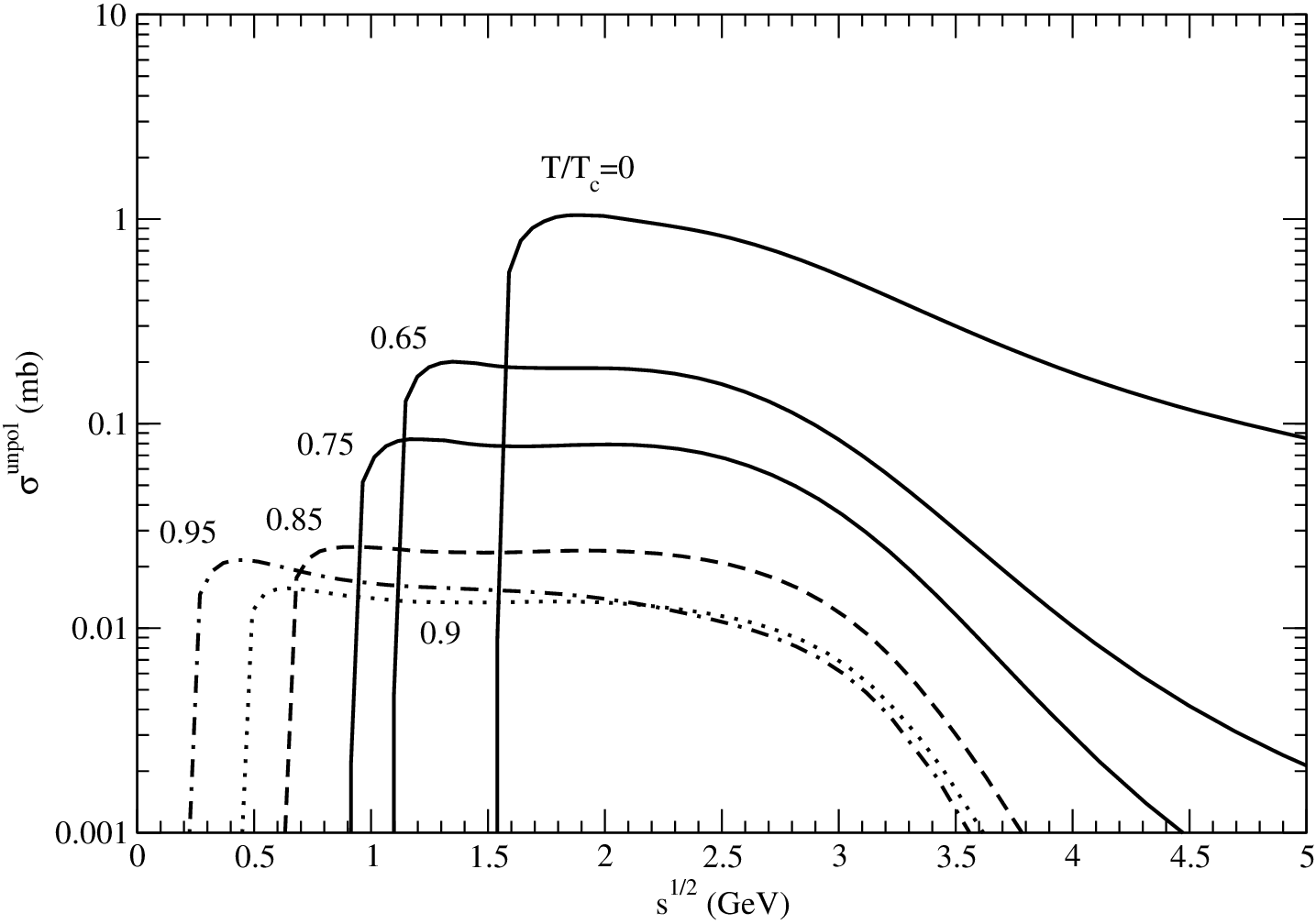}
\caption{Cross sections for $\pi \pi \to \rho \rho$ for $I=1$
at various temperatures.}
\label{fig5}
\end{figure}

\newpage
\begin{figure}[htbp]
  \centering
    \includegraphics[scale=0.65]{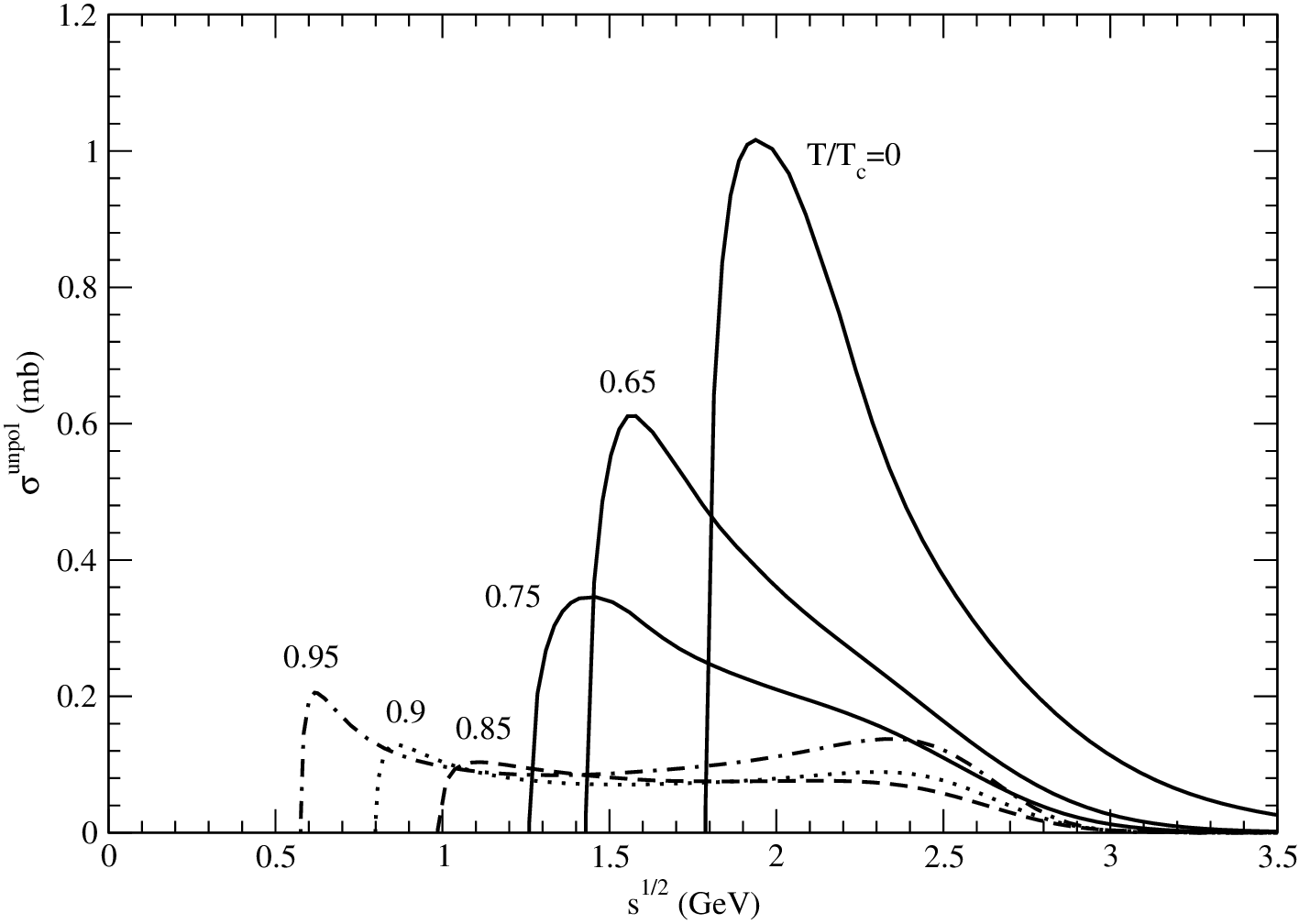}
\caption{Cross sections for $K \bar{K} \to K^\ast \bar{K}^\ast$ for $I=1$
at various temperatures.}
\label{fig6}
\end{figure}

\newpage
\begin{figure}[htbp]
  \centering
    \includegraphics[scale=0.65]{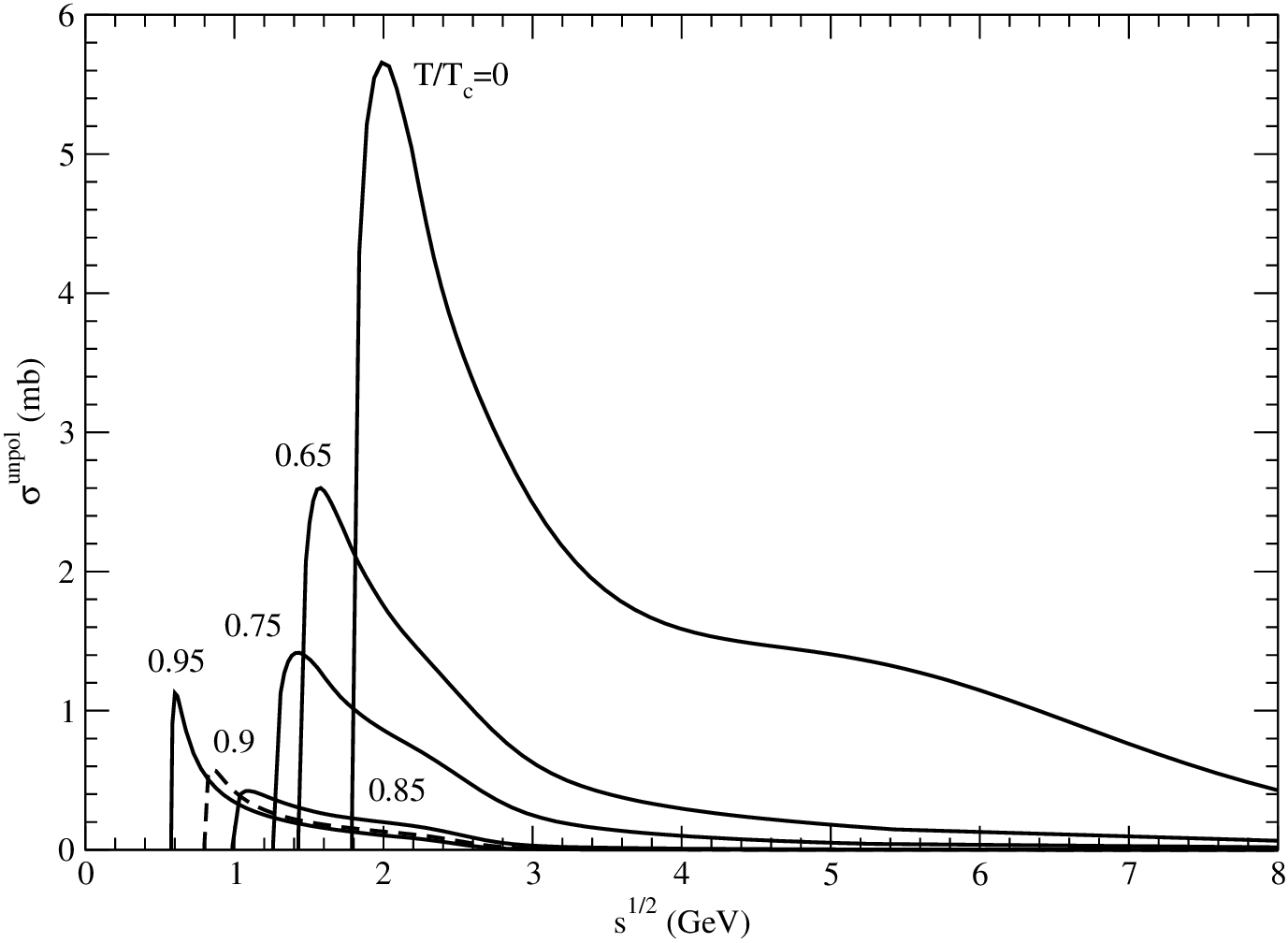}
\caption{Cross sections for $K \bar{K} \to K^\ast \bar{K}^\ast$ for $I=0$
at various temperatures.}
\label{fig7}
\end{figure}

\newpage
\begin{figure}[htbp]
  \centering
    \includegraphics[scale=0.65]{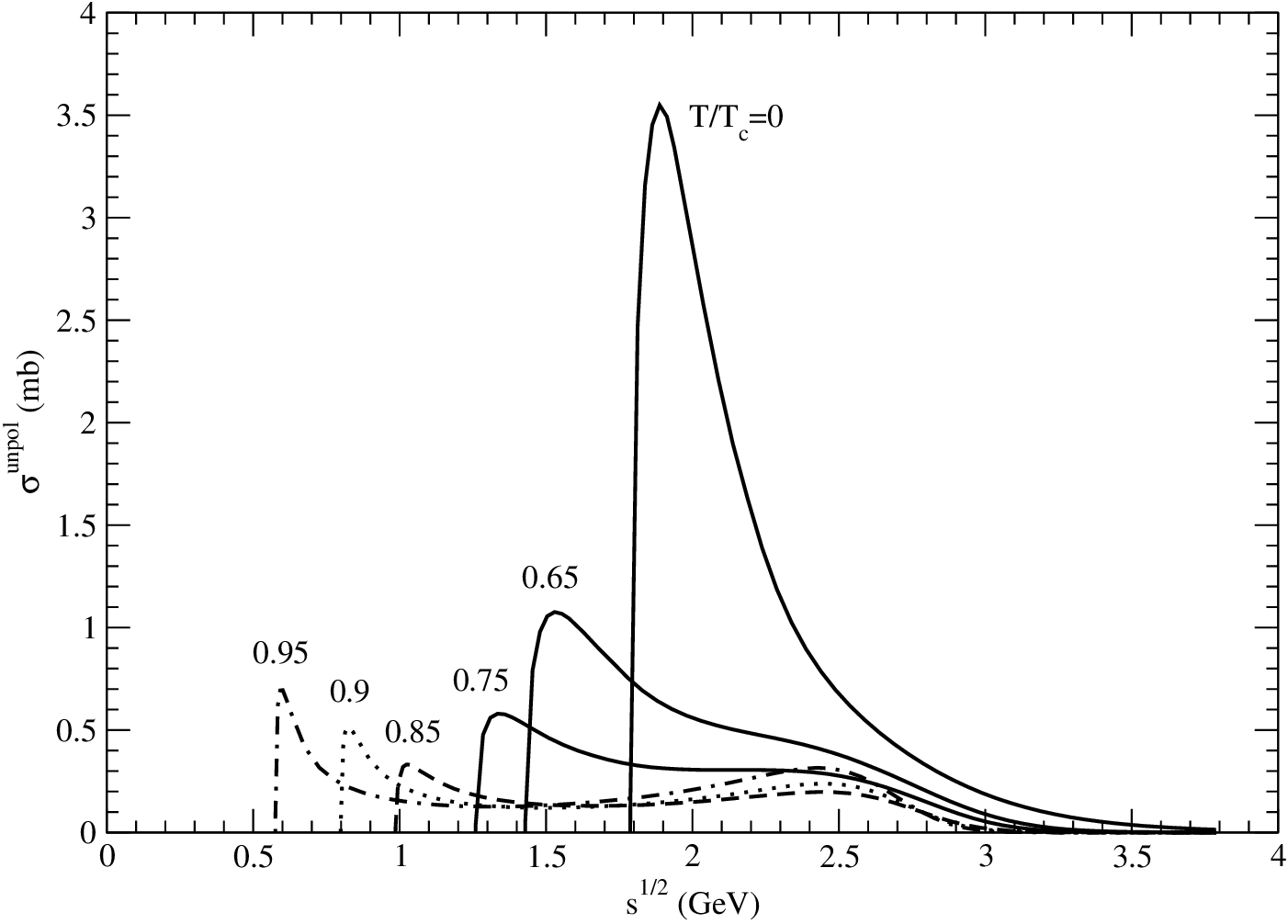}
\caption{Cross sections for $K \bar{K}^\ast \to K^\ast \bar{K}^\ast$ for $I=1$
at various temperatures.}
\label{fig8}
\end{figure}

\newpage
\begin{figure}[htbp]
  \centering
    \includegraphics[scale=0.65]{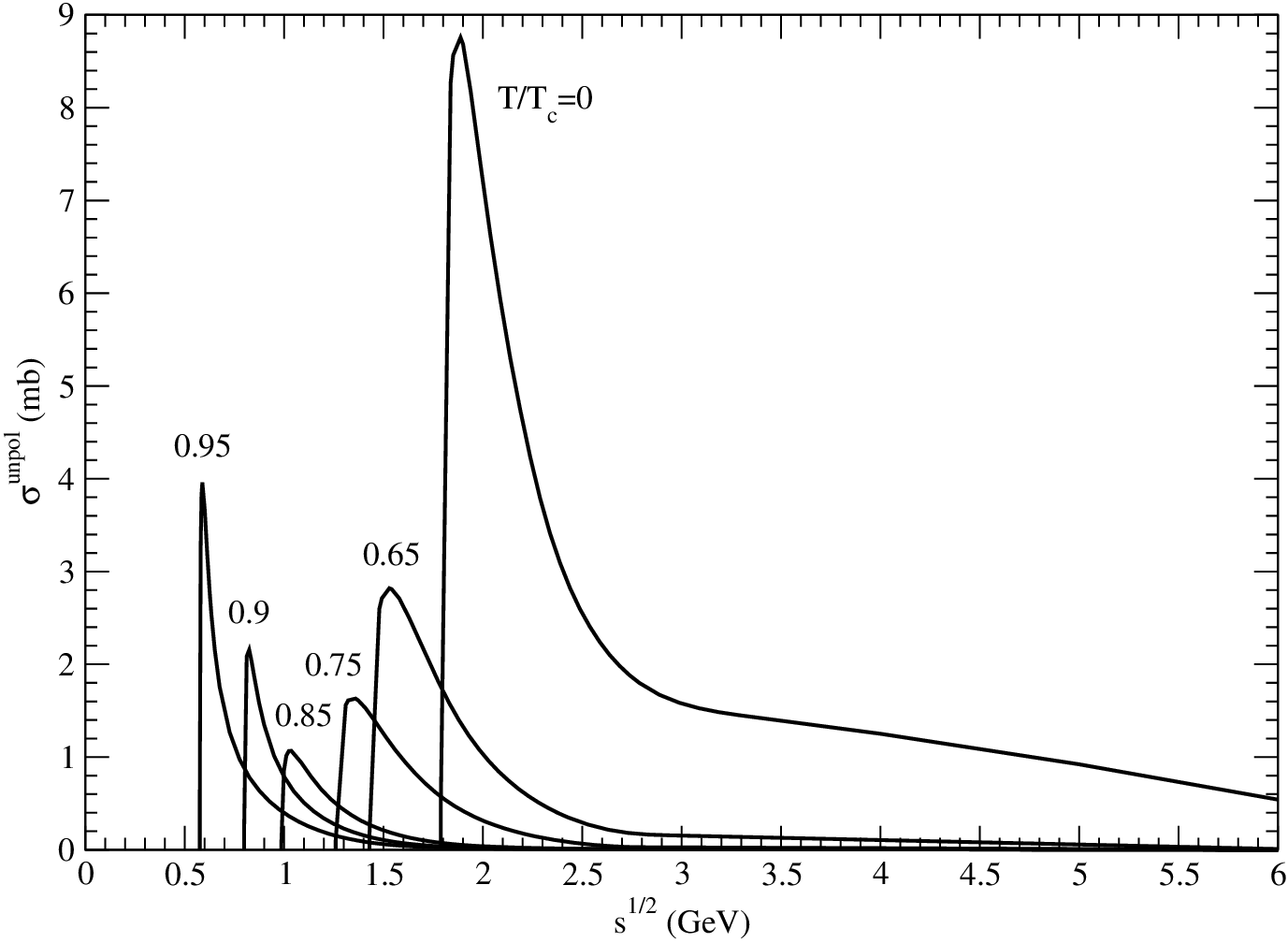}
\caption{Cross sections for $K \bar{K}^\ast \to K^\ast \bar{K}^\ast$ for $I=0$
at various temperatures.}
\label{fig9}
\end{figure}

\newpage
\begin{figure}[htbp]
  \centering
    \includegraphics[scale=0.65]{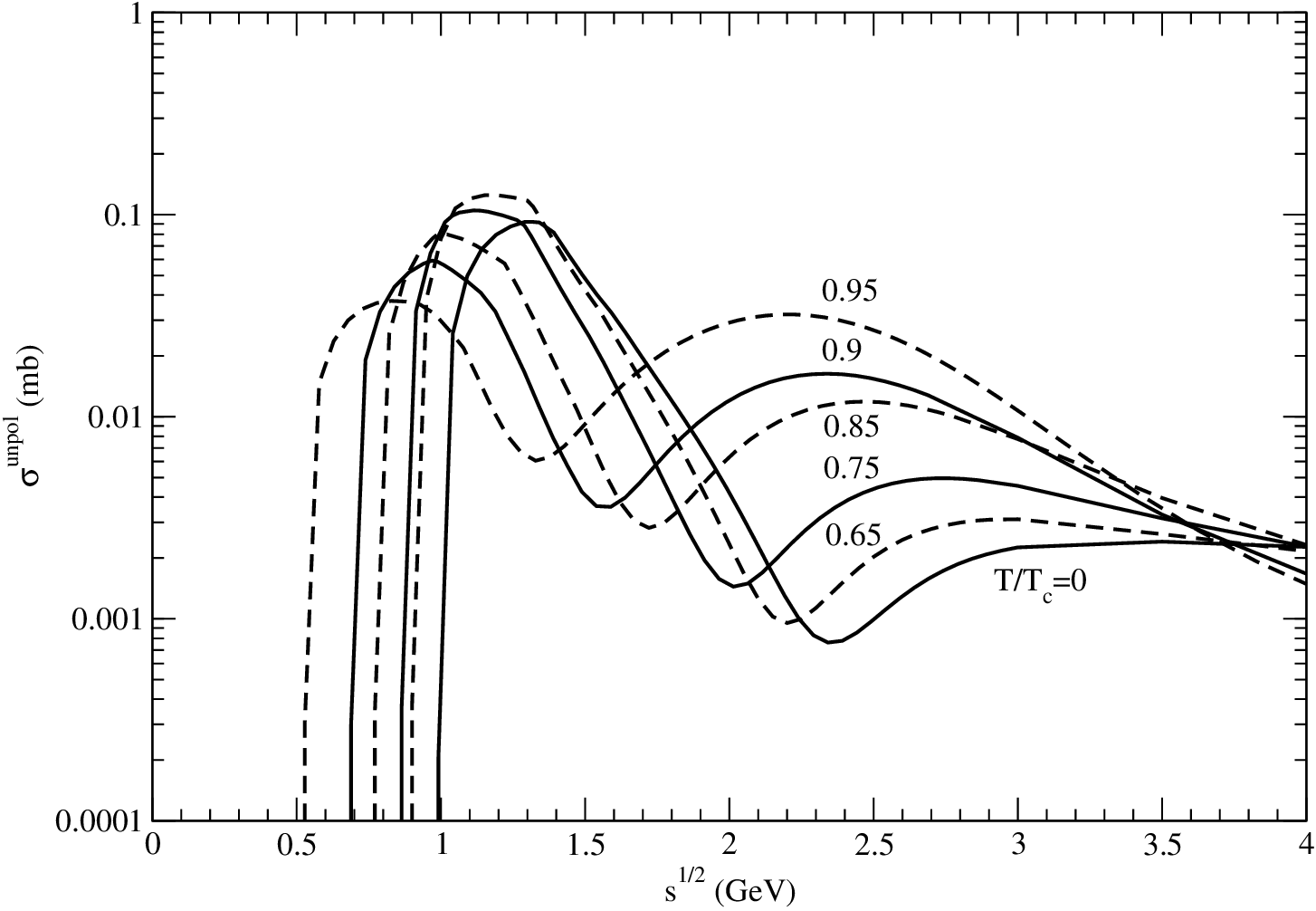}
\caption{Cross sections for $\pi \pi \to K \bar{K}$ for $I=1$
at various temperatures.}
\label{fig10}
\end{figure}

\newpage
\begin{figure}[htbp]
  \centering
    \includegraphics[scale=0.65]{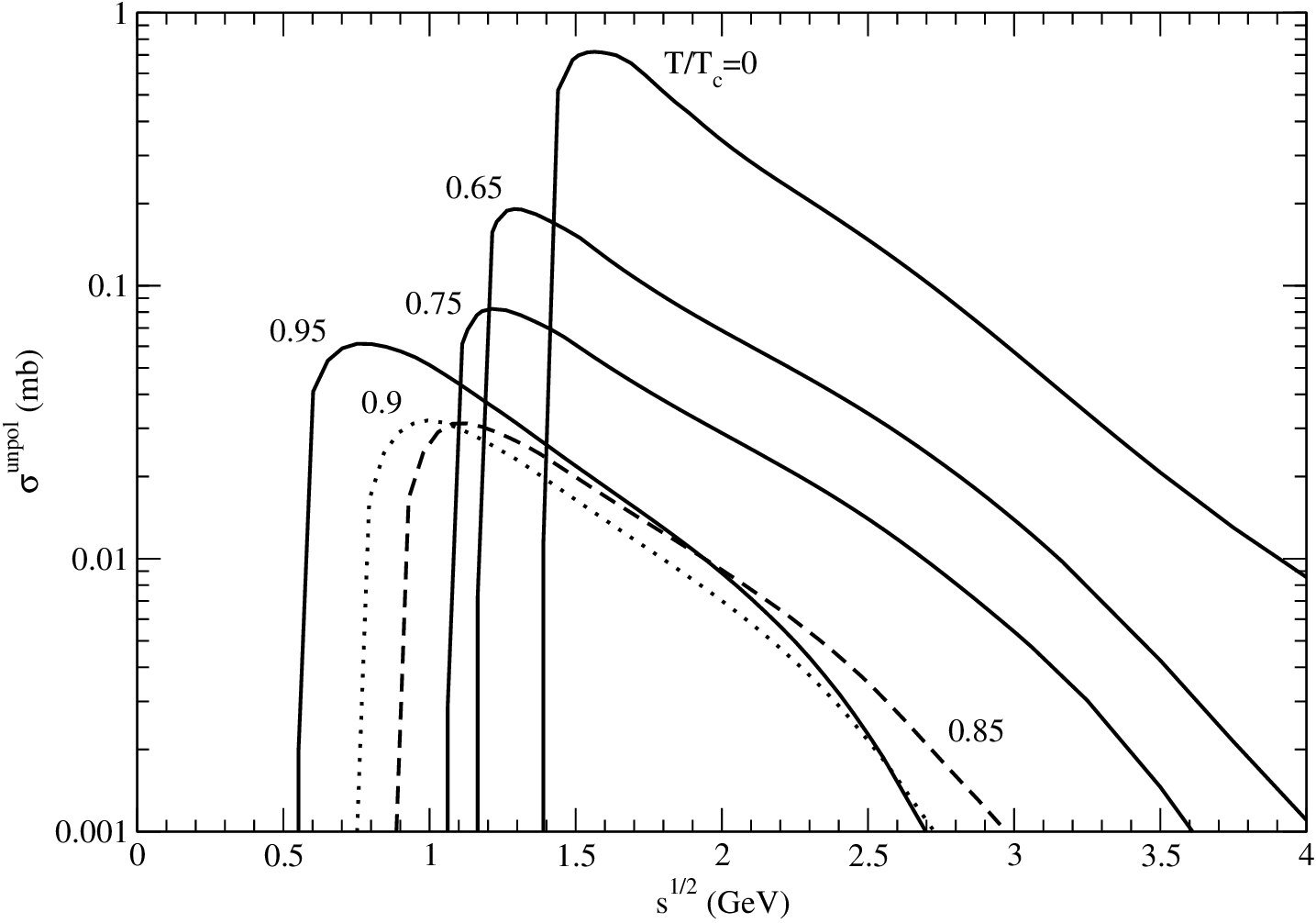}
\caption{Cross sections for $\pi \rho \to K \bar{K}^\ast$ for $I=1$
at various temperatures.}
\label{fig11}
\end{figure}

\newpage
\begin{figure}[htbp]
  \centering
    \includegraphics[scale=0.65]{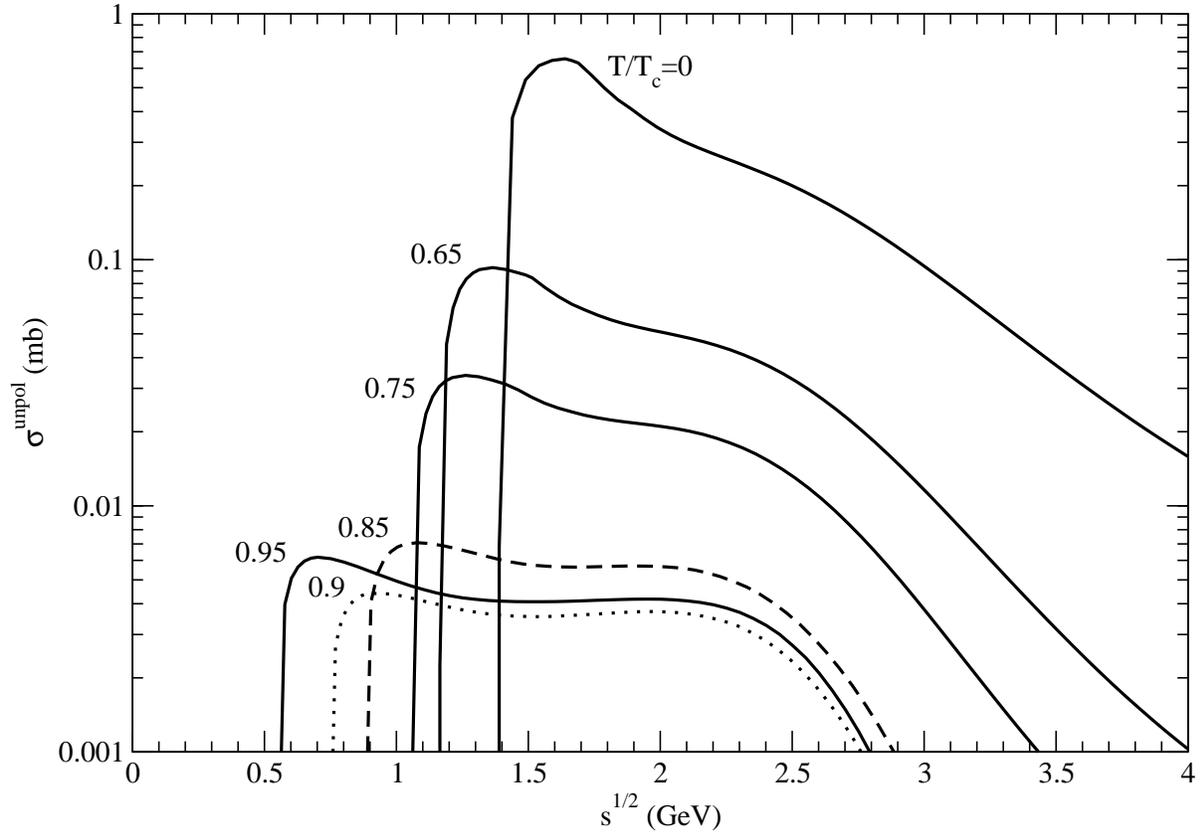}
\caption{Cross sections for $\pi \rho \to K^* \bar{K}$ for $I=1$
at various temperatures.}
\label{fig12}
\end{figure}

\newpage
\begin{figure}[htbp]
  \centering
    \includegraphics[scale=0.65]{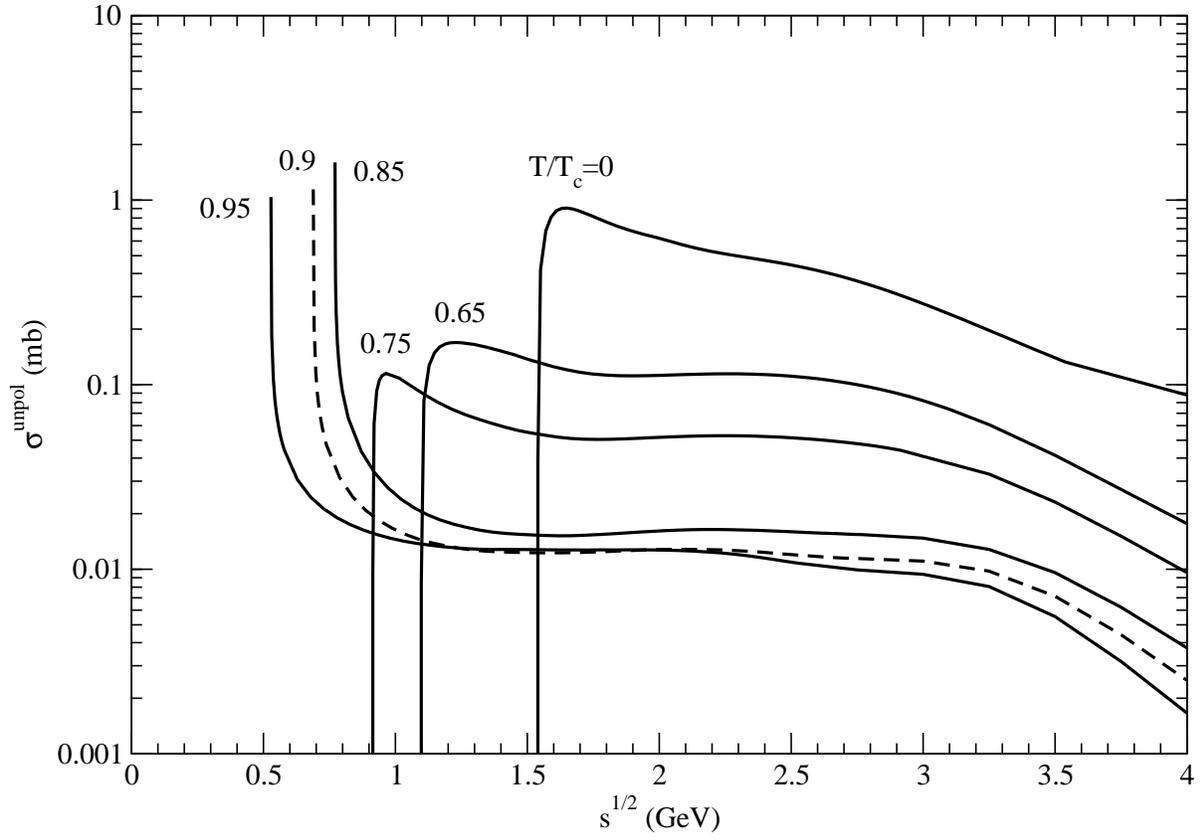}
\caption{Cross sections for $K \bar{K} \to \rho \rho$ for $I=1$
at various temperatures.}
\label{fig13}
\end{figure}

\newpage
\begin{table}
\caption{\label{table1} Spin matrix elements of 
${\cal M}_{{\rm a}q_1\bar{q}_2}$. The initial spin state is 
$\phi_{\rm iss}=\chi_{S_A S_{Az}} \chi_{S_B S_{Bz}}$, and 
the final spin state $\phi_{\rm fss}=\chi_{S_C S_{Cz}} \chi_{S_D S_{Dz}}$.
The second column corresponds to $S_A=S_B=S_C=S_D=0$, the third to fifth 
columns correspond to $S_A=S_B=S_C=0$, the sixth to eighth columns 
$S_A=S_B=S_D=0$, and the ninth to eleventh columns $S_A=S_C=S_D=0$.}
\begin{tabular}{lllllllllll}
\hline
$S_{Az}$ & 0 & 0  & 0  & 0 & 0  & 0 & 0 &  0 & 0 & 0 \\
$S_{Bz}$ & 0 & 0  & 0  & 0 & 0  & 0 & 0 & -1 & 0 & 1 \\
$S_{Cz}$ & 0 & 0  & 0  & 0 & -1 & 0 & 1 &  0 & 0 & 0 \\
$S_{Dz}$ & 0 & -1 & 0  & 1 & 0  & 0 & 0 &  0 & 0 & 0 \\
\hline
$\phi_{\rm fss}^+ \phi_{\rm iss}$ & $\frac {1}{2}$ & 0 & 0 & 0 & 0 & 0 & 0 
& 0 & 0 & 0 \\
$\phi_{\rm fss}^+ \sigma_1(21) \phi_{\rm iss}$ & 0 & -$\frac {1}{2\sqrt 2}$ & 0
& $\frac {1}{2\sqrt 2}$ & $\frac {1}{2\sqrt 2}$ & 0 & -$\frac {1}{2\sqrt 2}$ 
& -$\frac {1}{2\sqrt 2}$ & 0 & $\frac {1}{2\sqrt 2}$  \\
$\phi_{\rm fss}^+ \sigma_2(21) \phi_{\rm iss}$ & 0 & $\frac {1}{2\sqrt 2}i$ & 0
& $\frac {1}{2\sqrt 2}i$ & $\frac {1}{2\sqrt 2}i$ & 0 & $\frac {1}{2\sqrt 2}i$ 
& -$\frac {1}{2\sqrt 2}i$ & 0 & -$\frac {1}{2\sqrt 2}i$  \\
$\phi_{\rm fss}^+ \sigma_3(21) \phi_{\rm iss}$ & 0 & 0 & -$\frac {1}{2}$
& 0 & 0 & $\frac {1}{2}$ & 0 & 0 & -$\frac {1}{2}$ & 0  \\
$\phi_{\rm fss}^+ \sigma_1(34) \phi_{\rm iss}$ & 0 & -$\frac {1}{2\sqrt 2}$ & 0
& $\frac {1}{2\sqrt 2}$ & $\frac {1}{2\sqrt 2}$ & 0 & -$\frac {1}{2\sqrt 2}$ 
& -$\frac {1}{2\sqrt 2}$ & 0 & $\frac {1}{2\sqrt 2}$  \\
$\phi_{\rm fss}^+ \sigma_2(34) \phi_{\rm iss}$ & 0 & $\frac {1}{2\sqrt 2}i$ & 0
& $\frac {1}{2\sqrt 2}i$ & $\frac {1}{2\sqrt 2}i$ & 0 & $\frac {1}{2\sqrt 2}i$ 
& -$\frac {1}{2\sqrt 2}i$ & 0 & -$\frac {1}{2\sqrt 2}i$  \\
$\phi_{\rm fss}^+ \sigma_3(34) \phi_{\rm iss}$ & 0 & 0 & -$\frac {1}{2}$
& 0 & 0 & $\frac {1}{2}$ & 0 & 0 & -$\frac {1}{2}$ & 0  \\
\hline
$\phi_{\rm fss}^+ \sigma_1(21) \sigma_1(34) \phi_{\rm iss}$
& $\frac {1}{2}$ & 0 & 0 & 0 & 0 & 0 & 0 & 0 & 0 & 0 \\
$\phi_{\rm fss}^+ \sigma_1(21) \sigma_2(34) \phi_{\rm iss}$ & 0 & 0 
& -$\frac {1}{2}i$ & 0 & 0 & -$\frac {1}{2}i$ & 0 & 0 & -$\frac {1}{2}i$ & 0 \\
$\phi_{\rm fss}^+ \sigma_1(21) \sigma_3(34) \phi_{\rm iss}$ & 0 
& $\frac {1}{2\sqrt 2}$ & 0 & $\frac {1}{2\sqrt 2}$ & -$\frac {1}{2\sqrt 2}$ 
& 0 & -$\frac {1}{2\sqrt 2}$ & -$\frac {1}{2\sqrt 2}$ & 0 
& -$\frac {1}{2\sqrt 2}$ \\
$\phi_{\rm fss}^+ \sigma_2(21) \sigma_1(34) \phi_{\rm iss}$ & 0 & 0 
& $\frac {1}{2}i$ & 0 & 0 & $\frac {1}{2}i$ & 0 & 0 & $\frac {1}{2}i$ & 0 \\
$\phi_{\rm fss}^+ \sigma_2(21) \sigma_2(34) \phi_{\rm iss}$
& $\frac {1}{2}$ & 0 & 0 & 0 & 0 & 0 & 0 & 0 & 0 & 0 \\
$\phi_{\rm fss}^+ \sigma_2(21) \sigma_3(34) \phi_{\rm iss}$ & 0 
& -$\frac {1}{2\sqrt 2}i$ & 0 & $\frac {1}{2\sqrt 2}i$ 
& -$\frac {1}{2\sqrt 2}i$ & 0 & $\frac {1}{2\sqrt 2}i$ 
& -$\frac {1}{2\sqrt 2}i$ & 0 & $\frac {1}{2\sqrt 2}i$ \\
$\phi_{\rm fss}^+ \sigma_3(21) \sigma_1(34) \phi_{\rm iss}$ & 0 
& -$\frac {1}{2\sqrt 2}$ & 0 & -$\frac {1}{2\sqrt 2}$ & $\frac {1}{2\sqrt 2}$ 
& 0 & $\frac {1}{2\sqrt 2}$ & $\frac {1}{2\sqrt 2}$ & 0 
& $\frac {1}{2\sqrt 2}$ \\
$\phi_{\rm fss}^+ \sigma_3(21) \sigma_2(34) \phi_{\rm iss}$ & 0 
& $\frac {1}{2\sqrt 2}i$ & 0 & -$\frac {1}{2\sqrt 2}i$ 
& $\frac {1}{2\sqrt 2}i$ & 0 & -$\frac {1}{2\sqrt 2}i$ 
& $\frac {1}{2\sqrt 2}i$ & 0 & -$\frac {1}{2\sqrt 2}i$ \\
$\phi_{\rm fss}^+ \sigma_3(21) \sigma_3(34) \phi_{\rm iss}$
& $\frac {1}{2}$ & 0 & 0 & 0 & 0 & 0 & 0 & 0 & 0 & 0 \\
\hline
\end{tabular}
\end{table}

\newpage
\begin{table}
\caption{\label{table2} Spin matrix elements of 
${\cal M}_{{\rm a}q_1\bar {q}_2}$ at $S_A=S_B=0$ and $S_C=S_D=1$. 
The initial spin state is 
$\phi_{\rm iss}=\chi_{S_A S_{Az}} \chi_{S_B S_{Bz}}$, and 
the final spin state $\phi_{\rm fss}=\chi_{S_C S_{Cz}} \chi_{S_D S_{Dz}}$.}
\begin{tabular}{llllllllll}
\hline
$S_{Cz}$ & -1 & -1 & -1 & 0  & 0 & 0 &  1 & 1 & 1 \\
$S_{Dz}$ & -1 & 0  & 1  & -1 & 0 & 1 & -1 & 0 & 1 \\
\hline
$\phi_{\rm fss}^+ \phi_{\rm iss}$ & -$\frac {1}{2}$ & 0 & 0 & 0 
& -$\frac {1}{2}$ & 0 & 0 & 0 & -$\frac {1}{2}$  \\
$\phi_{\rm fss}^+ \sigma_1(21) \phi_{\rm iss}$ & 0 & $\frac {1}{2\sqrt 2}$ & 0
& $\frac {1}{2\sqrt 2}$ & 0 & $\frac {1}{2\sqrt 2}$ & 0 & $\frac {1}{2\sqrt 2}$
& 0  \\
$\phi_{\rm fss}^+ \sigma_2(21) \phi_{\rm iss}$ & 0 & $\frac {1}{2\sqrt 2}i$ 
& 0 & -$\frac {1}{2\sqrt 2}i$ & 0 & $\frac {1}{2\sqrt 2}i$ & 0 
& -$\frac {1}{2\sqrt 2}i$ & 0  \\
$\phi_{\rm fss}^+ \sigma_3(21) \phi_{\rm iss}$ & -$\frac {1}{2}$ & 0 & 0 & 0 
& 0 & 0 & 0 & 0 & $\frac {1}{2}$  \\
$\phi_{\rm fss}^+ \sigma_1(34) \phi_{\rm iss}$ & 0 & -$\frac {1}{2\sqrt 2}$ & 0
& -$\frac {1}{2\sqrt 2}$ & 0 & -$\frac {1}{2\sqrt 2}$ & 0 
& -$\frac {1}{2\sqrt 2}$ & 0  \\
$\phi_{\rm fss}^+ \sigma_2(34) \phi_{\rm iss}$ & 0 & -$\frac {1}{2\sqrt 2}i$ 
& 0 & $\frac {1}{2\sqrt 2}i$ & 0 & -$\frac {1}{2\sqrt 2}i$ & 0 
& $\frac {1}{2\sqrt 2}i$ & 0  \\
$\phi_{\rm fss}^+ \sigma_3(34) \phi_{\rm iss}$ & $\frac {1}{2}$ & 0 & 0
& 0 & 0 & 0 & 0 & 0 & -$\frac {1}{2}$  \\
\hline
$\phi_{\rm fss}^+ \sigma_1(21) \sigma_1(34) \phi_{\rm iss}$ & 0 & 0
& $\frac {1}{2}$ & 0 & $\frac {1}{2}$ & 0 & $\frac {1}{2}$ & 0 & 0  \\
$\phi_{\rm fss}^+ \sigma_1(21) \sigma_2(34) \phi_{\rm iss}$ & 0 & 0 
& $\frac {1}{2}i$ & 0 & 0 & 0 & -$\frac {1}{2}i$ & 0 & 0  \\
$\phi_{\rm fss}^+ \sigma_1(21) \sigma_3(34) \phi_{\rm iss}$ & 0 
& -$\frac {1}{2\sqrt 2}$ & 0 & -$\frac {1}{2\sqrt 2}$ & 0 
& $\frac {1}{2\sqrt 2}$ & 0 & $\frac {1}{2\sqrt 2}$ & 0  \\
$\phi_{\rm fss}^+ \sigma_2(21) \sigma_1(34) \phi_{\rm iss}$ & 0 & 0 
& $\frac {1}{2}i$ & 0 & 0 & 0 & -$\frac {1}{2}i$ & 0 & 0  \\
$\phi_{\rm fss}^+ \sigma_2(21) \sigma_2(34) \phi_{\rm iss}$ & 0 & 0
& -$\frac {1}{2}$ & 0 & $\frac {1}{2}$ & 0 & -$\frac {1}{2}$ & 0 & 0  \\
$\phi_{\rm fss}^+ \sigma_2(21) \sigma_3(34) \phi_{\rm iss}$ & 0 
& -$\frac {1}{2\sqrt 2}i$ & 0 & $\frac {1}{2\sqrt 2}i$ & 0
& $\frac {1}{2\sqrt 2}i$ & 0 & -$\frac {1}{2\sqrt 2}i$ & 0  \\
$\phi_{\rm fss}^+ \sigma_3(21) \sigma_1(34) \phi_{\rm iss}$ & 0 
& -$\frac {1}{2\sqrt 2}$ & 0 & -$\frac {1}{2\sqrt 2}$ & 0 
& $\frac {1}{2\sqrt 2}$ & 0 & $\frac {1}{2\sqrt 2}$ & 0  \\
$\phi_{\rm fss}^+ \sigma_3(21) \sigma_2(34) \phi_{\rm iss}$ & 0 
& -$\frac {1}{2\sqrt 2}i$ & 0 & $\frac {1}{2\sqrt 2}i$ & 0
& $\frac {1}{2\sqrt 2}i$ & 0 & -$\frac {1}{2\sqrt 2}i$ & 0  \\
$\phi_{\rm fss}^+ \sigma_3(21) \sigma_3(34) \phi_{\rm iss}$
& $\frac {1}{2}$ & 0 & 0 & 0 & -$\frac {1}{2}$ & 0 & 0 & 0 & $\frac {1}{2}$  \\
\hline
\end{tabular}
\end{table}

\newpage
\begin{table}
\caption{\label{table3} The same as Table 2 except for 
$S_A=S_C=0$ and $S_B=S_D=1$.}
\begin{tabular}{llllllllll}
\hline
$S_{Bz}$ & -1 & -1 & -1 & 0  & 0 & 0 &  1 & 1 & 1 \\
$S_{Dz}$ & -1 & 0  & 1  & -1 & 0 & 1 & -1 & 0 & 1 \\
\hline
$\phi_{\rm fss}^+ \phi_{\rm iss}$ & $\frac {1}{2}$ & 0 & 0 & 0 
& $\frac {1}{2}$ & 0 & 0 & 0 & $\frac {1}{2}$  \\
$\phi_{\rm fss}^+ \sigma_1(21) \phi_{\rm iss}$ & 0 & $\frac {1}{2\sqrt 2}$ & 0
& $\frac {1}{2\sqrt 2}$ & 0 & $\frac {1}{2\sqrt 2}$ & 0 & $\frac {1}{2\sqrt 2}$
& 0  \\
$\phi_{\rm fss}^+ \sigma_2(21) \phi_{\rm iss}$ & 0 & $\frac {1}{2\sqrt 2}i$ 
& 0 & -$\frac {1}{2\sqrt 2}i$ & 0 & $\frac {1}{2\sqrt 2}i$ & 0 
& -$\frac {1}{2\sqrt 2}i$ & 0  \\
$\phi_{\rm fss}^+ \sigma_3(21) \phi_{\rm iss}$ & -$\frac {1}{2}$ & 0 & 0 & 0 
& 0 & 0 & 0 & 0 & $\frac {1}{2}$  \\
$\phi_{\rm fss}^+ \sigma_1(34) \phi_{\rm iss}$ & 0 & $\frac {1}{2\sqrt 2}$ & 0
& $\frac {1}{2\sqrt 2}$ & 0 & $\frac {1}{2\sqrt 2}$ & 0 
& $\frac {1}{2\sqrt 2}$ & 0  \\
$\phi_{\rm fss}^+ \sigma_2(34) \phi_{\rm iss}$ & 0 & $\frac {1}{2\sqrt 2}i$ & 0
& -$\frac {1}{2\sqrt 2}i$ & 0 & $\frac {1}{2\sqrt 2}i$ & 0 
& -$\frac {1}{2\sqrt 2}i$ & 0  \\
$\phi_{\rm fss}^+ \sigma_3(34) \phi_{\rm iss}$ & -$\frac {1}{2}$ & 0 & 0
& 0 & 0 & 0 & 0 & 0 & $\frac {1}{2}$  \\
\hline
$\phi_{\rm fss}^+ \sigma_1(21) \sigma_1(34) \phi_{\rm iss}$ & $\frac {1}{2}$ 
& 0 & 0 & 0 & $\frac {1}{2}$ & 0 & 0 & 0 & $\frac {1}{2}$  \\
$\phi_{\rm fss}^+ \sigma_1(21) \sigma_2(34) \phi_{\rm iss}$ & -$\frac {1}{2}i$
& 0 & 0 & 0 & 0 & 0 & 0 & 0 & $\frac {1}{2}i$  \\
$\phi_{\rm fss}^+ \sigma_1(21) \sigma_3(34) \phi_{\rm iss}$ & 0 
& $\frac {1}{2\sqrt 2}$ & 0 & -$\frac {1}{2\sqrt 2}$ & 0 
& $\frac {1}{2\sqrt 2}$ & 0 & -$\frac {1}{2\sqrt 2}$ & 0  \\
$\phi_{\rm fss}^+ \sigma_2(21) \sigma_1(34) \phi_{\rm iss}$ & $\frac {1}{2}i$ 
& 0 & 0 & 0 & 0 & 0 & 0 & 0 & -$\frac {1}{2}i$  \\
$\phi_{\rm fss}^+ \sigma_2(21) \sigma_2(34) \phi_{\rm iss}$ & $\frac {1}{2}$ 
& 0 & 0 & 0 & $\frac {1}{2}$ & 0 & 0 & 0 & $\frac {1}{2}$  \\
$\phi_{\rm fss}^+ \sigma_2(21) \sigma_3(34) \phi_{\rm iss}$ & 0 
& $\frac {1}{2\sqrt 2}i$ & 0 & $\frac {1}{2\sqrt 2}i$ & 0
& $\frac {1}{2\sqrt 2}i$ & 0 & $\frac {1}{2\sqrt 2}i$ & 0  \\
$\phi_{\rm fss}^+ \sigma_3(21) \sigma_1(34) \phi_{\rm iss}$ & 0 
& -$\frac {1}{2\sqrt 2}$ & 0 & $\frac {1}{2\sqrt 2}$ & 0 
& -$\frac {1}{2\sqrt 2}$ & 0 & $\frac {1}{2\sqrt 2}$ & 0  \\
$\phi_{\rm fss}^+ \sigma_3(21) \sigma_2(34) \phi_{\rm iss}$ & 0 
& -$\frac {1}{2\sqrt 2}i$ & 0 & -$\frac {1}{2\sqrt 2}i$ & 0
& -$\frac {1}{2\sqrt 2}i$ & 0 & -$\frac {1}{2\sqrt 2}i$ & 0  \\
$\phi_{\rm fss}^+ \sigma_3(21) \sigma_3(34) \phi_{\rm iss}$
& $\frac {1}{2}$ & 0 & 0 & 0 & $\frac {1}{2}$ & 0 & 0 & 0 & $\frac {1}{2}$  \\
\hline
\end{tabular}
\end{table}

\newpage
\begin{table}
\caption{\label{table4} The same as Table 2 except for  $S_A=S_D=0$ and 
$S_B=S_C=1$.}
\begin{tabular}{llllllllll}
\hline
$S_{Bz}$ & -1 & -1 & -1 & 0  & 0 & 0 &  1 & 1 & 1 \\
$S_{Cz}$ & -1 & 0  & 1  & -1 & 0 & 1 & -1 & 0 & 1 \\
\hline
$\phi_{\rm fss}^+ \phi_{\rm iss}$ & 0 & 0 & $\frac {1}{2}$ & 0
& -$\frac {1}{2}$ & 0 & $\frac {1}{2}$ & 0 & 0  \\
$\phi_{\rm fss}^+ \sigma_1(21) \phi_{\rm iss}$ & 0 & -$\frac {1}{2\sqrt 2}$ & 0
& $\frac {1}{2\sqrt 2}$ & 0 & $\frac {1}{2\sqrt 2}$ & 0 
& -$\frac {1}{2\sqrt 2}$ & 0  \\
$\phi_{\rm fss}^+ \sigma_2(21) \phi_{\rm iss}$ & 0 & -$\frac {1}{2\sqrt 2}i$ 
& 0 & $\frac {1}{2\sqrt 2}i$ & 0 & -$\frac {1}{2\sqrt 2}i$ & 0 
& $\frac {1}{2\sqrt 2}i$ & 0  \\
$\phi_{\rm fss}^+ \sigma_3(21) \phi_{\rm iss}$ & 0 & 0 & -$\frac {1}{2}$ & 0 
& 0 & 0 & $\frac {1}{2}$ & 0 & 0  \\
$\phi_{\rm fss}^+ \sigma_1(34) \phi_{\rm iss}$ & 0 & $\frac {1}{2\sqrt 2}$ & 0
& -$\frac {1}{2\sqrt 2}$ & 0 & -$\frac {1}{2\sqrt 2}$ & 0 
& $\frac {1}{2\sqrt 2}$ & 0  \\
$\phi_{\rm fss}^+ \sigma_2(34) \phi_{\rm iss}$ & 0 & $\frac {1}{2\sqrt 2}i$ & 0
& -$\frac {1}{2\sqrt 2}i$ & 0 & $\frac {1}{2\sqrt 2}i$ & 0 
& -$\frac {1}{2\sqrt 2}i$ & 0  \\
$\phi_{\rm fss}^+ \sigma_3(34) \phi_{\rm iss}$ & 0 & 0 & $\frac {1}{2}$ & 0 & 0
& 0 & -$\frac {1}{2}$ & 0 & 0  \\
\hline
$\phi_{\rm fss}^+ \sigma_1(21) \sigma_1(34) \phi_{\rm iss}$ & -$\frac {1}{2}$ 
& 0 & 0 & 0 & $\frac {1}{2}$ & 0 & 0 & 0 & -$\frac {1}{2}$  \\
$\phi_{\rm fss}^+ \sigma_1(21) \sigma_2(34) \phi_{\rm iss}$ & -$\frac {1}{2}i$ 
& 0 & 0 & 0 & 0 & 0 & 0 & 0 & $\frac {1}{2}i$  \\
$\phi_{\rm fss}^+ \sigma_1(21) \sigma_3(34) \phi_{\rm iss}$ & 0 
& -$\frac {1}{2\sqrt 2}$ & 0 & -$\frac {1}{2\sqrt 2}$ & 0 
& $\frac {1}{2\sqrt 2}$ & 0 & $\frac {1}{2\sqrt 2}$ & 0  \\
$\phi_{\rm fss}^+ \sigma_2(21) \sigma_1(34) \phi_{\rm iss}$ & -$\frac {1}{2}i$ 
& 0 & 0 & 0 & 0 & 0 & 0 & 0 & $\frac {1}{2}i$  \\
$\phi_{\rm fss}^+ \sigma_2(21) \sigma_2(34) \phi_{\rm iss}$ & $\frac {1}{2}$ 
& 0 & 0 & 0 & $\frac {1}{2}$ & 0 & 0 & 0 & $\frac {1}{2}$  \\
$\phi_{\rm fss}^+ \sigma_2(21) \sigma_3(34) \phi_{\rm iss}$ & 0 
& -$\frac {1}{2\sqrt 2}i$ & 0 & -$\frac {1}{2\sqrt 2}i$ & 0
& -$\frac {1}{2\sqrt 2}i$ & 0 & -$\frac {1}{2\sqrt 2}i$ & 0  \\
$\phi_{\rm fss}^+ \sigma_3(21) \sigma_1(34) \phi_{\rm iss}$ & 0 
& -$\frac {1}{2\sqrt 2}$ & 0 & -$\frac {1}{2\sqrt 2}$ & 0 
& $\frac {1}{2\sqrt 2}$ & 0 & $\frac {1}{2\sqrt 2}$ & 0  \\
$\phi_{\rm fss}^+ \sigma_3(21) \sigma_2(34) \phi_{\rm iss}$ & 0 
& -$\frac {1}{2\sqrt 2}i$ & 0 & -$\frac {1}{2\sqrt 2}i$ & 0
& -$\frac {1}{2\sqrt 2}i$ & 0 & -$\frac {1}{2\sqrt 2}i$ & 0  \\
$\phi_{\rm fss}^+ \sigma_3(21) \sigma_3(34) \phi_{\rm iss}$ & 0 & 0
& -$\frac {1}{2}$ & 0 & -$\frac {1}{2}$ & 0 & -$\frac {1}{2}$ & 0 & 0  \\
\hline
\end{tabular}
\end{table}

\newpage
\begin{table}
\caption{\label{table5} The same as Table 2 except for $S_A=0$, $S_{Cz}=1$, and
$S_B=S_C=S_D=1$.}
\begin{tabular}{llllllllll}
\hline
$S_{Bz}$ & -1 & -1 & -1 & 0  & 0 & 0 &  1 & 1 & 1 \\
$S_{Dz}$ & -1 & 0  & 1  & -1 & 0 & 1 & -1 & 0 & 1 \\
\hline
$\phi_{\rm fss}^+ \phi_{\rm iss}$ & 0 & -$\frac {1}{2}$ & 0 & 0 & 0 
& -$\frac {1}{2}$ & 0 & 0 & 0 \\
$\phi_{\rm fss}^+ \sigma_1(21) \phi_{\rm iss}$ & 0 & 0 & 0 & 0 
& -$\frac {1}{2\sqrt 2}$ & 0 & 0 & 0 & -$\frac {1}{\sqrt 2}$  \\
$\phi_{\rm fss}^+ \sigma_2(21) \phi_{\rm iss}$ & 0 & 0 & 0 & 0 
& $\frac {1}{2\sqrt 2}i$ & 0 & 0 & 0 & $\frac {1}{\sqrt 2}i$  \\
$\phi_{\rm fss}^+ \sigma_3(21) \phi_{\rm iss}$ & 0 & $\frac {1}{2}$
& 0 & 0 & 0 & $\frac {1}{2}$ & 0 & 0 & 0  \\
$\phi_{\rm fss}^+ \sigma_1(34) \phi_{\rm iss}$ & -$\frac {1}{\sqrt 2}$ & 0 & 0
& 0 & -$\frac {1}{2\sqrt 2}$ & 0 & 0 & 0 & 0  \\
$\phi_{\rm fss}^+ \sigma_2(34) \phi_{\rm iss}$ & $\frac {1}{\sqrt 2}i$ & 0 & 0
& 0 & $\frac {1}{2\sqrt 2}i$ & 0 & 0 & 0 & 0  \\
$\phi_{\rm fss}^+ \sigma_3(34) \phi_{\rm iss}$ & 0 & -$\frac {1}{2}$
& 0 & 0 & 0 & -$\frac {1}{2}$ & 0 & 0 & 0  \\
\hline
$\phi_{\rm fss}^+ \sigma_1(21) \sigma_1(34) \phi_{\rm iss}$
& 0 & 0 & 0 & -$\frac {1}{2}$ & 0 & 0 & 0 & -$\frac {1}{2}$ & 0 \\
$\phi_{\rm fss}^+ \sigma_1(21) \sigma_2(34) \phi_{\rm iss}$ & 0 & 0 & 0
& $\frac {1}{2}i$ & 0 & 0 & 0 & $\frac {1}{2}i$ & 0  \\
$\phi_{\rm fss}^+ \sigma_1(21) \sigma_3(34) \phi_{\rm iss}$ & 0 & 0 & 0 & 0
& -$\frac {1}{2\sqrt 2}$ & 0 & 0 & 0 & -$\frac {1}{\sqrt 2}$  \\
$\phi_{\rm fss}^+ \sigma_2(21) \sigma_1(34) \phi_{\rm iss}$ & 0 & 0 & 0
& $\frac {1}{2}i$ & 0 & 0 & 0 & $\frac {1}{2}i$ & 0  \\
$\phi_{\rm fss}^+ \sigma_2(21) \sigma_2(34) \phi_{\rm iss}$ & 0 & 0 & 0
& $\frac {1}{2}$ & 0 & 0 & 0 & $\frac {1}{2}$ & 0  \\
$\phi_{\rm fss}^+ \sigma_2(21) \sigma_3(34) \phi_{\rm iss}$ & 0 & 0 & 0 & 0
& $\frac {1}{2\sqrt 2}i$ & 0 & 0 & 0 & $\frac {1}{\sqrt 2}i$  \\
$\phi_{\rm fss}^+ \sigma_3(21) \sigma_1(34) \phi_{\rm iss}$ 
& $\frac {1}{\sqrt 2}$ & 0 & 0 & 0 & $\frac {1}{2\sqrt 2}$ & 0 & 0 & 0 & 0  \\
$\phi_{\rm fss}^+ \sigma_3(21) \sigma_2(34) \phi_{\rm iss}$ 
& -$\frac {1}{\sqrt 2}i$ & 0 & 0 & 0 & -$\frac {1}{2\sqrt 2}i$ 
& 0 & 0 & 0 & 0 \\
$\phi_{\rm fss}^+ \sigma_3(21) \sigma_3(34) \phi_{\rm iss}$ & 0
& $\frac {1}{2}$ & 0 & 0 & 0 & $\frac {1}{2}$ & 0 & 0 & 0  \\
\hline
\end{tabular}
\end{table}

\newpage
\begin{table}
\caption{\label{table6} The same as Table 2 except for $S_A=0$, $S_{Cz}=0$, and
$S_B=S_C=S_D=1$.}
\begin{tabular}{llllllllll}
\hline
$S_{Bz}$ & -1 & -1 & -1 & 0  & 0 & 0 &  1 & 1 & 1 \\
$S_{Dz}$ & -1 & 0  & 1  & -1 & 0 & 1 & -1 & 0 & 1 \\
\hline
$\phi_{\rm fss}^+ \phi_{\rm iss}$ & -$\frac {1}{2}$ & 0 & 0 & 0 & 0 & 0 & 0 & 0
& $\frac {1}{2}$  \\
$\phi_{\rm fss}^+ \sigma_1(21) \phi_{\rm iss}$ & 0 & $\frac {1}{2\sqrt 2}$ & 0
& -$\frac {1}{2\sqrt 2}$ & 0 & $\frac {1}{2\sqrt 2}$ & 0 
& -$\frac {1}{2\sqrt 2}$ & 0  \\
$\phi_{\rm fss}^+ \sigma_2(21) \phi_{\rm iss}$ & 0 & $\frac {1}{2\sqrt 2}i$ 
& 0 & $\frac {1}{2\sqrt 2}i$ & 0 & $\frac {1}{2\sqrt 2}i$ & 0 
& $\frac {1}{2\sqrt 2}i$ & 0  \\
$\phi_{\rm fss}^+ \sigma_3(21) \phi_{\rm iss}$ & $\frac {1}{2}$ & 0 & 0 & 0 
& $\frac {1}{2}$ & 0 & 0 & 0 & $\frac {1}{2}$  \\
$\phi_{\rm fss}^+ \sigma_1(34) \phi_{\rm iss}$ & 0 & -$\frac {1}{2\sqrt 2}$ & 0
& $\frac {1}{2\sqrt 2}$ & 0 & -$\frac {1}{2\sqrt 2}$ & 0 
& $\frac {1}{2\sqrt 2}$ & 0  \\
$\phi_{\rm fss}^+ \sigma_2(34) \phi_{\rm iss}$ & 0 & -$\frac {1}{2\sqrt 2}i$ 
& 0 & -$\frac {1}{2\sqrt 2}i$ & 0 & -$\frac {1}{2\sqrt 2}i$ & 0 
& -$\frac {1}{2\sqrt 2}i$ & 0  \\
$\phi_{\rm fss}^+ \sigma_3(34) \phi_{\rm iss}$ & $\frac {1}{2}$ & 0 & 0
& 0 & $\frac {1}{2}$ & 0 & 0 & 0 & $\frac {1}{2}$  \\
\hline
$\phi_{\rm fss}^+ \sigma_1(21) \sigma_1(34) \phi_{\rm iss}$ & $\frac {1}{2}$ 
& 0 & 0 & 0 & 0 & 0 & 0 & 0 & -$\frac {1}{2}$  \\
$\phi_{\rm fss}^+ \sigma_1(21) \sigma_2(34) \phi_{\rm iss}$ & -$\frac {1}{2}i$
& 0 & 0 & 0 & -$\frac {1}{2}i$ & 0 & 0 & 0 & -$\frac {1}{2}i$  \\
$\phi_{\rm fss}^+ \sigma_1(21) \sigma_3(34) \phi_{\rm iss}$ & 0 
& $\frac {1}{2\sqrt 2}$ & 0 & $\frac {1}{2\sqrt 2}$ & 0 
& $\frac {1}{2\sqrt 2}$ & 0 & $\frac {1}{2\sqrt 2}$ & 0  \\
$\phi_{\rm fss}^+ \sigma_2(21) \sigma_1(34) \phi_{\rm iss}$ & $\frac {1}{2}i$
& 0 & 0 & 0 & $\frac {1}{2}i$ & 0 & 0 & 0 & $\frac {1}{2}i$  \\
$\phi_{\rm fss}^+ \sigma_2(21) \sigma_2(34) \phi_{\rm iss}$ & $\frac {1}{2}$ 
& 0 & 0 & 0 & 0 & 0 & 0 & 0 & -$\frac {1}{2}$  \\
$\phi_{\rm fss}^+ \sigma_2(21) \sigma_3(34) \phi_{\rm iss}$ & 0 
& $\frac {1}{2\sqrt 2}i$ & 0 & -$\frac {1}{2\sqrt 2}i$ & 0
& $\frac {1}{2\sqrt 2}i$ & 0 & -$\frac {1}{2\sqrt 2}i$ & 0  \\
$\phi_{\rm fss}^+ \sigma_3(21) \sigma_1(34) \phi_{\rm iss}$ & 0 
& $\frac {1}{2\sqrt 2}$ & 0 & $\frac {1}{2\sqrt 2}$ & 0 
& $\frac {1}{2\sqrt 2}$ & 0 & $\frac {1}{2\sqrt 2}$ & 0  \\
$\phi_{\rm fss}^+ \sigma_3(21) \sigma_2(34) \phi_{\rm iss}$ & 0 
& $\frac {1}{2\sqrt 2}i$ & 0 & -$\frac {1}{2\sqrt 2}i$ & 0
& $\frac {1}{2\sqrt 2}i$ & 0 & -$\frac {1}{2\sqrt 2}i$ & 0  \\
$\phi_{\rm fss}^+ \sigma_3(21) \sigma_3(34) \phi_{\rm iss}$
& -$\frac {1}{2}$ & 0 & 0 & 0 & 0 & 0 & 0 & 0 & $\frac {1}{2}$  \\
\hline
\end{tabular}
\end{table}

\newpage
\begin{table}
\caption{\label{table7} The same as Table 2 except for $S_A=0$, $S_{Cz}=-1$, 
and $S_B=S_C=S_D=1$.}
\begin{tabular}{llllllllll}
\hline
$S_{Bz}$ & -1 & -1 & -1 & 0  & 0 & 0 &  1 & 1 & 1 \\
$S_{Dz}$ & -1 & 0  & 1  & -1 & 0 & 1 & -1 & 0 & 1 \\
\hline
$\phi_{\rm fss}^+ \phi_{\rm iss}$ & 0 & 0 & 0 & $\frac {1}{2}$ & 0 & 0 & 0 
& $\frac {1}{2}$ & 0  \\
$\phi_{\rm fss}^+ \sigma_1(21) \phi_{\rm iss}$ & $\frac {1}{\sqrt 2}$ & 0 & 0 
& 0 & $\frac {1}{2\sqrt 2}$ & 0 & 0 & 0 & 0  \\
$\phi_{\rm fss}^+ \sigma_2(21) \phi_{\rm iss}$ & $\frac {1}{\sqrt 2}i$ & 0 & 0 
& 0 & $\frac {1}{2\sqrt 2}i$ & 0 & 0 & 0 & 0  \\
$\phi_{\rm fss}^+ \sigma_3(21) \phi_{\rm iss}$ & 0 & 0 & 0 & $\frac {1}{2}$
& 0 & 0 & 0 & $\frac {1}{2}$ & 0  \\
$\phi_{\rm fss}^+ \sigma_1(34) \phi_{\rm iss}$ & 0 & 0 & 0 & 0 
& $\frac {1}{2\sqrt 2}$ & 0 & 0 & 0 & $\frac {1}{\sqrt 2}$  \\
$\phi_{\rm fss}^+ \sigma_2(34) \phi_{\rm iss}$ & 0 & 0 & 0 & 0 
& $\frac {1}{2\sqrt 2}i$ & 0 & 0 & 0 & $\frac {1}{\sqrt 2}i$  \\
$\phi_{\rm fss}^+ \sigma_3(34) \phi_{\rm iss}$ & 0 & 0 & 0 & -$\frac {1}{2}$
& 0 & 0 & 0 & -$\frac {1}{2}$ & 0  \\
\hline
$\phi_{\rm fss}^+ \sigma_1(21) \sigma_1(34) \phi_{\rm iss}$
& 0 & $\frac {1}{2}$ & 0 & 0 & 0 & $\frac {1}{2}$ & 0 & 0 & 0  \\
$\phi_{\rm fss}^+ \sigma_1(21) \sigma_2(34) \phi_{\rm iss}$ & 0 
& $\frac {1}{2}i$ & 0 & 0 & 0 & $\frac {1}{2}i$ & 0 & 0 & 0  \\
$\phi_{\rm fss}^+ \sigma_1(21) \sigma_3(34) \phi_{\rm iss}$ 
& -$\frac {1}{\sqrt 2}$ & 0 & 0 & 0 & -$\frac {1}{2\sqrt 2}$ & 0 & 0 & 0 & 0 \\
$\phi_{\rm fss}^+ \sigma_2(21) \sigma_1(34) \phi_{\rm iss}$ & 0 
& $\frac {1}{2}i$ & 0 & 0 & 0 & $\frac {1}{2}i$ & 0 & 0 & 0  \\
$\phi_{\rm fss}^+ \sigma_2(21) \sigma_2(34) \phi_{\rm iss}$ & 0 
& -$\frac {1}{2}$ & 0 & 0 & 0 & -$\frac {1}{2}$ & 0 & 0 & 0  \\
$\phi_{\rm fss}^+ \sigma_2(21) \sigma_3(34) \phi_{\rm iss}$ 
& -$\frac {1}{\sqrt 2}i$ & 0 & 0 & 0 & -$\frac {1}{2\sqrt 2}i$ & 0 & 0 & 0 
& 0  \\
$\phi_{\rm fss}^+ \sigma_3(21) \sigma_1(34) \phi_{\rm iss}$ & 0 & 0 & 0 & 0
& $\frac {1}{2\sqrt 2}$ & 0 & 0 & 0 & $\frac {1}{\sqrt 2}$  \\
$\phi_{\rm fss}^+ \sigma_3(21) \sigma_2(34) \phi_{\rm iss}$ & 0 & 0 & 0 & 0
& $\frac {1}{2\sqrt 2}i$ & 0 & 0 & 0 & $\frac {1}{\sqrt 2}i$  \\
$\phi_{\rm fss}^+ \sigma_3(21) \sigma_3(34) \phi_{\rm iss}$ & 0 & 0 & 0
& -$\frac {1}{2}$ & 0 & 0 & 0 & -$\frac {1}{2}$ & 0  \\
\hline
\end{tabular}
\end{table}

\newpage
\begin{table}
\caption{\label{table8} The same as Table 2 except for $S_A=S_B=1$ and
$S_C=S_D=0$.}
\begin{tabular}{llllllllll}
\hline
$S_{Az}$ & -1 & -1 & -1 & 0  & 0 & 0 &  1 & 1 & 1 \\
$S_{Bz}$ & -1 & 0  & 1  & -1 & 0 & 1 & -1 & 0 & 1 \\
\hline
$\phi_{\rm fss}^+ \phi_{\rm iss}$ & -$\frac {1}{2}$ & 0 & 0 
& 0 & -$\frac {1}{2}$ & 0 & 0 & 0 & -$\frac {1}{2}$  \\
$\phi_{\rm fss}^+ \sigma_1(21) \phi_{\rm iss}$ & 0 & -$\frac {1}{2\sqrt 2}$ & 0
& -$\frac {1}{2\sqrt 2}$ & 0 & -$\frac {1}{2\sqrt 2}$ & 0 
& -$\frac {1}{2\sqrt 2}$ & 0  \\
$\phi_{\rm fss}^+ \sigma_2(21) \phi_{\rm iss}$ & 0 & $\frac {1}{2\sqrt 2}i$ 
& 0 & -$\frac {1}{2\sqrt 2}i$ & 0 & $\frac {1}{2\sqrt 2}i$ & 0 
& -$\frac {1}{2\sqrt 2}i$ & 0  \\
$\phi_{\rm fss}^+ \sigma_3(21) \phi_{\rm iss}$ & $\frac {1}{2}$ & 0 & 0 & 0 
& 0 & 0 & 0 & 0 & -$\frac {1}{2}$  \\
$\phi_{\rm fss}^+ \sigma_1(34) \phi_{\rm iss}$ & 0 & $\frac {1}{2\sqrt 2}$ & 0
& $\frac {1}{2\sqrt 2}$ & 0 & $\frac {1}{2\sqrt 2}$ & 0 
& $\frac {1}{2\sqrt 2}$ & 0  \\
$\phi_{\rm fss}^+ \sigma_2(34) \phi_{\rm iss}$ & 0 & -$\frac {1}{2\sqrt 2}i$ 
& 0 & $\frac {1}{2\sqrt 2}i$ & 0 & -$\frac {1}{2\sqrt 2}i$ & 0 
& $\frac {1}{2\sqrt 2}i$ & 0  \\
$\phi_{\rm fss}^+ \sigma_3(34) \phi_{\rm iss}$ & -$\frac {1}{2}$ & 0 & 0
& 0 & 0 & 0 & 0 & 0 & $\frac {1}{2}$  \\
\hline
$\phi_{\rm fss}^+ \sigma_1(21) \sigma_1(34) \phi_{\rm iss}$ & 0 & 0 
& $\frac {1}{2}$ & 0 & $\frac {1}{2}$ & 0 & $\frac {1}{2}$  & 0 & 0  \\
$\phi_{\rm fss}^+ \sigma_1(21) \sigma_2(34) \phi_{\rm iss}$ & 0 & 0
& -$\frac {1}{2}i$ & 0 & 0 & 0 & $\frac {1}{2}i$ & 0 & 0  \\
$\phi_{\rm fss}^+ \sigma_1(21) \sigma_3(34) \phi_{\rm iss}$ & 0 
& -$\frac {1}{2\sqrt 2}$ & 0 & -$\frac {1}{2\sqrt 2}$ & 0 
& $\frac {1}{2\sqrt 2}$ & 0 & $\frac {1}{2\sqrt 2}$ & 0  \\
$\phi_{\rm fss}^+ \sigma_2(21) \sigma_1(34) \phi_{\rm iss}$ & 0 & 0
& -$\frac {1}{2}i$ & 0 & 0 & 0 & $\frac {1}{2}i$ & 0 & 0  \\
$\phi_{\rm fss}^+ \sigma_2(21) \sigma_2(34) \phi_{\rm iss}$ & 0 & 0
& -$\frac {1}{2}$ & 0 & $\frac {1}{2}$ & 0 & -$\frac {1}{2}$ & 0 & 0  \\
$\phi_{\rm fss}^+ \sigma_2(21) \sigma_3(34) \phi_{\rm iss}$ & 0 
& $\frac {1}{2\sqrt 2}i$ & 0 & -$\frac {1}{2\sqrt 2}i$ & 0
& -$\frac {1}{2\sqrt 2}i$ & 0 & $\frac {1}{2\sqrt 2}i$ & 0  \\
$\phi_{\rm fss}^+ \sigma_3(21) \sigma_1(34) \phi_{\rm iss}$ & 0 
& -$\frac {1}{2\sqrt 2}$ & 0 & -$\frac {1}{2\sqrt 2}$ & 0 
& $\frac {1}{2\sqrt 2}$ & 0 & $\frac {1}{2\sqrt 2}$ & 0  \\
$\phi_{\rm fss}^+ \sigma_3(21) \sigma_2(34) \phi_{\rm iss}$ & 0 
& $\frac {1}{2\sqrt 2}i$ & 0 & -$\frac {1}{2\sqrt 2}i$ & 0
& -$\frac {1}{2\sqrt 2}i$ & 0 & $\frac {1}{2\sqrt 2}i$ & 0  \\
$\phi_{\rm fss}^+ \sigma_3(21) \sigma_3(34) \phi_{\rm iss}$
& $\frac {1}{2}$ & 0 & 0 & 0 & -$\frac {1}{2}$ & 0 & 0 & 0 & $\frac {1}{2}$  \\
\hline
\end{tabular}
\end{table}

\newpage
\begin{table}
\caption{\label{table9} The same as Table 2 except for $S_C=0$, $S_{Az}=1$, and
$S_A=S_B=S_D=1$.}
\begin{tabular}{llllllllll}
\hline
$S_{Bz}$ & -1 & -1 & -1 & 0  & 0 & 0 &  1 & 1 & 1 \\
$S_{Dz}$ & -1 & 0  & 1  & -1 & 0 & 1 & -1 & 0 & 1 \\
\hline
$\phi_{\rm fss}^+ \phi_{\rm iss}$ & 0 & 0 & 0 & -$\frac {1}{2}$ & 0 & 0 & 0 
& -$\frac {1}{2}$ & 0  \\
$\phi_{\rm fss}^+ \sigma_1(21) \phi_{\rm iss}$ & -$\frac {1}{\sqrt 2}$ & 0 & 0
& 0 & -$\frac {1}{2\sqrt 2}$ & 0 & 0 & 0 & 0  \\
$\phi_{\rm fss}^+ \sigma_2(21) \phi_{\rm iss}$ & -$\frac {1}{\sqrt 2}i$ & 0 & 0
& 0 & -$\frac {1}{2\sqrt 2}i$ & 0 & 0 & 0 & 0  \\
$\phi_{\rm fss}^+ \sigma_3(21) \phi_{\rm iss}$ & 0 & 0 & 0 & -$\frac {1}{2}$
& 0 & 0 & 0 & -$\frac {1}{2}$ & 0  \\
$\phi_{\rm fss}^+ \sigma_1(34) \phi_{\rm iss}$ & 0 & 0 & 0 & 0
& -$\frac {1}{2\sqrt 2}$ & 0 & 0 & 0 & -$\frac {1}{\sqrt 2}$  \\
$\phi_{\rm fss}^+ \sigma_2(34) \phi_{\rm iss}$ & 0 & 0 & 0 & 0
& -$\frac {1}{2\sqrt 2}i$ & 0 & 0 & 0 & -$\frac {1}{\sqrt 2}i$  \\
$\phi_{\rm fss}^+ \sigma_3(34) \phi_{\rm iss}$ & 0 & 0 & 0 & $\frac {1}{2}$
& 0 & 0 & 0 & $\frac {1}{2}$ & 0  \\
\hline
$\phi_{\rm fss}^+ \sigma_1(21) \sigma_1(34) \phi_{\rm iss}$
& 0 & -$\frac {1}{2}$ & 0 & 0 & 0 & -$\frac {1}{2}$ & 0 & 0 & 0  \\
$\phi_{\rm fss}^+ \sigma_1(21) \sigma_2(34) \phi_{\rm iss}$ & 0
& -$\frac {1}{2}i$ & 0 & 0 & 0 & -$\frac {1}{2}i$ & 0 & 0 & 0  \\
$\phi_{\rm fss}^+ \sigma_1(21) \sigma_3(34) \phi_{\rm iss}$ 
& $\frac {1}{\sqrt 2}$ & 0 & 0 & 0 & $\frac {1}{2\sqrt 2}$ & 0 & 0 & 0 & 0  \\
$\phi_{\rm fss}^+ \sigma_2(21) \sigma_1(34) \phi_{\rm iss}$ & 0 
& -$\frac {1}{2}i$ & 0 & 0 & 0 & -$\frac {1}{2}i$ & 0 & 0 & 0  \\
$\phi_{\rm fss}^+ \sigma_2(21) \sigma_2(34) \phi_{\rm iss}$ & 0 
& $\frac {1}{2}$ & 0 & 0 & 0 & $\frac {1}{2}$ & 0 & 0 & 0  \\
$\phi_{\rm fss}^+ \sigma_2(21) \sigma_3(34) \phi_{\rm iss}$ 
& $\frac {1}{\sqrt 2}i$ & 0 & 0 & 0 & $\frac {1}{2\sqrt 2}i$ & 0 & 0 & 0 & 0 \\
$\phi_{\rm fss}^+ \sigma_3(21) \sigma_1(34) \phi_{\rm iss}$ 
& 0 & 0 & 0 & 0 & -$\frac {1}{2\sqrt 2}$ & 0 & 0 & 0 & -$\frac {1}{\sqrt 2}$ \\
$\phi_{\rm fss}^+ \sigma_3(21) \sigma_2(34) \phi_{\rm iss}$ & 0 & 0 & 0 & 0
& -$\frac {1}{2\sqrt 2}i$ & 0 & 0 & 0 & -$\frac {1}{\sqrt 2}i$  \\
$\phi_{\rm fss}^+ \sigma_3(21) \sigma_3(34) \phi_{\rm iss}$ & 0 & 0 & 0
& $\frac {1}{2}$ & 0 & 0 & 0 & $\frac {1}{2}$ & 0  \\
\hline
\end{tabular}
\end{table}

\newpage
\begin{table}
\caption{\label{table10} The same as Table 2 except for $S_C=0$, $S_{Az}=0$,
and $S_A=S_B=S_D=1$.}
\begin{tabular}{llllllllll}
\hline
$S_{Bz}$ & -1 & -1 & -1 & 0  & 0 & 0 &  1 & 1 & 1 \\
$S_{Dz}$ & -1 & 0  & 1  & -1 & 0 & 1 & -1 & 0 & 1 \\
\hline
$\phi_{\rm fss}^+ \phi_{\rm iss}$ & -$\frac {1}{2}$ & 0 & 0 & 0 & 0 & 0 & 0 & 0
& $\frac {1}{2}$ \\
$\phi_{\rm fss}^+ \sigma_1(21) \phi_{\rm iss}$ & 0 & $\frac {1}{2\sqrt 2}$ & 0
& -$\frac {1}{2\sqrt 2}$ & 0 & $\frac {1}{2\sqrt 2}$ & 0 
& -$\frac {1}{2\sqrt 2}$ & 0  \\
$\phi_{\rm fss}^+ \sigma_2(21) \phi_{\rm iss}$ & 0 & $\frac {1}{2\sqrt 2}i$ 
& 0 & $\frac {1}{2\sqrt 2}i$ & 0 & $\frac {1}{2\sqrt 2}i$ & 0 
& $\frac {1}{2\sqrt 2}i$ & 0  \\
$\phi_{\rm fss}^+ \sigma_3(21) \phi_{\rm iss}$ & $\frac {1}{2}$ & 0 & 0 & 0 
& $\frac {1}{2}$ & 0 & 0 & 0 & $\frac {1}{2}$  \\
$\phi_{\rm fss}^+ \sigma_1(34) \phi_{\rm iss}$ & 0 & -$\frac {1}{2\sqrt 2}$ & 0
& $\frac {1}{2\sqrt 2}$ & 0 & -$\frac {1}{2\sqrt 2}$ & 0 
& $\frac {1}{2\sqrt 2}$ & 0  \\
$\phi_{\rm fss}^+ \sigma_2(34) \phi_{\rm iss}$ & 0 & -$\frac {1}{2\sqrt 2}i$ 
& 0 & -$\frac {1}{2\sqrt 2}i$ & 0 & -$\frac {1}{2\sqrt 2}i$ & 0 
& -$\frac {1}{2\sqrt 2}i$ & 0  \\
$\phi_{\rm fss}^+ \sigma_3(34) \phi_{\rm iss}$ & $\frac {1}{2}$ & 0 & 0
& 0 & $\frac {1}{2}$ & 0 & 0 & 0 & $\frac {1}{2}$  \\
\hline
$\phi_{\rm fss}^+ \sigma_1(21) \sigma_1(34) \phi_{\rm iss}$ & $\frac {1}{2}$ 
& 0 & 0 & 0 & 0 & 0 & 0 & 0 & -$\frac {1}{2}$  \\
$\phi_{\rm fss}^+ \sigma_1(21) \sigma_2(34) \phi_{\rm iss}$ & -$\frac {1}{2}i$
& 0 & 0 & 0 & -$\frac {1}{2}i$ & 0 & 0 & 0 & -$\frac {1}{2}i$  \\
$\phi_{\rm fss}^+ \sigma_1(21) \sigma_3(34) \phi_{\rm iss}$ & 0 
& $\frac {1}{2\sqrt 2}$ & 0 & $\frac {1}{2\sqrt 2}$ & 0 
& $\frac {1}{2\sqrt 2}$ & 0 & $\frac {1}{2\sqrt 2}$ & 0  \\
$\phi_{\rm fss}^+ \sigma_2(21) \sigma_1(34) \phi_{\rm iss}$ & $\frac {1}{2}i$ 
& 0 & 0 & 0 & $\frac {1}{2}i$ & 0 & 0 & 0 & $\frac {1}{2}i$  \\
$\phi_{\rm fss}^+ \sigma_2(21) \sigma_2(34) \phi_{\rm iss}$ & $\frac {1}{2}$ 
& 0 & 0 & 0 & 0 & 0 & 0 & 0 & -$\frac {1}{2}$  \\
$\phi_{\rm fss}^+ \sigma_2(21) \sigma_3(34) \phi_{\rm iss}$ & 0 
& $\frac {1}{2\sqrt 2}i$ & 0 & -$\frac {1}{2\sqrt 2}i$ & 0
& $\frac {1}{2\sqrt 2}i$ & 0 & -$\frac {1}{2\sqrt 2}i$ & 0  \\
$\phi_{\rm fss}^+ \sigma_3(21) \sigma_1(34) \phi_{\rm iss}$ & 0 
& $\frac {1}{2\sqrt 2}$ & 0 & $\frac {1}{2\sqrt 2}$ & 0 
& $\frac {1}{2\sqrt 2}$ & 0 & $\frac {1}{2\sqrt 2}$ & 0  \\
$\phi_{\rm fss}^+ \sigma_3(21) \sigma_2(34) \phi_{\rm iss}$ & 0 
& $\frac {1}{2\sqrt 2}i$ & 0 & -$\frac {1}{2\sqrt 2}i$ & 0
& $\frac {1}{2\sqrt 2}i$ & 0 & -$\frac {1}{2\sqrt 2}i$ & 0  \\
$\phi_{\rm fss}^+ \sigma_3(21) \sigma_3(34) \phi_{\rm iss}$
& -$\frac {1}{2}$ & 0 & 0 & 0 & 0 & 0 & 0 & 0 & $\frac {1}{2}$  \\
\hline
\end{tabular}
\end{table}

\newpage
\begin{table}
\caption{\label{table11} The same as Table 2 except for $S_C=0$, $S_{Az}=-1$, 
and $S_A=S_B=S_D=1$.}
\begin{tabular}{llllllllll}
\hline
$S_{Bz}$ & -1 & -1 & -1 & 0  & 0 & 0 &  1 & 1 & 1 \\
$S_{Dz}$ & -1 & 0  & 1  & -1 & 0 & 1 & -1 & 0 & 1 \\
\hline
$\phi_{\rm fss}^+ \phi_{\rm iss}$ & 0 & $\frac {1}{2}$ & 0 & 0  
& 0 & $\frac {1}{2}$ & 0 & 0 & 0 \\
$\phi_{\rm fss}^+ \sigma_1(21) \phi_{\rm iss}$ & 0 & 0 & 0 & 0 
& $\frac {1}{2\sqrt 2}$ & 0 & 0 & 0 & $\frac {1}{\sqrt 2}$  \\
$\phi_{\rm fss}^+ \sigma_2(21) \phi_{\rm iss}$ & 0 & 0 & 0 & 0 
& -$\frac {1}{2\sqrt 2}i$ & 0 & 0 & 0 & -$\frac {1}{\sqrt 2}i$  \\
$\phi_{\rm fss}^+ \sigma_3(21) \phi_{\rm iss}$ & 0 & -$\frac {1}{2}$
& 0 & 0 & 0 & -$\frac {1}{2}$ & 0 & 0 & 0  \\
$\phi_{\rm fss}^+ \sigma_1(34) \phi_{\rm iss}$ & $\frac {1}{\sqrt 2}$ & 0 & 0
& 0 & $\frac {1}{2\sqrt 2}$ & 0 & 0 & 0 & 0  \\
$\phi_{\rm fss}^+ \sigma_2(34) \phi_{\rm iss}$ & -$\frac {1}{\sqrt 2}i$ & 0 & 0
& 0 & -$\frac {1}{2\sqrt 2}i$ & 0 & 0 & 0 & 0  \\
$\phi_{\rm fss}^+ \sigma_3(34) \phi_{\rm iss}$ & 0 & $\frac {1}{2}$
& 0 & 0 & 0 & $\frac {1}{2}$ & 0 & 0 & 0  \\
\hline
$\phi_{\rm fss}^+ \sigma_1(21) \sigma_1(34) \phi_{\rm iss}$
& 0 & 0 & 0 & $\frac {1}{2}$ & 0 & 0 & 0 & $\frac {1}{2}$ & 0 \\
$\phi_{\rm fss}^+ \sigma_1(21) \sigma_2(34) \phi_{\rm iss}$ & 0 & 0 & 0
& -$\frac {1}{2}i$ & 0 & 0 & 0 & -$\frac {1}{2}i$ & 0  \\
$\phi_{\rm fss}^+ \sigma_1(21) \sigma_3(34) \phi_{\rm iss}$ & 0 & 0 & 0 & 0
& $\frac {1}{2\sqrt 2}$ & 0 & 0 & 0 & $\frac {1}{\sqrt 2}$  \\
$\phi_{\rm fss}^+ \sigma_2(21) \sigma_1(34) \phi_{\rm iss}$ & 0 & 0 & 0
& -$\frac {1}{2}i$ & 0 & 0 & 0 & -$\frac {1}{2}i$ & 0  \\
$\phi_{\rm fss}^+ \sigma_2(21) \sigma_2(34) \phi_{\rm iss}$ & 0 & 0 & 0
& -$\frac {1}{2}$ & 0 & 0 & 0 & -$\frac {1}{2}$ & 0  \\
$\phi_{\rm fss}^+ \sigma_2(21) \sigma_3(34) \phi_{\rm iss}$ & 0 & 0 & 0 & 0
& -$\frac {1}{2\sqrt 2}i$ & 0 & 0 & 0 & -$\frac {1}{\sqrt 2}i$  \\
$\phi_{\rm fss}^+ \sigma_3(21) \sigma_1(34) \phi_{\rm iss}$ 
& -$\frac {1}{\sqrt 2}$ & 0 & 0 & 0 & -$\frac {1}{2\sqrt 2}$ & 0 & 0 & 0 & 0 \\
$\phi_{\rm fss}^+ \sigma_3(21) \sigma_2(34) \phi_{\rm iss}$ 
& $\frac {1}{\sqrt 2}i$ & 0 & 0 & 0 & $\frac {1}{2\sqrt 2}i$ & 0 & 0 & 0 & 0 \\
$\phi_{\rm fss}^+ \sigma_3(21) \sigma_3(34) \phi_{\rm iss}$ & 0
& -$\frac {1}{2}$ & 0 & 0 & 0 & -$\frac {1}{2}$ & 0 & 0 & 0  \\
\hline
\end{tabular}
\end{table}

\newpage
\begin{table}
\caption{\label{table12} The same as Table 2 except for $S_D=0$, $S_{Az}=1$, 
and $S_A=S_B=S_C=1$.}
\begin{tabular}{llllllllll}
\hline
$S_{Bz}$ & -1 & -1 & -1 & 0  & 0 & 0 &  1 & 1 & 1 \\
$S_{Cz}$ & -1 & 0  & 1  & -1 & 0 & 1 & -1 & 0 & 1 \\
\hline
$\phi_{\rm fss}^+ \phi_{\rm iss}$ & 0 & 0 & 0 & 0 & 0
& -$\frac {1}{2}$ & 0 & $\frac {1}{2}$ & 0  \\
$\phi_{\rm fss}^+ \sigma_1(21) \phi_{\rm iss}$ & 0 & 0 
& -$\frac {1}{\sqrt 2}$ & 0 & $\frac {1}{2\sqrt 2}$ & 0 & 0 & 0 & 0  \\
$\phi_{\rm fss}^+ \sigma_2(21) \phi_{\rm iss}$ & 0 & 0 
& -$\frac {1}{\sqrt 2}i$ & 0 & $\frac {1}{2\sqrt 2}i$ & 0 & 0 & 0 & 0  \\
$\phi_{\rm fss}^+ \sigma_3(21) \phi_{\rm iss}$ & 0 & 0 & 0 & 0 & 0 
& -$\frac {1}{2}$ & 0 & $\frac {1}{2}$ & 0  \\
$\phi_{\rm fss}^+ \sigma_1(34) \phi_{\rm iss}$ & 0 & 0 & 0 & 0
& -$\frac {1}{2\sqrt 2}$ & 0 & 0 & 0 & $\frac {1}{\sqrt 2}$  \\
$\phi_{\rm fss}^+ \sigma_2(34) \phi_{\rm iss}$ & 0 & 0 & 0 & 0
& -$\frac {1}{2\sqrt 2}i$ & 0 & 0 & 0 & -$\frac {1}{\sqrt 2}i$  \\
$\phi_{\rm fss}^+ \sigma_3(34) \phi_{\rm iss}$ & 0 & 0 & 0 & 0 & 0 
& -$\frac {1}{2}$ & 0 & -$\frac {1}{2}$ & 0  \\
\hline
$\phi_{\rm fss}^+ \sigma_1(21) \sigma_1(34) \phi_{\rm iss}$ & 0 
& -$\frac {1}{2}$ & 0 & 0 & 0 & $\frac {1}{2}$ & 0 & 0 & 0  \\
$\phi_{\rm fss}^+ \sigma_1(21) \sigma_2(34) \phi_{\rm iss}$ & 0
& -$\frac {1}{2}i$ & 0 & 0 & 0 & -$\frac {1}{2}i$ & 0 & 0 & 0  \\
$\phi_{\rm fss}^+ \sigma_1(21) \sigma_3(34) \phi_{\rm iss}$ & 0 & 0
& -$\frac {1}{\sqrt 2}$ & 0 & -$\frac {1}{2\sqrt 2}$ & 0 
& 0 & 0 & 0  \\
$\phi_{\rm fss}^+ \sigma_2(21) \sigma_1(34) \phi_{\rm iss}$ & 0 
& -$\frac {1}{2}i$ & 0 & 0 & 0 & $\frac {1}{2}i$ & 0 & 0 & 0  \\
$\phi_{\rm fss}^+ \sigma_2(21) \sigma_2(34) \phi_{\rm iss}$ & 0 
& $\frac {1}{2}$ & 0 & 0 & 0 & $\frac {1}{2}$ & 0 & 0 & 0  \\
$\phi_{\rm fss}^+ \sigma_2(21) \sigma_3(34) \phi_{\rm iss}$ & 0 & 0
& -$\frac {1}{\sqrt 2}i$ & 0 & -$\frac {1}{2\sqrt 2}i$ & 0
& 0 & 0 & 0  \\
$\phi_{\rm fss}^+ \sigma_3(21) \sigma_1(34) \phi_{\rm iss}$ & 0 
& 0 & 0 & 0 & -$\frac {1}{2\sqrt 2}$ & 0 & 0 & 0 & $\frac {1}{\sqrt 2}$  \\
$\phi_{\rm fss}^+ \sigma_3(21) \sigma_2(34) \phi_{\rm iss}$ & 0 & 0
& 0 & 0 & -$\frac {1}{2\sqrt 2}i$ & 0 & 0 & 0 & -$\frac {1}{\sqrt 2}i$  \\
$\phi_{\rm fss}^+ \sigma_3(21) \sigma_3(34) \phi_{\rm iss}$ & 0 & 0 & 0 & 0 & 0
& -$\frac {1}{2}$ & 0 & -$\frac {1}{2}$ & 0  \\
\hline
\end{tabular}
\end{table}

\newpage
\begin{table}
\caption{\label{table13} The same as Table 2 except for $S_D=0$, $S_{Az}=0$,
and $S_A=S_B=S_C=1$.}
\begin{tabular}{llllllllll}
\hline
$S_{Bz}$ & -1 & -1 & -1 & 0  & 0 & 0 &  1 & 1 & 1 \\
$S_{Cz}$ & -1 & 0  & 1  & -1 & 0 & 1 & -1 & 0 & 1 \\
\hline
$\phi_{\rm fss}^+ \phi_{\rm iss}$ & 0 & 0 & -$\frac {1}{2}$ & 0 & 0 & 0 
& $\frac {1}{2}$ & 0 & 0 \\
$\phi_{\rm fss}^+ \sigma_1(21) \phi_{\rm iss}$ & 0 & -$\frac {1}{2\sqrt 2}$ & 0
& $\frac {1}{2\sqrt 2}$ & 0 & -$\frac {1}{2\sqrt 2}$ & 0 
& $\frac {1}{2\sqrt 2}$ & 0  \\
$\phi_{\rm fss}^+ \sigma_2(21) \phi_{\rm iss}$ & 0 & -$\frac {1}{2\sqrt 2}i$ 
& 0 & $\frac {1}{2\sqrt 2}i$ & 0 & $\frac {1}{2\sqrt 2}i$ & 0 
& -$\frac {1}{2\sqrt 2}i$ & 0  \\
$\phi_{\rm fss}^+ \sigma_3(21) \phi_{\rm iss}$ & 0 & 0 & $\frac {1}{2}$ & 0
& -$\frac {1}{2}$ & 0 & $\frac {1}{2}$ & 0 & 0  \\
$\phi_{\rm fss}^+ \sigma_1(34) \phi_{\rm iss}$ & 0 & -$\frac {1}{2\sqrt 2}$ & 0
& -$\frac {1}{2\sqrt 2}$ & 0 & $\frac {1}{2\sqrt 2}$ & 0 
& $\frac {1}{2\sqrt 2}$ & 0  \\
$\phi_{\rm fss}^+ \sigma_2(34) \phi_{\rm iss}$ & 0 & -$\frac {1}{2\sqrt 2}i$ 
& 0 & -$\frac {1}{2\sqrt 2}i$ & 0 & -$\frac {1}{2\sqrt 2}i$ & 0 
& -$\frac {1}{2\sqrt 2}i$ & 0  \\
$\phi_{\rm fss}^+ \sigma_3(34) \phi_{\rm iss}$ & 0 & 0 & -$\frac {1}{2}$ & 0 
& -$\frac {1}{2}$ & 0 & -$\frac {1}{2}$ & 0 & 0  \\
\hline
$\phi_{\rm fss}^+ \sigma_1(21) \sigma_1(34) \phi_{\rm iss}$ & -$\frac {1}{2}$ 
& 0 & 0 & 0 & 0 & 0 & 0 & 0 & $\frac {1}{2}$  \\
$\phi_{\rm fss}^+ \sigma_1(21) \sigma_2(34) \phi_{\rm iss}$ & -$\frac {1}{2}i$
& 0 & 0 & 0 & -$\frac {1}{2}i$ & 0 & 0 & 0 & -$\frac {1}{2}i$  \\
$\phi_{\rm fss}^+ \sigma_1(21) \sigma_3(34) \phi_{\rm iss}$ & 0 
& -$\frac {1}{2\sqrt 2}$ & 0 & -$\frac {1}{2\sqrt 2}$ & 0 
& -$\frac {1}{2\sqrt 2}$ & 0 & -$\frac {1}{2\sqrt 2}$ & 0  \\
$\phi_{\rm fss}^+ \sigma_2(21) \sigma_1(34) \phi_{\rm iss}$ & -$\frac {1}{2}i$ 
& 0 & 0 & 0 & $\frac {1}{2}i$ & 0 & 0 & 0 & -$\frac {1}{2}i$  \\
$\phi_{\rm fss}^+ \sigma_2(21) \sigma_2(34) \phi_{\rm iss}$ & $\frac {1}{2}$ 
& 0 & 0 & 0 & 0 & 0 & 0 & 0 & -$\frac {1}{2}$  \\
$\phi_{\rm fss}^+ \sigma_2(21) \sigma_3(34) \phi_{\rm iss}$ & 0 
& -$\frac {1}{2\sqrt 2}i$ & 0 & -$\frac {1}{2\sqrt 2}i$ & 0
& $\frac {1}{2\sqrt 2}i$ & 0 & $\frac {1}{2\sqrt 2}i$ & 0  \\
$\phi_{\rm fss}^+ \sigma_3(21) \sigma_1(34) \phi_{\rm iss}$ & 0 
& $\frac {1}{2\sqrt 2}$ & 0 & -$\frac {1}{2\sqrt 2}$ & 0 
& -$\frac {1}{2\sqrt 2}$ & 0 & $\frac {1}{2\sqrt 2}$ & 0  \\
$\phi_{\rm fss}^+ \sigma_3(21) \sigma_2(34) \phi_{\rm iss}$ & 0 
& $\frac {1}{2\sqrt 2}i$ & 0 & -$\frac {1}{2\sqrt 2}i$ & 0
& $\frac {1}{2\sqrt 2}i$ & 0 & -$\frac {1}{2\sqrt 2}i$ & 0  \\
$\phi_{\rm fss}^+ \sigma_3(21) \sigma_3(34) \phi_{\rm iss}$ & 0 & 0
& $\frac {1}{2}$ & 0 & 0 & 0 & -$\frac {1}{2}$ & 0 & 0  \\
\hline
\end{tabular}
\end{table}

\newpage
\begin{table}
\caption{\label{table14} The same as Table 2 except for $S_D=0$, $S_{Az}=-1$, 
and $S_A=S_B=S_C=1$.}
\begin{tabular}{llllllllll}
\hline
$S_{Bz}$ & -1 & -1 & -1 & 0  & 0 & 0 &  1 & 1 & 1 \\
$S_{Cz}$ & -1 & 0  & 1  & -1 & 0 & 1 & -1 & 0 & 1 \\
\hline
$\phi_{\rm fss}^+ \phi_{\rm iss}$ & 0 & -$\frac {1}{2}$ & 0 & $\frac {1}{2}$ 
& 0 & 0 & 0 & 0 & 0 \\
$\phi_{\rm fss}^+ \sigma_1(21) \phi_{\rm iss}$ & 0 & 0 & 0 & 0 
& -$\frac {1}{2\sqrt 2}$ & 0 & $\frac {1}{\sqrt 2}$ & 0 & 0  \\
$\phi_{\rm fss}^+ \sigma_2(21) \phi_{\rm iss}$ & 0 & 0 & 0 & 0 
& $\frac {1}{2\sqrt 2}i$ & 0 & -$\frac {1}{\sqrt 2}i$ & 0 & 0  \\
$\phi_{\rm fss}^+ \sigma_3(21) \phi_{\rm iss}$ & 0 & $\frac {1}{2}$ & 0
& -$\frac {1}{2}$ & 0 & 0 & 0 & 0 & 0  \\
$\phi_{\rm fss}^+ \sigma_1(34) \phi_{\rm iss}$ & -$\frac {1}{\sqrt 2}$ & 0 & 0
& 0 & $\frac {1}{2\sqrt 2}$ & 0 & 0 & 0 & 0  \\
$\phi_{\rm fss}^+ \sigma_2(34) \phi_{\rm iss}$ & -$\frac {1}{\sqrt 2}i$ & 0 & 0
& 0 & -$\frac {1}{2\sqrt 2}i$ & 0 & 0 & 0 & 0  \\
$\phi_{\rm fss}^+ \sigma_3(34) \phi_{\rm iss}$ & 0 & -$\frac {1}{2}$
& 0 & -$\frac {1}{2}$ & 0 & 0 & 0 & 0 & 0  \\
\hline
$\phi_{\rm fss}^+ \sigma_1(21) \sigma_1(34) \phi_{\rm iss}$
& 0 & 0 & 0 & -$\frac {1}{2}$ & 0 & 0 & 0 & $\frac {1}{2}$ & 0 \\
$\phi_{\rm fss}^+ \sigma_1(21) \sigma_2(34) \phi_{\rm iss}$ & 0 & 0 & 0
& -$\frac {1}{2}i$ & 0 & 0 & 0 & -$\frac {1}{2}i$ & 0  \\
$\phi_{\rm fss}^+ \sigma_1(21) \sigma_3(34) \phi_{\rm iss}$ & 0 & 0 & 0 & 0
& -$\frac {1}{2\sqrt 2}$ & 0 & -$\frac {1}{\sqrt 2}$ & 0 & 0  \\
$\phi_{\rm fss}^+ \sigma_2(21) \sigma_1(34) \phi_{\rm iss}$ & 0 & 0 & 0
& $\frac {1}{2}i$ & 0 & 0 & 0 & -$\frac {1}{2}i$ & 0  \\
$\phi_{\rm fss}^+ \sigma_2(21) \sigma_2(34) \phi_{\rm iss}$ & 0 & 0 & 0
& -$\frac {1}{2}$ & 0 & 0 & 0 & -$\frac {1}{2}$ & 0  \\
$\phi_{\rm fss}^+ \sigma_2(21) \sigma_3(34) \phi_{\rm iss}$ & 0 & 0 & 0 & 0
& $\frac {1}{2\sqrt 2}i$ & 0 & $\frac {1}{\sqrt 2}i$ & 0 & 0  \\
$\phi_{\rm fss}^+ \sigma_3(21) \sigma_1(34) \phi_{\rm iss}$ 
& $\frac {1}{\sqrt 2}$ & 0 & 0 & 0 & -$\frac {1}{2\sqrt 2}$ & 0 & 0 & 0 & 0  \\
$\phi_{\rm fss}^+ \sigma_3(21) \sigma_2(34) \phi_{\rm iss}$ 
& $\frac {1}{\sqrt 2}i$ & 0 & 0 & 0 & $\frac {1}{2\sqrt 2}i$ 
& 0 & 0 & 0 & 0 \\
$\phi_{\rm fss}^+ \sigma_3(21) \sigma_3(34) \phi_{\rm iss}$ & 0
& $\frac {1}{2}$ & 0 & $\frac {1}{2}$ & 0 & 0 & 0 & 0 & 0  \\
\hline
\end{tabular}
\end{table}

\newpage
\begin{table*}
\caption{\label{table15} Flavor matrix elements.}
\begin{tabular}{ccc}
\hline
Channel & ${\cal M}_{{\rm a}q_1\bar{q}_2\rm f}$ & 
${\cal M}_{{\rm a}\bar{q}_1 q_2\rm f}$ \\
\hline
$I=1~ \pi \pi \to \rho \rho$ & 1 & 1  \\
$I=0~ \pi \pi \to \rho \rho$ & $\frac{3}{2}$ & $\frac{3}{2}$  \\
$I=1~ K \bar {K} \to K^* \bar {K}^\ast$ & 0 & 1  \\
$I=0~ K \bar {K} \to K^* \bar {K}^\ast$ & 2 & 1  \\
$I=1~ K \bar {K}^\ast \to K^* \bar {K}^\ast$ & 0 & 1  \\
$I=0~ K \bar {K}^\ast \to K^* \bar {K}^\ast$ & 2 & 1  \\
$I=1~ K^* \bar {K} \to K^* \bar {K}^\ast$ & 0 & 1  \\
$I=0~ K^* \bar {K} \to K^* \bar {K}^\ast$ & 2 & 1  \\
$I=1~ \pi \pi \to K \bar K$ & 0 & -1  \\
$I=0~ \pi \pi \to K \bar K$ & 0 & -$\frac {\sqrt 6}{2}$  \\
$I=1~ \pi \rho \to K \bar {K}^\ast$ & 0 & -1  \\
$I=0~ \pi \rho \to K \bar {K}^\ast$ & 0 & -$\frac {\sqrt 6}{2}$  \\
$I=1~ \pi \rho \to K^* \bar {K}$ & 0 & -1  \\
$I=0~ \pi \rho \to K^* \bar {K}$ & 0 & -$\frac {\sqrt 6}{2}$  \\
$I=1~ K \bar {K} \to \rho \rho$ & 0 & -1  \\
$I=0~ K \bar {K} \to \rho \rho$ & 0 & -$\frac {\sqrt 6}{2}$  \\
\hline
\end{tabular}
\end{table*}

\newpage
\begin{table*}
\caption{\label{table16}Values of the parameters. $a_1$ and $a_2$ are 
in units of millibarns; $b_1$, $b_2$, $d_0$, and $\sqrt{s_{\rm z}}$ are 
in units of GeV; $c_1$ and $c_2$ are dimensionless.}
\tabcolsep=5pt
\begin{tabular}{cccccccccc}
  \hline
  \hline
Reactions & $T/T_{\rm c} $ & $a_1$ & $b_1$ & $c_1$ & $a_2$ & $b_2$ & $c_2$ &
$d_0$ & $\sqrt{s_{\rm z}} $\\ 
\hline
 $I=1~\pi\pi\to\rho\rho$
  &  0     & 0.19  & 1.4  & 4.2   & 1.08  & 0.35  & 0.67  & 0.35 & 7.76\\
  &  0.65  & 0.11  & 1.3  & 6.8   & 0.2   & 0.27  & 0.54  & 0.25 & 5.05\\
  &  0.75  & 0.042 & 1.46 & 10.55 & 0.086 & 0.34  & 0.51  & 0.25 & 4.57\\
  &  0.85  & 0.015 & 1.69 & 12.6  & 0.027 & 0.35  & 0.48  & 0.3  & 4.13\\
  &  0.9   & 0.0085 & 1.84 & 9.13  & 0.0166& 0.3   & 0.45  & 0.2  & 4.01\\
  &  0.95  & 0.0089& 1.9  & 6.9   & 0.022 & 0.29  & 0.48  & 0.25 & 3.87\\
  \hline
 $ I=1~K\bar{K}\to K^* \bar{K}^{*} $
  &  0     & 0.95  & 0.13  & 0.53 & 0.17  & 0.48 & 1.07 & 0.15   & 3.81\\
  &  0.65  & 0.6   & 0.142 & 0.52 & 0.133 & 0.79 & 4.54 & 0.15   & 3.28\\
  &  0.75  & 0.049 & 1.02  & 9.24 & 0.349 & 0.2  & 0.49 & 0.2    & 3.19\\
  &  0.85  & 0.109 & 0.145 & 0.47 & 0.068 & 1.12 & 6.27 & 0.125  & 3.19\\
  &  0.9   & 0.121 & 0.069 & 0.45 & 0.082 & 1.28 & 1.39 & 0.071  & 3.15\\
  &  0.95  & 0.181 & 0.052 & 0.43 & 0.158 & 5.89 & 0.5  & 0.0418 & 3.03\\
  \hline
 $ I=0~K\bar{K}\to K^{*} \bar{K}^{*} $
  &  0     & 1.74 & 1.3   & 0.42 & 4.71 & 0.186 & 0.55 & 0.2    & 10.36\\
  &  0.65  & 1.45 & 0.54  & 0.99 & 1.87 & 0.09  & 0.51 & 0.15   &  9.22\\
  &  0.75  & 1.42 & 0.155 & 0.51 & 0.47 & 0.98  & 4.04 & 0.175  &  8.4\\
  &  0.85  & 0.2  & 0.83  & 2.74 & 0.42 & 0.1   & 0.49 & 0.1126 &  4.73\\
  &  0.9   & 0.38 & 0.051 & 0.61 & 0.27 & 0.24  & 0.31 & 0.0418 &  3.3\\
  &  0.95  & 0.75 & 0.026 & 0.55 & 0.51 & 0.132 & 0.33 & 0.025  &  2.77\\
  \hline
  \hline
\end{tabular}
\end{table*}

\newpage
\begin{table*}
\caption{\label{table17}The same as Table 16, but for three other reactions.}
\tabcolsep=5.3pt
\begin{tabular}{ccccccccccc}
  \hline
  \hline
  Reactions & $T/T_{\rm c} $ & $a_1$ & $b_1$ & $c_1$ & $a_2$ & $b_2$ & $c_2$ &
  $d_0$  & $\sqrt{s_{\rm z}} $\\
  \hline
  $I=1~K\bar{K}^{*}\to K^{*} \bar{K}^{*} $
  &  0     & 1.4  & 0.08  & 0.8  & 2.19  & 0.13  & 0.45 & 0.1    & 3.52\\
  &  0.65  & 0.36 & 0.87  & 4.82 & 1.08  & 0.11  & 0.49 & 0.1    & 3.33\\
  &  0.75  & 0.6  & 0.11  & 0.46 & 0.29  & 1.01  & 6.85 & 0.075  & 3.28\\
  &  0.85  & 0.33 & 0.056 & 0.43 & 0.173 & 1.35  & 1.56 & 0.0418 & 3.18\\
  &  0.9   & 0.48 & 0.041 & 0.4  & 0.22  & 2.69  & 0.9  & 0.025  & 3.16\\
  &  0.95  & 0.61 & 0.036 & 0.36 & 0.29  & 3.67  & 0.92 & 0.025  & 3.02\\
  \hline
  $I=0~K\bar{K}^{*}\to K^{*} \bar{K}^{*} $
   & 0     & 1.4 & 0.9 & 2.65 & 8.9 & 0.1 & 0.5 & 0.1 & 7.96\\
   & 0.65  & 0.35 & 0.05 & 0.72 & 2.5 & 0.13 & 0.45 & 0.1 & 5.6\\
   & 0.75  & 0.1 & 0.61 & 4.58 & 1.64 & 0.1 & 0.47 & 0.1 & 4.15\\
   & 0.85  & 1.08 & 0.06 & 0.44 & 0.09 & 0.5 & 2.15 & 0.05 & 2.46\\
   & 0.9   & 0.79 & 0.15 & 1.49 & 2.04 & 0.019 & 0.47 & 0.025 & 1.99\\
   & 0.95  & 0.78 & 0.18 & 0.63 & 3.6 & 0.017 & 0.41 & 0.0126 & 1.62\\
  \hline
  $I=1~\pi\pi\to K\bar{K}$
   & 0     & 0.02   & 0.068 & 0.82 & 0.088  & 0.29 & 2.83 & 0.3   & 2.26\\
   & 0.65  & 0.03   & 0.06  & 0.8  & 0.13   & 0.28 & 2.76 & 0.28  & 2.1\\
   & 0.75  & 0.04   & 0.11  & 0.8  & 0.09   & 0.3  & 3.5  & 0.25  & 9.1\\
   & 0.85  & 0.064  & 0.3   & 3.89 & 0.031  & 0.11 & 0.76 & 0.219 & 7.26\\
   & 0.9   & 0.021  & 0.092 & 0.74 & 0.051  & 0.3  & 3.72 & 0.28  & 6.32\\
   & 0.95  & 0.0136 & 0.038 & 0.75 & 0.0407 & 0.29 & 2.83 & 0.3   & 6.17\\
   \hline
   \hline
\end{tabular}
\end{table*}

\newpage
\begin{table*}
\caption{\label{table18}The same as Table 16, but for three other reactions.}
\begin{tabular}{cccccccccc}
  \hline
  \hline
  Reactions & $T/T_{\rm c} $ & $a_1$ & $b_1$ & $c_1$ & $a_2$ & $b_2$ & $c_2$ &
  $d_0$ & $\sqrt{s_{\rm z}} $\\ 
  \hline
  $I=1~\pi \rho\to K\bar{K}^{*}$
  &  0     & 0.1    & 0.96 & 0.71 & 0.66  & 0.16 & 0.64 & 0.175 & 4.08\\
  &  0.65  & 0.048  & 0.62 & 0.98 & 0.17  & 0.12 & 0.52 & 0.125 & 3.81\\
  &  0.75  & 0.02   & 0.43 & 0.38 & 0.064 & 0.16 & 0.57 & 0.15  & 3.69\\
  &  0.85  & 0.004 & 0.54 & 0.27 & 0.027 & 0.25 & 0.84 & 0.25  & 3.36\\
  &  0.9   & 0.0076 & 0.33 & 0.38 & 0.024 & 0.25 & 0.92 & 0.25  & 3.15\\
  &  0.95  & 0.008  & 0.22 & 1.21 & 0.053 & 0.22 & 0.49 & 0.2   & 3.05\\
  \hline
  $I=1~\pi\rho\to K^*\bar{K}$
   & 0     & 0.37   & 0.17  & 1.34 & 0.32  & 0.36 & 0.54 & 0.25  & 4.67\\
   & 0.65  & 0.034  & 0.12  & 1.01 & 0.065 & 0.35 & 0.47 & 0.2   & 4.05\\
   & 0.75  & 0.015  & 0.09  & 0.5  & 0.024 & 0.5  & 0.71 & 0.2   & 3.86\\
   & 0.85  & 0.0074 & 0.21  & 0.49 & 0.004 & 1.21 & 8.98 & 0.2   & 3.41\\
   & 0.9   & 0.0045 & 0.19  & 0.48 & 0.003  & 1.29 & 6.6  & 0.2   & 3.25\\
   & 0.95  & 0.0062 & 0.162 & 0.48 & 0.0038 & 1.39 & 4.95 & 0.15  & 3.18\\
  \hline
  $I=1~K \bar{K}\to \rho \rho$
   & 0     & 0.77   & 0.08   & 0.53 & 0.49   & 0.64 & 1.29 & 0.11 & 5.99\\
   & 0.65  & 0.175  & 0.133  & 0.49 & 0.108  & 1.27 & 3.6  & 0.13 & 5.52\\
   & 0.75  & 0.112  & 0.065  & 0.44 & 0.056   & 1.08 & 1.5  & 0.05 & 5.28\\
   & 0.85  & 0.021  & 0.116  & 0.45 & 0.0173  & 1.55 & 3.44 & 0.1  & 4.83\\
   & 0.9   & 0.0117 & 0.15   & 0.47 & 0.0122 & 1.65 & 2.89 & 0.15  & 4.77\\
   & 0.95  & 0.0068  & 0.08  & 0.47 & 0.0118 & 1.2  & 0.79 & 0.3 & 4.7\\
   \hline
   \hline
\end{tabular}
\end{table*}

\end{document}